\documentclass[11pt,a4paper]{article}
\usepackage{jheppub,bm,booktabs,multirow}

\usepackage{amsmath}
\allowdisplaybreaks[4]
\usepackage{amssymb}
\usepackage{graphicx}
\usepackage{booktabs}
\usepackage{bm}
\usepackage{psfrag}
\usepackage[normalem]{ulem}
\usepackage{color}
\usepackage{overpic}
\usepackage[utf8x]{inputenc}

\makeatletter
\def\@fpheader{~}
\makeatother

\usepackage[Q=yes,pverb-linebreak=no]{examplep}

\def\as{\alpha_s}
\def\e{\epsilon}
\def\nno{\nonumber}
\def\nb{\bar{n}}
\def\nslash{\rlap{\hspace{0.02cm}/}{n}}
\def\nbslash{\rlap{\hspace{0.02cm}/}{\bar n}}

\title{Factorization and Resummation for Jet Processes}
\author[a]{Thomas Becher,}
\author[b,c]{Matthias Neubert,}
\author[d]{Lorena Rothen,}
\author[a]{and Ding Yu Shao}
\affiliation[a]{Albert Einstein Center for Fundamental Physics, Institut f\"ur Theoretische Physik, Universit\"at Bern,
  Sidlerstrasse 5, CH-3012 Bern, Switzerland}
\affiliation[b]{PRISMA Cluster of Excellence \& Mainz Institute for Theoretical Physics, 
Johannes Gutenberg University, 55099 Mainz, Germany}
\affiliation[c]{Department of Physics, LEPP, Cornell University, Ithaca, NY 14853, U.S.A.}
\affiliation[d]{Theory Group, Deutsches Elektronen-Synchrotron (DESY), Notkestrasse 85, D-22607 Hamburg, Germany}

\emailAdd{becher@itp.unibe.ch}
\emailAdd{matthias.neubert@uni-mainz.de}
\emailAdd{lorena.rothen@desy.de}
\emailAdd{shao@itp.unibe.ch}

\date{\today}

\preprint{\begin{flushright}
DESY\phantom{/}16-052\\
MITP/16-042
\end{flushright}}

\abstract
{From a detailed analysis of cone-jet cross sections in effective field theory, we obtain novel factorization theorems which separate the physics associated with different energy scales present in such processes. The relevant low-energy physics is encoded in Wilson lines along the directions of the energetic particles inside the jets. This multi-Wilson-line structure is present even for narrow-cone jets due to the relevance of small-angle soft radiation. We discuss the renormalization-group equations satisfied by these operators. Their solution resums all logarithmically enhanced contributions to such processes, including non-global logarithms. Such logarithms arise in many observables, in particular whenever hard phase-space constraints are imposed, and are not captured with standard resummation techniques. Our formalism provides the basis for higher-order logarithmic resummations of jet and other non-global observables. As a nontrivial consistency check, we use it to obtain explicit two-loop results for all logarithmically enhanced terms in cone-jet cross sections and verify those against numerical fixed-order computations.}

\begin{document}
\maketitle

\section{Introduction\label{sec:introduction}}

A crucial ingredient for the factorization of high-energy processes is the simple form of soft emissions. Their eikonal structure forms the basis of the resummation of soft-photon effects in QED \cite{Bloch:1937pw,Yennie:1961ad,Weinberg:1965nx} and proved to be equally relevant when analyzing factorization in QCD \cite{Ciafaloni:1981nm,Collins:1981uk,Bassetto:1984ik}. In position space, soft emissions are described by Wilson lines along the classical trajectories of the energetic particles \cite{Korchemsky:1985xj,Korchemsky:1993uz} and the renormalization properties of the corresponding Wilson-line matrix elements \cite{Polyakov:1980ca,Brandt:1981kf,Korchemskaya:1992je,Becher:2009cu,Becher:2009qa,Dixon:2009ur} play an important role when resumming soft-gluon effects.

An important property of soft radiation is that wide-angle radiation off a jet of collinear particles does not resolve the individual energetic partons. Instead of probing the charge distribution inside the jet, the long-wavelength soft radiation is sensitive only to the total charge, because the eikonal factors associated with the collinear momenta $p^\mu_i$ can be expanded around a common 
reference vector $n^\mu$ pointing along the jet direction,
\begin{equation}\label{eikExp}
\frac{p_i \cdot \epsilon}{p_i \cdot k} \approx  \frac{n \cdot \epsilon}{n \cdot k} \,,
\end{equation}
where $k$ and $\epsilon$ are the momentum and polarization of the emitted soft gluon. This approximation fails for a soft gluon which is collinear to the jet, but typically this small region of phase space does not give an unsuppressed contribution to the cross section. As long as this is true, the soft function for a dijet cross section involves only two Wilson lines, accounting for soft radiation emitted from the two jets.  The cross section then factorizes into a product (in the appropriate space; more generally a convolution) of this soft function $S$, jet functions $J$ and $\overline{J}$  for the collinear radiation inside the two jets, and a hard function $H$ encoding the virtual corrections to the production of the two leading partons:
\begin{equation}\label{HJJS}
\sigma = H \cdot  J \cdot \overline{J} \cdot S\,.
\end{equation}
Computations based on this structure were successfully used to resum large logarithms in many event-shape variables, in particular thrust, heavy jet mass or the $C$-parameter \cite{Dasgupta:2003iq,Luisoni:2015xha}. In the meantime, using renormalization-group (RG) methods in Soft-Collinear Effective Theory (SCET) \cite{Bauer:2000yr,Bauer:2001yt,Beneke:2002ph} (see \cite{Becher:2014oda} for a review), the resummations for selected observables has been carried out up to next-to-next-to-next-to-leading logarithmic accuracy \cite{Becher:2008cf,Chien:2010kc,Hoang:2014wka}. 

However, it was found that certain observables contain single-logarithmic terms which are not captured by resummation techniques based on \eqref{HJJS}. The classic example is the light-jet mass event shape in $e^+e^-$ collisions, or equivalently, the jet mass in a single hemisphere. In \cite{Dasgupta:2001sh}, Dasgupta and Salam traced the problem back to the fact that these observables are insensitive to radiation in certain regions of phase space (the right hemisphere in the example of the left-hemisphere mass). The additional logarithms first appear at two-loop order and are referred to as non-global logarithms (NGLs). From an analysis of soft emissions, Dasgupta and Salam extracted the leading two-loop logarithm analytically and gave an algorithm to numerically compute the leading higher-order logarithms in the large-$N_c$ limit. Based on the simple form of amplitudes in the strongly energy-ordered limit (see e.g.\ \cite{Bassetto:1984ik,Schwartz:2014wha}), Banfi, Marchesini and Smye (BMS) derived a non-linear integral equation which can be used to perform the resummation of the leading NGLs in the large-$N_c$ limit \cite{Banfi:2002hw}. However, since strong energy ordering is a crucial ingredient for the BMS equation, its logarithmic accuracy cannot easily be increased. What has been achieved is leading-logarithmic resummation of these logarithms for $N_c=3$ \cite{Weigert:2003mm,Hatta:2013iba}. It turns out that the corrections to the large-$N_c$ limit are numerically quite small \cite{Hagiwara:2015bia}.

Since the vast majority of collider observables include hard phase-space cuts or, more generally, regions of phase space in which radiation is not restricted, the presence of NGLs severely limits the applicability of higher-order resummation techniques. For this reason, a lot of effort was put into trying to get a better understanding of these types of logarithms. For example, several groups computed hemisphere soft functions up to two-loops to obtain the full result for the non-global structure at this order \cite{Kelley:2011ng,Hornig:2011iu,Kelley:2011aa,vonManteuffel:2013vja}. Also, using the BMS equation, the analytic result for the leading-logarithmic terms up to five-loop order was extracted \cite{Schwartz:2014wha}. This analysis was also extended beyond the large-$N_c$ limit by computing the higher-order terms directly from strongly-ordered soft amplitudes \cite{Khelifa-Kerfa:2015mma}. While these fixed-order computations provide important insights into the form of NGLs, ultimately one is interested in their all-order structure. Steps towards a resummation of such terms were recently taken in \cite{Larkoski:2015zka,Neill:2015nya}. The authors claim that the NGLs arise from soft subjets near phase-space boundaries and propose a set of factorization theorems which resum global logarithms in the presence of subjets. They then argue that this resummation will capture a large part of the NGLs in more inclusive cross sections. They propose an expansion in the number of soft subjets, which they call ``dressed gluons''. Since the dressed gluons include Sudakov factors, the expansion in dressed gluons does not suffer from the same divergence as standard perturbation theory when the logarithms become large. However, an arbitrary number of soft subjets contributes even to the leading NGLs, and it is not clear what expansion parameter governs the expansion in subjets and whether there is any parametric suppression of the higher-multiplicity terms.\footnote{We were informed by an author of \cite{Larkoski:2015zka} that a paper addressing these questions is in preparation.} 
From a numerical point of view (see Figure~8 in \cite{Larkoski:2015zka}), the expansion in dressed gluons appears to provide only a modest improvement over a pure fixed-order treatment for moderately large values of the logarithms.  A second interesting proposal to go beyond leading-logarithmic resummation is the functional RG of Caron-Huot \cite{Caron-Huot:2015bja}. We will comment in more detail on both approaches and their relation to our results below. 

An important example of non-global observables are jet cross sections, in particular those involving cone jets, which are insensitive to radiation inside the jet cone. In the present paper we analyze dijet cross sections in $e^+ e^-$ collisions. In addition to the narrow-jet case analyzed in our previous work \cite{Becher:2015hka}, we also treat the case where the opening angle of the jet cone is large. For brevity, we will refer to these as ``wide-angle jets''. We find that in both situations, the simple factorization theorem \eqref{HJJS} is {\em incorrect}. This is immediately obvious for wide-angle jets since the jet opening angle is as large as the typical angle of the soft radiation. The approximation \eqref{eikExp} is therefore not appropriate and each hard parton inside the jet produces its own Wilson line. In the next section we show that the relevant factorization theorem for the cross section takes the form
\begin{equation}\label{wide}
\sigma=  \sum_{m=2}^\infty \big\langle \bm{\mathcal{H}}_{m}\otimes \bm{\mathcal{S}}_{m} \big\rangle\,.
\end{equation}
The function $\bm{\mathcal{H}}_{m}$ is the squared amplitude for having $m$ particles inside the two jets, integrated over their energies but at fixed angles. The function $\bm{\mathcal{S}}_{m}$ contains soft Wilson lines along the directions of the $m$ hard partons. The symbol $\otimes$ indicates an integral over the angles specifying these directions. The functions $\bm{\mathcal{H}}_{m}$ and $\bm{\mathcal{S}}_{m}$ are matrices in the color space of the $m$ hard partons and the angle brackets indicate that one has to take the trace of their product to get the cross section. One could naively expect to recover the form \eqref{HJJS} in the narrow-jet case, but this turns out not to be true because small-angle soft radiation gives a leading-power contribution to the cross section. The relevance of a new mode describing small-angle soft radiation was demonstrated in \cite{Becher:2015hka,Chien:2015cka}. Since the radiation is simultaneously collinear and soft, we call it ``coft'' to distinguish it from the former two types of radiation. In the narrow-jet case we find that, for $k$ energetic partons inside the left jet and $l$ inside the right one, the additional factorization
\begin{equation}\label{facHS}
\bm{\mathcal{H}}_{k+l} = H\cdot  \bm{\mathcal{J}}_{\!\!l} \cdot \overline{\bm{\mathcal{J}}}_{\!\!k}\,, \qquad
\widetilde{\bm{\mathcal{S}}}_{k+l} = \widetilde{S}\cdot \,\widetilde{\bm{\mathcal{U}}}_l \cdot \,\widetilde{\overline{\bm{\mathcal{U}}}}_k\,
\end{equation}
is valid, up to corrections suppressed by the small cone angle. The first equation is the usual collinear factorization, the statement that the hard function factorizes into a hard function for the two jets times collinear splitting functions $\bm{\mathcal{J}}_{\!\!l}$ and $\overline{\bm{\mathcal{J}}}_{\!\!k}$ for the partons inside the jets. The second equation states that the soft radiation splits up into wide-angle soft radiation consisting of two Wilson lines and coft radiation in the direction of the left or right jet, which resolves the individual energetic partons inside the jets. The tildes in this equation indicate that the product form holds in Laplace space; in momentum space the factorization would involve a convolution. The coft function $\bm{\mathcal{U}}_l$ contains $l$ Wilson lines from particles inside the right jet and a single Wilson line along the left jet, and analogously for $\widetilde{\overline{\bm{\mathcal{U}}}}_k$. We use the method of regions \cite{Beneke:1997zp,Smirnov:2002pj} and consistently expand both the amplitudes and the phase-space constraints in small momentum components to avoid double counting between different momentum modes. The expansion of the phase-space constraints is an important ingredient of our approach, which leads to complete scale separation and avoids the necessity of performing zero-bin subtractions.  

The renormalization of the operators $\bm{\mathcal{S}}_{m}$ is quite interesting and nontrivial \cite{Becher:2015hka}.  The factorization theorem \eqref{wide} splits the cross section into a sum of ingredients with a fixed number of hard partons. Scattering amplitudes with a fixed number of partons are not finite, because the virtual corrections to lower multiplicities are needed to cancel the infrared (IR) divergences of the real-emission amplitudes. In the effective field-theory framework we are using, the IR divergences of the hard functions are in one-to-one correspondence to the ultraviolet (UV) divergences of the Wilson-line matrix elements $\bm{\mathcal{S}}_{m}$, see e.g.\ \cite{Becher:2009cu,Becher:2009qa}. The fact that the cancellation of divergences involves different multiplicities then implies that the renormalization matrix $\bm{Z}_{km}$ which absorbs the divergences of  $\bm{\mathcal{S}}_{m}$ is not diagonal, or more specifically, that higher-multiplicity Wilson-line operators mix into lower-multiplicity ones under renormalization. The reason for this mixing are UV divergences arising from soft emissions inside the jets, which are unconstrained. The scale dependence of the different components of the factorization theorem is driven by anomalous dimensions obtained from the respective $Z$-factors. Solving the relevant RG equations resums all large logarithms in cone-jet processes \cite{Becher:2015hka}. 

The focus of the present paper is on the derivation of the factorization theorems and the RG equations which govern its ingredients. A large fraction of our paper is devoted to computing the ingredients for the next-to-next-to-leading (NNLO) jet cross section in both the narrow-jet and wide-angle cases. These computations provide us with nontrivial consistency conditions on our results and show that our formalism correctly predicts all logarithmic terms. Moreover, we establish that RG evolution in the  effective theory reproduces the known results at leading logarithmic accuracy in the large-$N_c$ limit. A detailed study of the RG equation and its solution beyond this accuracy is left for future work. Given that all jet observables as well as most other experimental measurements are non-global, there are many potential applications for our framework, once methods have been developed to solve the associated RG evolution equations.

The structure of the paper is as follows. In Section \ref{sec:factorization}, we analyze the factorization properties of cone-jet cross sections, both in the wide-angle case and for narrow jets. We also discuss the renormalization of the operators which appear in these formulas. As a check of the factorization theorem for narrow jets, we compute in Section \ref{sec:twoloopNarrow} all ingredients to two-loop accuracy. The cancellation of divergences provides an important consistency check and we verify that our analytical result is compatible with numerical fixed-order results.  In Section \ref{sec:twoLoopWide}, we perform the analogous two-loop analysis for wide-angle jets. We again check our results against numerical computations and then verify their factorization properties in the limit small-angle limit. We give explicit results for the one-loop anomalous dimensions matrices which govern the resummation in Section \ref{sec:renorm}. We also verify that for the leading NGLs at large $N_c$ our results reduce to the ones obtained from the BMS equation. After a short discussion of methods to solve the RG equations and a comparison to the approaches of \cite{Caron-Huot:2015bja,Larkoski:2015zka}, we conclude. Some lengthy two-loop expressions are relegated into appendices. Appendix \ref{NarrowCone} provides expressions relevant for the narrow-cone case, while those for the wide-angle case are given in Appendix \ref{widejetapp}.

\section{Factorization of jet cross sections\label{sec:factorization}}

In this section we will derive a factorization formula for the cross section for $e^+ e^- \to 2\, {\rm jets}$ at center-of-mass energy $\sqrt{s}=Q$. We use the thrust axis, defined as the unit-vector $\vec{n}$ in the direction of maximum momentum flow, as the jet axis and define two light-like vectors $n^\mu=(1,\vec{n})$ and $\bar{n}^\mu=(1,-\vec{n})$ along the jets.
 Using these vectors, we can rewrite any four-momentum as
\begin{equation}
p^{\mu}=\bar{n}\cdot p\,\frac{n^{\mu}}{2}+n\cdot p\,\frac{\bar{n}^{\mu}}{2}+p_{\perp}^{\mu} \,.
\end{equation}
We use the thrust axis to split each event in two hemispheres and call particles with $n\cdot p<\bar{n}\cdot p $ right-moving. The definition of the thrust axis implies that the total transverse momentum in each hemisphere vanishes.  Particles are considered to be part of the right jet if $n\cdot p<\delta^2\,\bar{n}\cdot p$, where the parameter $\delta$ is related to the opening angle $\alpha$ of the jet via
\begin{equation}
\delta=\tan\frac{\alpha}{2} \,,
\end{equation}
see Figure 1. To define the cross section, we impose that the total energy emitted outside the left and right jet cones, fulfills the condition $2 E_{\rm out} < \beta Q$. Except for the choice of the jet axis, our definition is identical to the one in the seminal paper of Sterman and Weinberg \cite{Sterman:1977wj}. Using the thrust vector as the jet axis leads to a simpler form of the phase-space constraints and enables us to use existing two-loop results for the cone-jet soft function obtained in \cite{Kelley:2011aa,vonManteuffel:2013vja}.

\begin{figure}[t!]
\begin{center}
\begin{overpic}[scale=0.6]{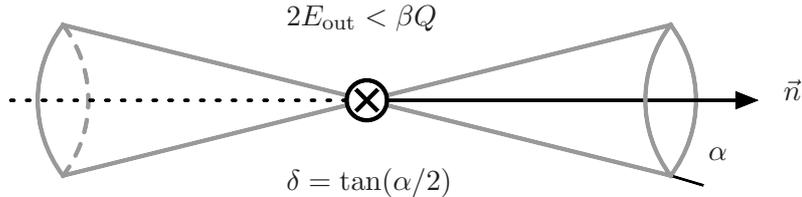}
\put(103,12){$\vec{n}$}
\put(93,4){$\alpha$}
\put(37,0){$\delta=\tan(\alpha/2)$}
\put(37,22){$2E_{\rm out}<\beta Q$}
\end{overpic}
\end{center}
\vspace{-0.3cm}
\caption{Definition of the parameters $\delta$ and $\beta$ of the dijet cross section. We use the thrust axis $\vec{n}$ as the jet axis. \label{fig:cone}}
\end{figure}

\subsection{Wide-angle jets}
\label{subsec:Wideanglejets}

Let us first consider wide-angle jets with $\delta\sim 1$. In this case the effective theory contains only two relevant momentum regions, whose components $(n\cdot p,\bar{n}\cdot p,p_\perp)$ scale as follows:
\begin{equation}
\begin{aligned}
  \mbox{hard:}\quad  p_{h} &\sim \phantom{\beta}Q \,(1,1, 1)  \,,  \\
  \mbox{soft:}\quad     p_{ s} &\sim  Q\beta  \,(1, 1,1) \,.
\end{aligned}
\end{equation}
The hard mode describes the energetic particles inside the jet.  Since we are dealing with wide jets, the energetic radiation inside the jet covers a large angular range. It is thus not collinear to $\vec{n}$ but has a homogenous scaling of all components. Given their large energy, these particles can never go outside the jet, in contrast to the soft partons which can be emitted inside or outside. Since there are no collinear singularities for large cone size, the cross section is single-logarithmic, i.e.\ the leading logarithms have the form $\alpha_s^n \ln^n\! \beta$.

The factorization of an amplitude with $m$ hard partons and an arbitrary number of soft partons is of course well known. Each   hard parton gets dressed with a Wilson line along its direction. For an outgoing particle in the color representation $\bm{T}_i$ propagating along the direction $n_i$, the appropriate Wilson line is given by the path-ordered exponential
\begin{equation}\label{eq:Si}
\bm{S}_i(n_i) = {\bf P} \exp\left( ig_s \int_0^{\infty}\!ds\,
   n_i\cdot A_s^a( s n_i)\,\bm{T}_i^a \right).
\end{equation}
The Wilson line $\bm{S}_i$ is a matrix in color space, which acts on the color index of particle $i$. The operator for the emission from an amplitude with $m$ hard partons then takes the form
\begin{equation}\label{eq:WilsonSoft}
 \bm{S}_1(n_1) \, \bm{S}_2(n_2) \,  \dots\,  {\bm S}_m(n_m)\,|\mathcal{M}_m(\{\underline{p}\})\rangle \,,
\end{equation}
where $n_i^\mu = p_i^\mu/E_i$, and we use the compact notation $\{\underline{p}\}\equiv\{p_1,p_2,\dots, p_m\}$. This equation is analogous to the factorization for amplitudes with coft particles \cite{Becher:2015hka}, but while the coft case involves splitting amplitudes, we are now dealing with ordinary amplitudes $|\mathcal{M}_m(\{\underline{p}\})\rangle$. 
In writing \eqref{eq:WilsonSoft} we use the color-space formalism of \cite{Catani:1996jh,Catani:1996vz}, in which amplitudes are treated as $n$-dimensional vectors in color space. Since they act on different particles, the different generators trivially commute $[\bm{T}_i^a, \bm{T}_j^b]=0$ for $i \neq j$. The same is therefore true for the associated Wilson lines. Note that the gluon fields in the product of Wilson lines are time-ordered, but for brevity, we do not indicate this explicitly. Since they commute, Wilson lines along common directions immediately combine into single Wilson lines, for example
\begin{equation}\label{eq:softcombine}
\bm{S}_1(n)\, \bm{S}_2(n)  =  {\bf P} \exp\left( ig_s \int_0^{\infty}\,ds\,
   n\cdot A_s^a( s n)\,(\bm{T}_1^a + \bm{T}_2^a) \right).
\end{equation}
This property ensures that collinear particles only produce a single Wilson line carrying the total color charge. However, since we deal with large-angle jets, the individual Wilson lines do not combine in our example.

To derive formula \eqref{eq:WilsonSoft} in the effective field theory we introduce a separate collinear field for each of the energetic particles in the final state, i.e.\ we write down the SCET operators for processes with $m$ jets. This is possible since on the amplitude level there is no difference between collinear and hard on-shell particles. The relevant purely collinear SCET Lagrangian consists of $m$ copies of the ordinary QCD Lagrangian. Operators in the effective theory are conveniently expressed in terms of gauge-invariant fields $\chi_i$ and ${\cal A}_{i\perp}^\mu$, which are related to the usual quark and gluon fields via \cite{Hill:2002vw}
\begin{equation}\label{eq:building}
   \chi_i(0) = W_i^\dagger(\bar{n}_i)\,\frac{\nslash_i\nbslash_i}{4}\,
    \psi_i(0) \,, \qquad 
   {\cal A}_{i\perp}^\mu(0) = W_i^\dagger(\bar{n}_i)\,[iD_\perp^\mu\,W_i(\bar{n}_i)] \,.
\end{equation}
The $i$-collinear Wilson lines in the fundamental representation are defined analogously to the soft Wilson lines in \eqref{eq:Si} as
\begin{equation}\label{eq:Whc}
   W_{i}(\bar{n}_i) = {\rm\bf P}\,\exp\left(ig_s\int_{-\infty}^0\!ds\,
   \nb_i\cdot A^a_i(s\nb_i) t^a \right) .
\end{equation}
The argument denotes the direction of the Wilson line, which is conjugate to the direction $n_i$ of the collinear particle. These Wilson lines ensure that these fields are invariant under collinear gauge transformations in each sector \cite{Bauer:2000yr,Bauer:2001yt}. 

At leading order in power counting, $m$-jet operators in this effective theory involve exactly one collinear field $\Phi_i\in \{\chi_i, \bar\chi_i, {\cal A}_{i\perp}^\mu \}$ from each sector $i=1,\dots,m$. Performing the usual decoupling transformation
\begin{equation}
   \Phi_i =  {\bm S}_i(n_i)\,\Phi_i^{(0)} ,
\end{equation}
with the appropriate color representation $\bm{T}_i$ for each field, yields the Wilson-line structure shown in \eqref{eq:WilsonSoft}. Finally, one evaluates the matrix element of the operator with one collinear particle in each sector, using 
\begin{equation}\label{eq:jetfun}
\begin{aligned}
   \langle 0|\,\chi_j^{(0)}(0)\,|p_i\rangle 
   &= \delta_{ij}\,u(p_i) \,, \\
   \langle 0|\,{\cal A}_{j\perp}^{\mu,a(0)}(0)\,|p_i;a\rangle
   &= \delta_{ij}\,\epsilon^\mu(p_i) \,.
\end{aligned}
\end{equation}
Along with the Wilson coefficient of the $m$-jet operator this gives the amplitude $|\mathcal{M}_m(\{\underline{p}\})\rangle$, see \cite{Becher:2009qa} for details. Since the particles are on the mass shell, the higher-order corrections to the relations \eqref{eq:jetfun} are all scaleless and vanish.

To get the amplitude for the emission of $l$ soft partons in the final state with momenta $k_1,\dots,k_l$, one computes the matrix element
\begin{equation}\label{eq:softmatrix}
\langle k_1, \dots, k_l | \, \bm{S}_1(n_1) \, \bm{S}_2(n_2) \,  \dots\,  {\bm S}_m(n_m)\, | 0 \rangle
\end{equation}
of the Wilson-line operator. To obtain the contribution of an arbitrary number of soft partons to the jet cross section, one first defines the squared matrix element for the emissions from $m$ partons as
\begin{equation}
\bm{\mathcal{S}}_m(\{\underline{n}\},Q \beta,\delta) = \int\limits_{X_s}\hspace{-0.55cm}\sum \,\langle  0 |\, \bm{S}_1^\dagger(n_1) \,  \dots\,  {\bm S}_m^\dagger(n_m)\,  |X_s \rangle\langle  X_s | \,\bm{S}_1(n_1) \,  \dots\,  {\bm S}_m(n_m) \, |0 \rangle \, \theta( Q\beta - 2E_{\rm \, out}) \,.\label{eq:Sn}
\end{equation}
This is the same as the coft function which arises for narrow-angle jets \cite{Becher:2015hka}, up to the fact that the constraint now acts on the out-of-cone energy $E_{\rm \, out}$ of the soft radiation, as opposed to $\bar{n}\cdot p_{\rm \, out}$, the large component of the total momentum of the coft fields. Since the soft function depends on the outside energy, it depends on the cone size $\delta$. In terms of the matrix element \eqref{eq:Sn}, the jet cross section takes the form
\begin{align}\label{eq:sigmaWideFirst}
\sigma(\beta,\delta) = \frac{1}{2Q^2}\,\,\sum_{m=2}^\infty \,\prod_{i=1}^m &\int \!\!  \frac{d^{d-1}p_i}{(2\pi)^{d-1} 2 E_i}
 \,\langle  \mathcal{M}_m(\{\underline{p}\}) |\, \bm{\mathcal{S}}_m(\{\underline{n}\}) \, |\mathcal{M}_m(\{\underline{p}\}) \rangle \nonumber \\
&  \times (2\pi)^d \,\delta(Q-E_{\rm tot}) \,\delta^{(d-1)}(\vec{p}_{\rm tot}) \, {\Theta }^{n\bar{n}}_{\rm in}\!\left(\left\{\underline{p}\right\}\right) ,
\end{align}
up to terms suppressed by powers of $\beta$. The integration is over the $m$-dimensional phase-space of the hard partons, which are all constrained to lie inside the two jet cones. The function ${\Theta }_{\rm in}^{n\bar{n}}\left(\left\{\underline{p}\right\}\right)$ ensures that the hard partons are either inside the right jet along the direction $n$  or the left jet along $\bar{n}$. In the narrow-cone case, we will encounter constraints which involve only one of the jets. Note that, due to the multipole expansion, the contribution of soft particles must be neglected in the momentum-conservation $\delta$-functions. 

In order to write the cross section in a more transparent way, we now define hard functions which are obtained by integrating over the energies of the hard particles subject to the constraint that their sum is equal to the center-of-mass energy $Q$, while keeping their directions $n_i^\mu$ fixed,
\begin{align}\label{eq:Hm}
\bm{\mathcal{H}}_m(\{\underline{n}\},Q,\delta) =\frac{1}{2Q^2} \sum_{\rm spins}
\prod_{i=1}^m & \int \! \frac{dE_i \,E_i^{d-3} }{(2\pi)^{d-2}} \, |\mathcal{M}_m(\{\underline{p}\}) \rangle \langle  \mathcal{M}_m(\{\underline{p}\}) |\nonumber \\
&\times (2\pi)^d \,\delta\Big(Q - \sum_{i=1}^m E_i\Big) \,\delta^{(d-1)}(\vec{p}_{\rm tot})\,{\Theta }^{n\bar{n}}_{\rm in}\!\left(\left\{\underline{p}\right\}\right) .
\end{align}
These hard functions are distribution-valued in the angles of the particles, since they contain additional divergences which arise when particles become collinear. These real-emission divergences get cancelled by the divergences associated with the virtual corrections to amplitudes with fewer legs. In contrast, the soft function \eqref{eq:Sn} is regular in the angles. The function $\bm{\mathcal{H}}_2(\{\underline{n}\},Q)=\sigma_0\, H(Q^2) \bm{1}$, where $H(Q^2)=|C_V(-Q^2-i\epsilon)|^2$ is the familiar dijet hard function and $\sigma_0$ the Born-level cross section. The functions $\bm{\mathcal{H}}_{2+k}(\{\underline{n}\})$ are of order $\mathcal{O}(\alpha_s^k)$ since they involve $k$ emissions. The full cross section has the form
\begin{align}\label{sigbarefinal}
\sigma(\beta,\delta) &=  \sum_{m=2}^\infty \big\langle \bm{\mathcal{H}}_m(\{\underline{n}\},Q,\delta) \otimes \bm{\mathcal{S}}_m(\{\underline{n}\},Q \beta,\delta) \big\rangle \,,
\end{align}
where the angular brackets denote the color trace  $\langle M \rangle = \frac{1}{N_c}\, {\rm tr}(M)$. The symbol $ \otimes $ indicates that one has to integrate over the directions of the $m$ hard partons in $\bm{\mathcal{H}}_m(\{\underline{n}\},Q,\delta)$ which are the same as the directions of the Wilson lines in $\bm{\mathcal{S}}_m(\{\underline{n}\},Q\beta,\delta)$, i.e.\
\begin{equation}\label{eq:otimes}
 \bm{\mathcal{H}}_m(\{\underline{n}\},Q,\delta) \otimes \bm{\mathcal{S}}_m(\{\underline{n}\},Q \beta,\delta)  = \prod_{i=2}^m \int \! \frac{d\Omega(n_i)}{4\pi} \,  \bm{\mathcal{H}}_m(\{\underline{n}\},Q,\delta)\, \bm{\mathcal{S}}_m(\{\underline{n}\},Q \beta,\delta)   \,.
\end{equation}

In contrast to the standard formula \eqref{HJJS}, the factorization formula \eqref{sigbarefinal} does not involve jet functions. The reason is that there is no collinear scale in our problem. The collinear matrix elements in \eqref{eq:jetfun} are scaleless and do not receive higher-order corrections. In dimensional regularization, there is thus a cancellation between IR and UV singularities in these functions. When added to the hard functions, the IR divergences in the collinear matrix elements cancel against IR divergences of the hard functions, so that the net effect is to convert the IR divergences of the hard function into UV divergences. It would be possible to separate IR and UV singularities at each step by introducing parton masses or additional subjet resolution parameters into our analysis, however this would only complicate the problem in an unnecessary way.

We have obtained the formula \eqref{sigbarefinal} from an analysis of QCD amplitudes in the hard and soft momentum regions. In the context of a low-energy effective theory, the hard momentum modes are integrated out and the functions $\bm{\mathcal{H}}_m(\{\underline{n}\},Q,\delta)$ are Wilson coefficients of the soft operators $\bm{\mathcal{S}}_m(\{\underline{n}\},Q \beta,\delta)$. Because all components of the soft modes are smaller than their hard counterparts, all hard-soft interactions in the Lagrangian are power suppressed. As in standard SCET, the soft Lagrangian is the same as the QCD Lagrangian. To resum large logarithms of $\beta$, one has to solve the RG equations of the Wilson coefficients. This will be discussed in detail below, after we have analyzed the factorization properties for the narrow-jet case.

\begin{figure}[t!]
\begin{center}
\hspace{-2cm}
\begin{overpic}[scale=0.5]{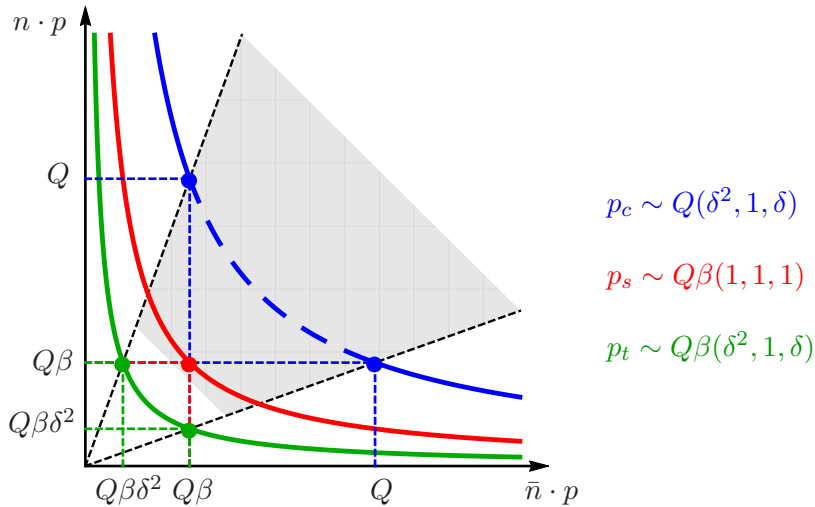}
\put(110,55){\color{blue}{$p_c\sim Q (\delta^2,1,\delta)$}}
\put(110,40){\color{red}{$p_s\sim Q\beta (1,1,1)$}}
\put(110,25){\color[rgb]{0,0.6,0}$p_t\sim Q\beta (\delta^2,1,\delta)$}
\put(61,-5){$Q$}
\put(-6,61){$Q$}
\put(20,-5){$Q\beta$}
\put(-9,22){$Q\beta$}
\put(4,-5){$Q\beta\delta^2$}
\put(-14,9){$Q\beta\delta^2$}
\put(93,-5){$\bar{n}\cdot p$}
\put(-13,93){$n\cdot p$}
\end{overpic}
\end{center}
\vspace{2ex}
\caption{Momentum regions relevant for narrow-angle jet production. The plot shows the scaling of the light-cone components $n\cdot p$ and $\bar{n}\cdot p$, and we assume that $\beta\ll \delta$ (we use $\beta\sim\delta^2$ in the narrow-jet case to ensure this condition). The meshed gray area shows the veto in the out-of-jet region which forbids the presence of energetic modes. In the wide-angle limit $\delta \sim 1$, soft and coft modes coincide and the collinear and hard scalings are the same.}
\label{fig:widejet_mommodes}
\end{figure}

In our derivation of the factorization theorem we first analyzed the factorization properties on the amplitude level and then computed the cross section with the factorized amplitudes. As we discussed, the relevant factorization of the amplitude shown in \eqref{eq:WilsonSoft} can easily be obtained from SCET with collinear fields along the directions of the energetic on-shell particles. However, in standard SCET derivations one is usually working directly on the level of the cross section. For our approach to be valid, it was important to know the relevant momentum regions in the cross section, so that the appropriate expansion of the amplitude could be performed. In other words, it was important that the phase-space integrations indeed only involved the regions we considered for the expansion of the amplitude.

\subsection{Narrow jets and coft radiation}

In the small-angle limit $\delta \ll 1$, two additional types of momentum regions need to be included in our effective theory, besides the hard and soft regions \cite{Becher:2015hka}. First of all, we need the usual collinear modes to describe the energetic collimated radiation inside the jets
\begin{equation}
\begin{aligned}
  \mbox{collinear:}\quad   p_{c} &\sim Q\,( \delta^2, 1, \delta)  \,, \\
  \mbox{anti-collinear:}\quad    p_{ \bar{c}} &\sim Q\,(1,\delta^2, \delta) \,,
\end{aligned}
\end{equation}
but in addition, we need modes which describe small-angle soft radiation
\begin{equation}
\begin{aligned}
  \mbox{coft:}\quad   p_{t} &\sim Q \beta \,(\delta^2,1, \delta)  \,, \\
  \mbox{anti-coft:}\quad    p_{ \bar{t}} &\sim Q \beta \,(1, \delta^2,\delta) \,. \label{eq:coftscaling}
\end{aligned}
\end{equation}
In Figure~\ref{fig:widejet_mommodes} we show the corresponding momentum regions. Modes which simultaneously have soft and collinear scaling have arisen in other SCET applications, see e.g.\ \cite{Becher:2003qh,Bauer:2011uc,Larkoski:2015zka}. What is special about the present case is that every single component of the coft momentum is suppressed compared to its collinear counterpart, while in standard applications the small component of the collinear mode is commensurate with the soft scaling. In our case, the relative scaling is the same as for the soft and hard modes present in the wide-angle case. Because of this fact, the coft Wilson lines associated with different collinear particles cannot be combined and we end up with multi-Wilson-line operators. 

We emphasize that the coft modes have very low virtuality $p_t^2=\Lambda_t^2=(Q\beta\delta)^2$, much lower than the virtuality of the collinear and soft modes. The presence of this low physical scale might have important implications for the relevance of non-perturbative effects. These are suppressed by the ratio $\Lambda_{\rm QCD}/\Lambda_t$, where $\Lambda_{\rm QCD}\sim 0.5\,{\rm GeV}$ is a scale associated with strong QCD dynamics. Non-perturbative corrections to jet processes can thus be much larger than the naive expectation $\Lambda_{\rm QCD}/Q$. For example, for a jet opening angle $\alpha=10^\circ$ ($\delta\approx 0.09$) and 5\% of the collision energy outside the jets ($\beta=0.1$), one obtains $\Lambda_t\approx 1\,{\rm GeV}$ for $Q= 100\,{\rm GeV}$. It would be interesting to explore phenomenological consequences of this low-scale physics. 

One way to construct the effective theory containing coft modes is to first match QCD onto standard SCET containing collinear fields along the jet directions and soft fields. One then matches the QCD quark vector current onto the vector current in the effective theory and obtains 
\begin{equation}\label{eq:current}
   \bar{\psi}\gamma^\mu \psi \to C_V(-Q^2-i\varepsilon,\mu)\, 
    \bar\chi_{c}^{(0)} S({n})\,\gamma^\mu S^\dagger(\bar{n})\,\chi_{\bar{c}}^{(0)}
    + (c \leftrightarrow \bar{c}) \,.
\end{equation}
The field $\chi_{c}^{(0)}$ ($\chi_{\bar{c}}^{(0)}$) describes an energetic collinear quark propagating in the $n$ direction ($\bar{n}$ direction), as defined in \eqref{eq:building}. The Wilson coefficient $C_V$ contains matching corrections from hard virtual particles. In the above equation we have already decoupled the soft fields which gives rise to the product $S(n)\,S^\dagger(\bar{n})$ of two soft Wilson lines in the fundamental representation. Note that, as in \eqref{eq:Si}, we use outgoing soft Wilson lines. After this decoupling the soft and collinear fields no longer interact. From now on we drop the superscript ``(0)'' on the fields, since we will need to perform a second decoupling transformation below.

In a next step, one splits the collinear field into two submodes 
\begin{equation}\label{split}
   A_c^\mu \to A_c^\mu + A_t^\mu \,, \qquad
   A_{\bar{c}}^\mu \to A_{\bar{c}}^\mu + A_{\bar{t}}^\mu \,,
\end{equation}
and analogously for the quark fields. More explicitly, one matches the purely collinear theory onto a second effective theory, which distinguishes genuine collinear from coft momenta. Note that the relative scaling of the momentum components of the collinear and coft particles is exactly the same as the one of the hard and soft modes analyzed in the wide-angle jet case. Indeed, if we perform a boost $n\cdot p \to n\cdot p/\delta$, $\bar{n}\cdot p \to \bar{n}\cdot p\,\delta$ and identify $\hat{Q}=Q\delta$ with the hard scale, we end up with exactly the same configuration as in the wide-angle case discussed in the previous subsection: after the boost the collinear fields become hard fields at the scale $\hat{Q}$ and the coft fields correspond to soft particles with momentum scale $\hat{Q}\beta$. We can thus analyze the problem with the same method used in section~\ref{subsec:Wideanglejets}. For the case with $m$ particles inside the jet, we introduce collinear fields along their directions with scaling $p^2= (\hat{Q}\beta)^2$ and then write down the corresponding leading-power operators. One important difference is that the ``full theory'' expression for this second matching step contains collinear quark fields which contain collinear Wilson lines. Specifically, let us consider the second term in \eqref{eq:current} which contains the field $\chi_{c}$. This field involves a Wilson line $W_c^\dagger(\bar{n})$ along the $\bar{n}$ direction, see \eqref{eq:building}. Since the full theory contains gluons in the direction $n_0\equiv\bar{n}$, we also need to introduce a corresponding collinear field along the $n_0$ direction in the effective theory. We thus define a new collinear building block 
\begin{equation}
   {\cal W}_0 = W_0^\dagger(n_0)\,W_0(\bar{n}_0) \,.
\end{equation}
This building block is the remnant of the anti-quark field (labeled as particle ``0'') in the original theory and behaves like an anti-quark under the decoupling transformation. It is of $\mathcal{O}(1)$ in SCET power counting, while the fields $\chi_0$ and ${\cal A}_{0\perp}$ count as $\mathcal{O}(\beta)$. As a consequence, the general leading-power operator for $m$ partons inside the jet is
\begin{equation}
   {\cal W}_0\,\Phi_1\,\dots\,\Phi_m \,,
\end{equation}
with $\Phi_i \in \{\chi_i, \bar{\chi}_i, {\cal A}_{i\perp}^\mu \}$. After performing the decoupling transformations 
\begin{equation}
   \Phi_i = {\bm U}_i(n_i)\,\Phi_i^{(0)} \,, \qquad 
   {\cal W}_0 = {\bm U}_0(n_0)\,{\cal W}_0^{(0)} \,,
\end{equation}
we end up with
\begin{equation}\label{eq:Wilson}
   \bm{U}_0(\bar{n})\,\bm{U}_1(n_1)\,\dots\,{\bm U}_m(n_m)\,
   |\mathcal{M}_m(p_0;\{\underline{p}\})\rangle \,,
\end{equation}
in complete analogy to expression \eqref{eq:WilsonSoft} obtained in the wide-angle case. The relevant collinear amplitude is
\begin{equation}
   |\mathcal{M}_m(p_0;\{\underline{p}\})\rangle
   = \langle \{\underline{p}\}|\,\chi_c(0)\,|0\rangle \,,
\end{equation}
where $p_0=\sum_{i=1}^m p_i$. We have added the argument $p_0$ to indicate that these matrix elements are the splitting amplitudes of the collinear quark into the given final state. Since they connect the color of the quark field to the color of the final-state particles, they are strictly speaking matrices in color space, and a more precise but also somewhat cumbersome notation would be to denote them by ${\bf Sp}_m({p}_0;\{\underline{p}\})$. 

In order to check these arguments diagrammatically and to confirm the factorization of soft, collinear and coft fields shown in \eqref{eq:current} and \eqref{eq:Wilson}, we have explicitly expanded the squared $\gamma^*(q)\to q(p_1)\,\bar{q}(p_2)\,g(p_3)\,g(p_4)$ tree-level QCD amplitude in all relevant momentum regions and compared the results with \eqref{eq:current} and \eqref{eq:Wilson}. For the leading contributions in the different regions, we obtain
\begin{equation}\label{eq:facttwogluon}
\begin{aligned}
   |\mathcal{M}(q;p_1,p_2,p_3^s,p_4^s)|^2 &\to |\mathcal{M}^{(0)}(q;p_1,p_2)|^2\,
    \mathcal{A}^{(2)}(\{n,\bar{n}\},p_3^s,p_4^s) \,, \\
   |\mathcal{M}(q;p_1,p_2,p_3^t,p_4^t)|^2 &\to |\mathcal{M}^{(0)}(q;p_1,p_2)|^2\,
    \mathcal{A}^{(2)}(\{n,\bar{n}\},p_3^t,p_4^t) \,, \\
   |\mathcal{M}(q;p_1,p_2,p_3^c,p_4^c)|^2 &\to |\mathcal{M}^{(0)}(q;\tilde{p},p_2)|^2\,
    \mathcal{F}^{(2)}(\tilde{p}; p_1,p_3^c,p_4^c) \,, \\
   |\mathcal{M}(q;p_1,p_2,p_3^c,p_4^s)|^2 &\to |\mathcal{M}^{(0)}(q;\tilde{p},p_2)|^2\,
    \mathcal{F}^{(1)}(\tilde{p};p_1,p_3^c)\,\mathcal{A}^{(1)}(\{n,\bar{n}\},p_4^s) \,, \\
   |\mathcal{M}(q;p_1,p_2,p_3^c,p_4^t)|^2 &\to |\mathcal{M}^{(0)}(q;\tilde{p},p_2)|^2\,
    \mathcal{F}^{(1)}(\tilde{p};p_1,p_3^c)\,\mathcal{A}^{(1)}(\{n_1,\bar{n},n_3\},p_4^t) \, ,\\
    |\mathcal{M}(q;p_1,p_2,p_3^s,p_4^t)|^2 &\to |\mathcal{M}^{(0)}(q;p_1,p_2)|^2\,
    \mathcal{A}^{(1)}(\{n,\bar{n}\},p_3^s)\,\mathcal{A}^{(1)}(\{n,\bar{n}\},p_4^t) \,, 
\end{aligned}
\end{equation}
where $\tilde{p}=q-p_2$. We have indicated the assumed scaling of the gluon momenta as a superscript, while the (anti-)quark momentum has always (anti-)collinear scaling. The functions $\mathcal{A}^{(1)}(\{\underline{n}\},k_1)$ and $\mathcal{A}^{(2)}(\{\underline{n}\},k_1,k_2)$ are squared one- and two-gluon matrix elements of an operator with Wilson lines along the directions $\{\underline{n}\}= \{n_1 ,\dots, n_m\}$. The one-gluon matrix elements $\mathcal{A}^{(1)}$ are given by the expressions shown in \eqref{eq:softOneGluon} below, while the two-gluon matrix elements $\mathcal{A}^{(2)}$ can be found e.g.\ in Appendix~C of \cite{Becher:2012qc}. The splitting functions $\mathcal{F}^{(1)}$ and $\mathcal{F}^{(2)}$ describe the emission of one or two collinear gluons from the quark with momentum $\tilde{p}$. The function $\mathcal{F}^{(1)}(\tilde{p};p_1,p_3^c)$  is given by the integrand in \eqref{eq:J2} below.

We can now compare the results in \eqref{eq:facttwogluon} with the factorized structures derived above. The first three relations are quite simple. The matrix elements factorize into the Born-level matrix element times the squared two-gluon amplitudes for each sector. The next two factorization relations are more interesting. They demonstrate that for two energetic particles with momenta $p_1$ and $p_3$ inside the right jet one ends up with a two-Wilson-line matrix element for a soft gluon, as in \eqref{eq:current}, but a three-Wilson-line matrix element for a coft gluon, confirming our relation \eqref{eq:Wilson}. Finally, for the case of $\mathcal{M}(q;p_1,p_2,p_3^s,p_4^t)$ one ends up with a product of a soft and a coft matrix element, demonstrating the coft-soft factorization. We see that all relations in \eqref{eq:facttwogluon} are fully compatible with the structure of \eqref{eq:current} and \eqref{eq:Wilson}.

Squaring the factorized amplitude and expanding the phase-space constraints in each region, we obtain a factorization theorem for the narrow-angle jet cross section. However, before presenting its final form, let us discuss an alternative way to derive the mode factorization in the narrow-jet case. To this end, we start with the wide-angle result (\ref{sigbarefinal}) and narrow the jet angle. To take the limit $\delta\to 0$, we can group the final-state particles into $k$ left and $l$ right ones, $\{\underline{p}\} = \{\underline{p}_L\} \cup  \{\underline{p}_R \}$. For small cone size, the amplitude factors into the amplitude for the production of a quark and an anti-quark, multiplied by splitting amplitudes
\begin{align}
 |\mathcal{M}_{k+l}(\{\underline{p}\}) \rangle 
 &=  {\bf Sp}_k({p_L}_0;\{\underline{p}_L\})\,{\bf Sp}_l({p_R}_0;\{\underline{p}_R\})\,  |\mathcal{M}_2(\{{p_L}_0,{p_R}_0\})\rangle  \nonumber \\
&\equiv |\mathcal{M}_k({p_L}_0;\{\underline{p}_L\})\rangle \, | \mathcal{M}_l({p_R}_0;\{\underline{p}_R\})\rangle \,|\mathcal{M}_2(\{{p_L}_0,{p_R}_0\})\rangle \,.
 \end{align}
The hard functions are the square of these amplitudes, and hence we obtain the factorized result 
\begin{align}\label{Hfac}
\bm{\mathcal{H}}_{k+l}(\{\underline{n}\},Q,\delta) =  \bm{\mathcal{H}}_2(Q) \, \bm{\mathcal{J}}_{\!\!l}(\{\underline{n}_R\}, Q\delta) \, \bm{\mathcal{\overline{J}}}_{\!\!k}(\{\underline{n}_L\}, Q\delta)\,,
\end{align}
where  $\bm{\mathcal{H}}_2(Q)= \sigma_0 \,H(Q^2)\, \bm{1}$. While the hard function $\bm{\mathcal{H}}_{k+l}$  depends both on the cone angle $\delta$ and the collision energy $Q$, this dependence is factorized on the right-hand side. The jet functions only depend on the collinear scale $Q\delta$ and for $l$ partons they are defined as
\begin{align}\label{jetfun}
   \frac{n\!\!\!/}{2}\,\bm{\mathcal{J}}_{\!\!l}(\{\underline{n}\},Q\delta) 
   = \sum_{\rm spins}  \prod_{i=1}^l &\int \! \frac{dE_i \,E_i^{d-3} }{(2\pi)^{d-2}} \,
    |\mathcal{M}_l(p_0;\{\underline{p}\})\rangle \langle\mathcal{M}_l(p_0;\{\underline{p}\})| \nonumber \\
  & \times  2\,(2\pi)^{d-1}\, \delta(Q-\bar{n}\cdot p_{X_c})\,\delta^{(d-2)}(p_{X_c}^\perp)\, 
    {\Theta }^{n}_{\rm in}\!\left(\left\{\underline{p}\right\}\right) ,
\end{align}
where $p_{X_c}=\sum_{i=1}^m p_i$ is the total momentum of the collinear particles. The theta function ${\Theta }^{n}_{\rm in}$ ensures that every collinear parton is inside the right jet, while the analogous constraint in the definition of the hard function \eqref{eq:Hm} was allowing for particles inside both jets. Again, we only integrate over the energies of the particles, keeping their directions fixed. The integrals over the directions will be performed after multiplication with the coft matrix elements. The function $\bm{\mathcal{\overline{J}}}_{\!\!k}$ describes the energetic particles inside the left jet and is defined exactly as \eqref{jetfun} but with $n\leftrightarrow\bar{n}$.

A similar factorization holds also  for the soft functions $\bm{\mathcal{S}}_m$ in \eqref{eq:Sn}. However, because the energy deposited outside the jet cones is shared between the soft and coft modes, we need 
a Laplace transformation to factorize the corresponding phase-space constraints. We define the Laplace-transformed functions as
\begin{align}\label{SoftLaplace}
\widetilde{\bm{\mathcal{S}}}_m(\{\underline{n}\},Q\tau,\delta)
= \int_0^\infty\! d\beta \, e^{-\beta/(\tau e^{\gamma_E})} \,\frac{d}{d\beta}\, \bm{\mathcal{S}}_m(\{\underline{n}\},Q\beta,\delta)\, .
\end{align}
Note that $\tau\sim \beta$ for power-counting purposes. The derivative in this definition is present because we introduce the Laplace transform for the differential soft function, defined with a fixed energy, rather than the integrated function $\bm{\mathcal{S}}_m$ given in \eqref{eq:Sn}. We can then factorize the Laplace transformed soft function as
\begin{align}\label{Sfac}
\widetilde{\bm{\mathcal{S}}}_{k+l}(\{\underline{n}\},Q\tau,\delta) = \widetilde{S}(Q\tau) \,\, \widetilde{\bm{\mathcal{U}}}_l(\{\underline{n}_R\},Q\delta\tau) \, \, \widetilde{\overline{\bm{\mathcal{U}}}}_k(\{\underline{n}_L\},Q\delta\tau)\, ,
\end{align}
where the soft function $\widetilde{S}(Q\tau)$ only includes two Wilson lines along the jet directions. To arrive at this factorized expression one starts with the original soft function and splits up the soft field as $A_s\to A_s + A_t$. One can then decouple the coft field using the usual field redefinition. This redefinition yields a new soft field $A_s'$, which is split up into $A_s'\to A_s'+A_{\bar{t}}$. The anti-coft-soft interactions can then be decoupled in turn. The $l$ soft Wilson lines in the right hemisphere can then be expanded around a common reference vector $n$ and combined using \eqref{eq:softcombine}, and similarly for the soft Wilson lines in the left hemisphere. This yields the same Wilson-line structure as in \eqref{eq:current}. Squaring the soft amplitudes yields
\begin{equation}\label{eq:S2}
S(Q\beta)\,\bm{1} =  \,\int\limits_{X_s}\hspace{-0.58cm} \sum\,
   \langle 0|\,S^\dagger(\bar{n}) \,S(n) \,
    |X_s\rangle \langle X_s|\, S^\dagger(n)\, S(\bar{n}) \, |0\rangle  
 \,  \theta(Q\beta-2 E_{X_s}) \,.
\end{equation}
Because the soft radiation has parametrically large angle, it is always outside the jet and the energy constraint is imposed on the total energy $E_{X_s}$. The coft function $\bm{\mathcal{U}}_m(\{\underline{n}_R\},Q\delta\beta)$ with $m$ Wilson lines is given by 
\begin{multline}\label{eq:Um}
 \bm{\mathcal{U}}_m(\{\underline{n}_R\},Q\delta \beta) \\= \int\limits_{X_t}\hspace{-0.58cm} \sum\,
   \langle 0|\, \bm{U}_0^\dagger(\bar{n})\,\bm{U}_1^\dagger(n_1)\dots {\bm U}_m^\dagger(n_m) 
   \, |X_t\rangle \langle X_t| \,\bm{U}_0(\bar{n})\dots {\bm U}_m(n_m)\, |0\rangle\, 
   \theta(Q\beta-\bar{n}\cdot p_{{\rm\,out}})\,. 
    \end{multline}
The right-moving coft particles are always outside the left jet in the sense that the out-of-jet constraint is always fulfilled after the multipole expansion, independent of the angle of the coft particle. The momentum  $p_{{\rm\,out}}$ is therefore the total momentum outside the right jet. The anti-coft function $\widetilde{\overline{\bm{\mathcal{U}}}}_k$ has the Wilson line $\bm{U}_0$ along the $n$ instead of the $\bar{n}$ direction and the constraint is placed on $n\cdot p_{\rm\,out}$. 

\begin{figure}[t!]
\begin{center}
\hspace{-1.0cm}
\begin{overpic}[scale=0.4]{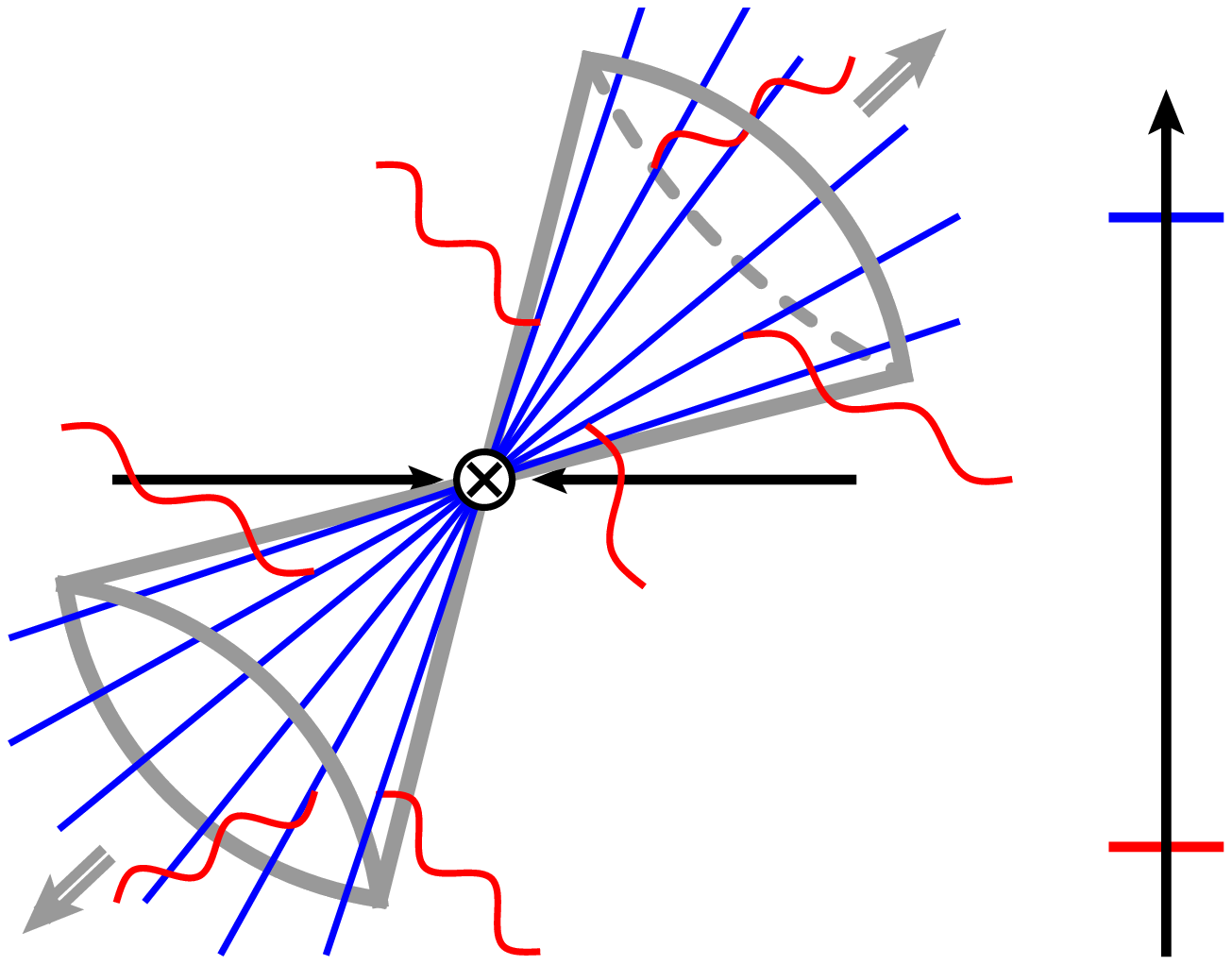}
\put(-4,33){\color{black}{$e^-$}}
\put(71,33){\color{black}{$e^+$}}
\put(102,59){\color{blue}{$Q$}}
\put(102,8){\color{red}{$Q\beta$}}
\end{overpic}
\hspace{1.5cm}
\begin{overpic}[scale=0.4]{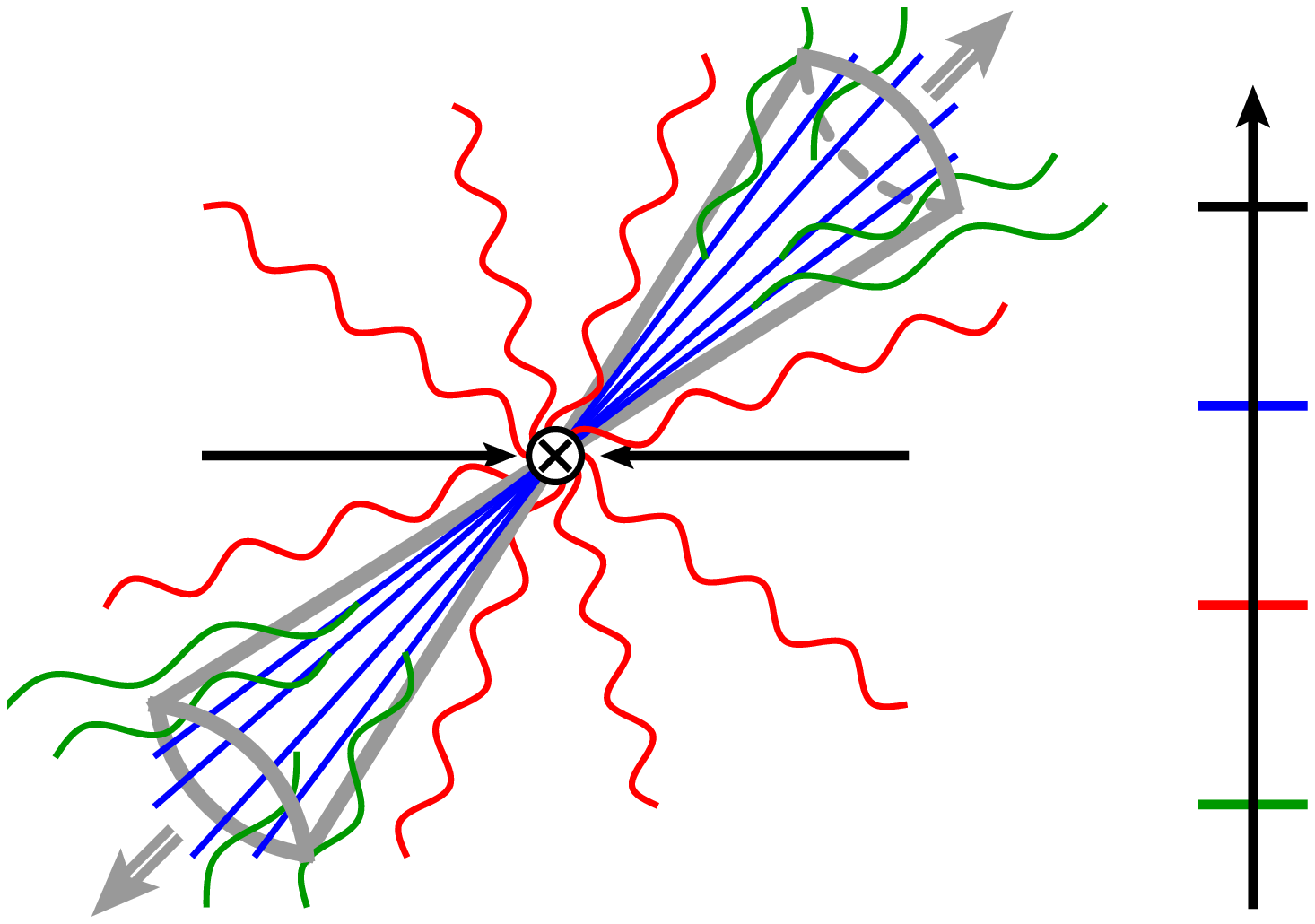}
\put(102,53){$Q$}
\put(102,37){\color{blue}{$Q\delta$}}
\put(102,21){\color{red}{$Q\beta$}}
\put(102,6){\color[rgb]{0,0.6,0}$Q\delta\beta$}
\end{overpic}
\end{center}
\caption{Momentum modes and associated scales for wide-angle (left) and narrow-angle (right) jet production.}
\label{fig:wide_narrow_fac}
\end{figure}

Putting these ingredients together, the cross section in Laplace space takes the form~\cite{Becher:2015hka}
\begin{align}\label{factformula}
\widetilde\sigma(\tau,\delta) = \sigma_0\,H(Q)\,\widetilde{S}(Q\tau)\! \left[\, \sum_{m=1}^{\infty} 
    \big\langle \bm{\mathcal{J}}_{\!\!m}(Q\delta) \otimes 
    \widetilde{\bm{\mathcal{U}}}_m(Q\delta\tau) \big\rangle \right]^2 ,
\end{align}
where we have used the fact that both jets give an identical contribution. In Figure~\ref{fig:wide_narrow_fac} we show a pictorial representation of the structure of the factorization formula and the different types of radiation relevant in both the wide-angle and narrow-jet cases.

\subsection{Renormalization and resummation\label{sec:renormFirst}}

The factorization theorems we have obtained involve operators with an arbitrary number of Wilson lines, both in the wide-angle and narrow-jet case. We now discuss the renormalization of these operators. The associated RG equations form the basis for the resummation of logarithmically-enhanced contributions to all orders in perturbation theory. This resummation is achieved by evolving the Wilson coefficients of these operators from the high scale $\mu \sim Q$ down to the scale where the low-energy physics takes place. Let us first discuss the wide-angle cross section for which the factorization theorem has been given in \eqref{sigbarefinal}. In our effective theory, the hard functions $\bm{\mathcal{H}}_m$ are the Wilson coefficients of the Wilson-line matrix elements $\bm{\mathcal{S}}_m$ and we regularize both quantities in $d=4-2\e$ dimensions. The effective field theory matrix elements contain UV divergences since the short-distance structure of the full theory is not resolved. The corresponding $1/\e$ poles can be removed by renormalizing the hard Wilson coefficients according to
\begin{equation}\label{eq:renH}
\bm{\mathcal{H}}_m(\{\underline{n}\},Q,\delta,\epsilon)= \sum_{l=2}^m\, \bm{\mathcal{H}}_l (\{\underline{n}\},Q,\delta,\mu)\, \bm{Z}^H_{lm}(\{\underline{n}\},Q,\delta,\epsilon,\mu)\, .
\end{equation}
In practice, it is easiest to obtain the bare Wilson coefficients from on-shell matching calculations, where the poles arise from IR divergences. However, these IR poles are in one-to-one correspondence to UV divergences since the effective-theory loop-integrals in such matching computations are scaleless, see e.g.\ \cite{Becher:2009qa} for a detailed explanation of this point within SCET. We have discussed this correspondence after \eqref{sigbarefinal}. It implies that we can understand the UV divergences of $\bm{\mathcal{H}}_m$ from the structure of the IR divergences in the real and virtual diagrams which contribute to these quantities. Given that the coefficients $\bm{\mathcal{H}}_m$ are fixed-multiplicity QCD amplitudes squared, integrated over energy, it is clear that the matrix $\bm{Z}^H_{lm}(\{\underline{n}\},Q,\delta,\epsilon,\mu)$ cannot be diagonal: lower-multiplicity virtual diagrams are needed to cancel the divergences of real-emission diagrams. In order to achieve this cancellation, the renormalization matrix must have the form 
\begin{equation}\label{eq:ZHstruct}
\bm{Z}^H(\{\underline n\},Q,\delta, \epsilon,\mu) \sim  \left(
\begin{array}{ccccc}
 1   &~  \alpha_s   &~  \alpha_s^2  &~  \alpha_s^3   &~ \hdots \\
 0 &~ 1  &~ \alpha_s  &~ \alpha_s^2   &~ \hdots \\
0 &~ 0  &~ 1  &~  \alpha_s  &~   \hdots \\
 0&~ 0&~ 0 &~ 1  & \hdots \\
 \vdots & \vdots & \vdots & \vdots &
   \ddots \\
\end{array}
\right),
\end{equation}
where we indicate the perturbative order of each element. At each higher order in perturbation theory, more off-diagonal contributions fill in. We have anticipated the upper diagonal structure of the matrix in \eqref{eq:renH} by restricting the sum to $l\leq m$. Note that $\bm{Z}^H_{lm}(\{\underline{n}\},Q,\delta,\epsilon,\mu)$ has logarithmic $Q$ dependence, because the fixed-multiplicity amplitudes involve both soft and collinear divergences. This dependence is a familiar feature of Sudakov-type processes.

By consistency, the matrix $\bm{Z}^H$ must render the soft functions finite, i.e.\ we must find that the functions
\begin{equation}\label{eq:softRen}
\bm{\mathcal{S}}_l(\{\underline{n}\},Q \beta,\delta,\mu) = \sum_{m= l}^{\infty} \bm{Z}^H_{lm}(\{\underline{n}\},Q,\delta,\epsilon,\mu) \,\hat{\otimes}\, \bm{\mathcal{S}}_m(\{\underline{n}\},Q \beta,\delta,\e)
\end{equation}
are finite for $\e\to 0$. The structure of this result is at first sight quite surprising, since Wilson-line matrix elements can usually be renormalized multiplicatively. However, in the present case additional UV divergences in the real-emission diagrams arise because the soft radiation is not constrained inside the jet. It is precisely those types of divergences which lead to NGLs. Furthermore, the upper triangular form of $\bm{Z}^H_{lm}$ implies that {\em higher-multiplicity} soft functions are needed to absorb the divergences of matrix elements with fewer Wilson lines. The symbol $\hat{\otimes}$ indicates that in \eqref{eq:softRen} one has to integrate over the $(m-l)$ additional directions of the unresolved partons on which the bare function $\bm{\mathcal{S}}_m$ depends.

The scale dependence of the renormalized hard and soft functions is governed by the RG equations
\begin{align}\label{eq:hrdRG}
\frac{d}{d\ln\mu}\,\bm{\mathcal{H}}_m(\{\underline{n} \},Q,\delta,\mu)  &= - \sum_{l =2}^{m}  \bm{\mathcal{H}}_l(\{\underline{n} \},Q,\delta,\mu) \, \bm{\Gamma}^H_{lm}(\{\underline{n}\},Q ,\delta,\mu) \, , \\
\frac{d}{d\ln\mu}\,\bm{\mathcal{S}}_l(\{\underline{n} \},Q\beta,\delta,\mu)  &=\sum_{m= l}^\infty \bm{\Gamma}^H_{lm}(\{\underline{n}\},Q,\delta ,\mu)\, \hat{\otimes}\, \bm{\mathcal{S}}_m(\{\underline{n} \},Q\beta,\delta,\mu) \, , \label{eq:sftRG}
\end{align}
which ensure that the cross section \eqref{sigbarefinal} is scale independent. The anomalous-dimension matrix is obtained from the standard relation
\begin{align}\label{eq:gammaH}
\frac{d}{d\ln\mu}\, \bm{Z}_{km}^{H}(\{\underline{n}\},Q,\delta,\e ,\mu) = \sum_{l=k}^m \bm{Z}^{H}_{kl}(\{\underline{n}\},Q,\delta,\e ,\mu) \,\hat{\otimes}\, \bm{\Gamma}^H_{lm}\left(\{\underline{n}\} ,Q,\delta,\mu\right) ,
\end{align}
and it has linear dependence on $\ln(Q/\mu)$ as is familiar from Sudakov-type problems. However, the wide-angle cross section we consider contains only a single large logarithm at each order. The Sudakov double logarithms must cancel in the sum over multiplicities in \eqref{sigbarefinal}. A related observation is that the RG equation \eqref{eq:sftRG} for the soft functions is only consistent if the $Q$-dependence of the anomalous dimension drops out after the integrals over the unresolved partons have been performed, since the expression on the left-hand side only involves the soft scale $Q\beta$.  This implies a set of highly nontrivial consistency relations among the entries of the anomalous-dimension matrix. At one-loop order this will be studied in Section \ref{sec:renorm}.

Solving the RG equations (\ref{eq:hrdRG}) and (\ref{eq:sftRG}) one can resum all large logarithms in the wide-angle jet cross section (\ref{sigbarefinal}). At the soft scale $\mu_s\approx Q\beta$ the soft functions do not involve large logarithms, and hence they can be calculated in a perturbative series in powers of $\alpha_s(\mu_s)$. Likewise, at the hard scale $\mu_h\approx Q$ the hard functions do not involve large logarithms, and hence they can be calculated in a perturbative series in powers of $\alpha_s(\mu_h)$. The large logarithms of the scale ratio $\mu_h/\mu_s$ are resummed by evolving the soft functions up to the hard scale (or vice versa), 
\begin{equation}\label{softRG}
   \bm{\mathcal{S}}_l(\{\underline{n}\},Q\beta,\delta,\mu_h) 
   = \sum_{m\geq l} \bm{U}^S_{lm}(\{\underline{n}\},\delta,\mu_s,\mu_h)\,\hat{\otimes}\, 
    \bm{\mathcal{S}}_m(\{\underline{n}\},Q\beta,\delta,\mu_s) \,,
\end{equation}
with an evolution matrix of the form 
\begin{equation}\label{eq:US}
   \bm{U}^S(\{\underline{n}\},\delta,\mu_s,\mu_h) 
   = {\bf P} \exp\left[\, \int_{\mu_s}^{\mu_h} \frac{d\mu}{\mu}\,
    \bm{\Gamma}^H(\{\underline{n}\},\delta,\mu) \right] .
\end{equation}
The path-ordering symbol ${\bf P}$ is necessary since $\bm{\Gamma}^H$ is a matrix. The resummed cross section is then given by
\begin{equation}
   \sigma(\beta,\delta) 
   = \sum_{l=2}^\infty \big\langle \bm{\mathcal{H}}_l(\{\underline{n}\},Q,\delta,\mu_h) 
    \otimes \sum_{m\geq l} \bm{U}^S_{lm}(\{\underline{n}\},\delta,\mu_s,\mu_h)\,\hat{\otimes}\, 
    \bm{\mathcal{S}}_m(\{\underline{n}\},Q\beta,\delta,\mu_s) \,.
\end{equation}

The renormalization procedure in the narrow-jet case is quite similar, except that the underlying factorization formula \eqref{factformula} is more complicated.  The RG evolution of the functions $H$ and $S$ is well known. Both are renormalized multiplicatively, 
\begin{equation}
H(Q,\e) = Z_H(Q,\epsilon,\mu) \,H(Q,\mu) \,, \qquad
 \widetilde{S}(Q\tau,\e) =   Z_S(Q\tau,\epsilon,\mu) \,\widetilde{S}(Q\tau,\mu)\,,
\end{equation}    
and the associated RG equations take the form \cite{Becher:2007ty}
\begin{equation}\label{RGsofthard}
\begin{aligned}
\frac{d}{d\ln\mu}\, H(Q,\mu) &= 2\left[\Gamma_{\rm cusp}(\as) \ln\frac{Q^2}{\mu^2} + \gamma_V(\as)\right]H(Q,\mu), \\
\frac{d}{d\ln\mu}\, \widetilde{S}(Q\tau,\mu) & = - 2\left[\Gamma_{\rm cusp}(\as) \ln\frac{Q^2\tau^2}{\mu^2} + \gamma_W(\as)\right]\widetilde{S}(Q\tau,\mu)\,,
\end{aligned}
\end{equation} 
and explicit expressions for $\gamma_V$, $\gamma_W$ and the cusp anomalous dimension $\Gamma_{\rm cusp}$ at three-loop order can be found in the same reference.   The jet functions $\bm{\mathcal{J}}_m$ for a fixed number of partons cannot be renormalized multiplicatively, as was the case for $\bm{\mathcal{H}}_m$. In analogy to \eqref{eq:renH}, we write
\begin{equation}\label{eq:ZJdef}
   \bm{\mathcal{J}}_{\!\!m}(\{\underline{n} \},Q\delta,\epsilon) 
   =  \sum_{l=1}^m \bm{\mathcal{J}}_{\!\!l}(\{\underline{n} \},Q\delta,\mu)\,\bm{Z}^{J}_{lm}(\{\underline{n} \},Q\delta,\epsilon,\mu) \,,
\end{equation}
where $\bm{Z}^{J}$ is an upper triangular matrix with the same structure as in \eqref{eq:ZHstruct}. RG invariance of the cross section in \eqref{factformula} implies that the product
\begin{equation}\label{eq:ZU}
   \bm{Z}^U(\{\underline{n} \},Q,Q\delta,Q\tau,\epsilon,\mu) \equiv Z_H^{1/2}(Q,\epsilon,\mu)\,
    Z_S^{1/2}(Q\tau,\epsilon,\mu)\,\bm{Z}^J(\{\underline{n} \},Q\delta,\epsilon,\mu) 
\end{equation}
of renormalization factors must render the coft matrix elements finite, i.e.\ 
\begin{equation}\label{eq:renCond}
   \widetilde{\bm{\mathcal{U}}}_l(\{\underline{n} \},Q\delta\tau,\mu) 
   = \sum_{m=l}^\infty \bm{Z}^U_{lm}(\{\underline{n} \},Q,Q\delta,Q\tau,\epsilon,\mu)\,
    \hat\otimes\,\,\widetilde{\bm{\mathcal{U}}}_m(\{\underline{n} \},Q\delta\tau,\epsilon) \,.
\end{equation}
From this one can derive the evolution equation
\begin{equation}
\frac{d}{d\ln\mu}\,\, \widetilde{\bm{\mathcal{U}}}_l(\{\underline{n} \},Q \delta\tau,\mu)  =\sum_{m = l}^\infty \bm{\Gamma}^U_{lm}(\{\underline{n}\},Q\delta,\tau, \mu)\, \hat{\otimes}\, \, \widetilde{\bm{\mathcal{U}}}_m(\{\underline{n} \},Q\delta\tau,\mu) \, , \label{eq:cftRG}
\end{equation}
where
\begin{equation}\label{eq:gammaU}
\bm{\Gamma}^U(\{\underline{n}\},Q\delta,\tau, \mu) = \bm{\Gamma}^J(\{\underline{n} \},Q\delta,\mu) +2\Gamma_{\rm cusp}(\as) \ln\tau +2\gamma_\phi(\as)\,,
\end{equation}
where the anomalous dimension $\bm{\Gamma}^J$ is defined in analogy to \eqref{eq:gammaH}, and $2\gamma_\phi = \gamma_W-\gamma_V$ governs the non-cusp part of the RG evolution of the quark parton distribution function near $x\to 1$ \cite{Becher:2006mr,Moch:2004pa}. Similarly to \eqref{eq:sftRG}, the RG equation for the coft functions implies nontrivial consistency conditions on $\bm{\Gamma}^J$. The dependence of this function on the jet scale $Q\delta$ must contain a universal piece proportional to $\Gamma_{\rm cusp}$, which can be factored out and conspires with $\ln\tau$ term in \eqref{eq:gammaU} to produce a logarithm of the coft scale $Q\delta\tau$. The remaining dependence on the jet scale drops out once $\bm{Z}^U$ is applied to the coft functions. The resummation of large logarithms works in an analogous way to the wide-angle case discussed earlier. In practice, it is most convenient to evolve the coft functions from the coft scale $\mu_t\approx Q\delta\tau$ to the jet scale $\mu_c\approx Q\delta$ by means of an equation analogous to (\ref{softRG}). The evolution of the hard and soft functions to the jet scale is readily obtained by solving the standard evolution equations \eqref{RGsofthard}.

The explicit computations presented below allow us to verify that the renormalization of the operators indeed works as we have described. In the next section, we demonstrate at two-loop order that \eqref{eq:renCond} renders the coft function $\widetilde{\bm{\mathcal{U}}}_1$ finite. In Section \ref{sec:renorm}, we verify at one-loop order that the matrix $\bm{Z}_{lm}^{H}$, which renormalizes the Wilson coefficients $\bm{\mathcal{H}}_l$, indeed renders all soft functions $\bm{\mathcal{S}}_m$ in \eqref{eq:softRen} finite. In Appendix \ref{sec:oneloopRenNarrow}, we perform the same check also for the coft functions $\widetilde{\bm{\mathcal{U}}}_m$ and relate the one-loop $Z$-factors for the two cases. Also in Section \ref{sec:renorm} we show that at large $N_c$ the RG equation \eqref{eq:sftRG} is equivalent to the BMS equation at the leading logarithmic level.

\section{Two-loop result for the narrow-jet cross section\label{sec:twoloopNarrow}}

In the following two sections we derive all ingredients entering the factorization theorems \eqref{sigbarefinal} and \eqref{factformula} at NNLO and verify that we reproduce the correct cross sections at this order. While the cross sections are finite after coupling renormalization, the different component functions in the factorization theorems contain divergences, which we regularize dimensionally. The $1/\epsilon^n$ poles cancel when the various components are combined, and hence we can test the factorization theorems working with bare functions. We start with the analysis of the narrow-jet cross section, for which our NNLO expression can be compared with numerical results obtained using the event generator {\sc Event2} \cite{Catani:1996vz}. The wide-angle case will be studied in Section~\ref{sec:twoLoopWide}.

\subsection{Ingredients of the factorization theorem}

At two-loop order in perturbation theory the factorization formula (\ref{factformula}) simplifies, since the jet functions $\bm{\mathcal{J}}_{\!\!m}\sim\alpha_s^{m-1}$. Expressing the cross section in terms of bare functions, we have
\begin{align}\label{NA_NNLO}
   \widetilde{\sigma}(\tau,\delta) &= \sigma_0\,H(Q,\e)\,\widetilde{S}(Q\tau,\e)\, 
    \big\langle \bm{\mathcal{J}}_{\!\!1}(\{n_1\},Q\delta,\e)\otimes\,
     \widetilde{\bm{\mathcal{U}}}_{1}(\{n_1\},Q\delta\tau,\e) \nonumber\\
   &\quad + \bm{\mathcal{J}}_{\!\!2}(\{n_1,n_2\},Q\delta,\e)\otimes\,
     \widetilde{\bm{\mathcal{U}}}_{2}(\{n_1,n_2\},Q\delta\tau,\e) 
    + \bm{\mathcal{J}}_{\!\!3}(\{n_1,n_2,n_3\},Q\delta,\e) \otimes \bm{1}
    + \dots \big\rangle^2\, ,
\end{align}
where we have used that at lowest order $\widetilde{\bm{\mathcal{U}}}_{m}=\bm{1}+{\cal O}(\alpha_s)$. Recall that the symbol $\otimes$ is defined as in (\ref{eq:otimes}) as an integral over the particle directions with measure $\Pi_i\,d\Omega(n_i)/(4\pi)$. For each parton in the jet, we introduce a polar angle $\theta_i$ with respect to the jet axis $\vec{n}$ and an azimuthal angle $\phi_i$ in the plane orthogonal to $\vec{n}$. Rotational invariance in the transverse plane allows us to choose one of the azimuthal angles at will, and we choose to set $\phi_1=0$ for the first parton. For the small-angle case it will be convenient to rescale the polar angles to new variables $\hat\theta_i\equiv\theta_i/(2\delta)\in[0,1]$ (see below). We will thus parameterize the directions $\{n_1,\dots,n_m\}$ with the angular variables $\{\hat\theta_1,\dots,\hat\theta_m,\phi_2,\dots,\phi_m\}$ and introduce new jet and soft functions (with the same names but different variables) such that
\begin{equation}\label{JUnewdef}
   \bm{\mathcal{J}}_{\!\!m}(\{\underline{n}\},Q\delta)\otimes\,
    \widetilde{\bm{\mathcal{U}}}_m(\{\underline{n}\},Q\delta\tau) 
   \equiv \prod_{i=1}^m \int_0^1\!d\hat\theta_i\,\prod_{i=2}^m \int_0^{2\pi}\!d\phi_i\,
    \bm{\mathcal{J}}_{\!\!m}(\underline{\hat\theta},\underline{\phi},Q\delta)\,\,
    \widetilde{\bm{\mathcal{U}}}_m(\underline{\hat\theta},\underline{\phi},Q\delta\tau) \,.
\end{equation}
Note that for $m>2$ there are additional azimuthal angles in $d\ne 4$ dimensions, see \eqref{eq:azimuthal} below.

The function $H(Q,\e)$ is the standard hard function for two-jet processes and can be obtained from a matching calculation of the QCD vector current onto the corresponding SCET current operator at time-like momentum transfer. The renormalized two-loop expression can be found, for example, in \cite{Becher:2006mr,Becher:2007ty}. The bare function is identical to the two-loop on-shell quark form factor, which has the form \cite{Matsuura:1987wt,Matsuura:1988sm,Kramer:1986sg}
\begin{equation}{\label{twoloopH}}
   H(Q,\epsilon) = 1 + \frac{\alpha_0}{4\pi}\,e^{-2\epsilon L_h}\,C_F h_F
    + \left( \frac{\alpha_0}{4\pi} \right)^2 e^{-4\epsilon L_h} 
    \left(  C_F^2 h_{2F} + C_F C_A h_A + C_F T_F n_f h_f \right) + \dots \,,
\end{equation}
with $L_h=\ln\frac{Q}{\mu}$. The Laurent series in $\epsilon$ of the coefficients $h_F$, $h_{2F}$, $h_A$ and $h_f$ to the required order is given in Appendix~\ref{NarrowCone}. Note that we expand the bare functions in the bare coupling constant $\alpha_0=Z_{\alpha}\,\as$, with
\begin{equation}\label{eq:coupRen}
   Z_\alpha = 1- \frac{\alpha_s}{4\pi} \frac{\beta_0}{\epsilon} + \dots \,, 
   \quad \text{and}\quad \beta_0 = \frac{11}{3}\,C_A - \frac{4}{3}\,T_F n_f \,,
\end{equation}
where $\alpha_s\equiv \alpha_s(\mu)$ is the renormalized coupling constant. We write the $d$-dimensional coupling as $\alpha_0\,\tilde{\mu}^{2\e}$ to make $\alpha_0$ dimensionless, where the scale $\tilde{\mu}^{2}=\mu^2 e^{\gamma_E}/(4\pi)$ is chosen such that the subtraction of divergences leads to renormalization in the $\overline{\rm MS}$ scheme.

The soft function $\widetilde{S}(Q\tau,\e)$ defined in \eqref{eq:S2} is the same as the soft function for threshold resummation in Drell-Yan production, except that in our case the Wilson lines are outgoing instead of incoming. Since the soft function only depends on the product of the two light-light vectors, its result is unchanged under the simultaneous change $n\to -n$ and $\bar{n}\to -\bar{n}$ (see \cite{Kang:2015moa} for a detailed discussion of relations among time-like and space-like soft functions). We can therefore simply use the Drell-Yan soft function at NNLO obtained in
\cite{Belitsky:1998tc,Becher:2007ty}. It reads 
\begin{equation}
   \widetilde{S}(Q\tau,\epsilon) = 1 + \frac{\alpha_0}{4\pi}\,e^{-2\epsilon L_s}\,C_F W_F
    + \left( \frac{\alpha_0 }{4\pi} \right)^2 e^{-4\epsilon L_s} 
    \left(  C_F^2 W_{2F} + C_F C_A W_A + C_F T_F n_f W_f \right) + \dots \,,
\end{equation}
where $L_s=\ln\frac{Q\tau}{\mu}$. The coefficients $W_i$ can again be found in Appendix~\ref{NarrowCone}.

We now turn to the new ingredients in the factorization theorem, namely the coft and jet functions. For convenience we write their perturbative expansions in the form 
\begin{equation}
   \widetilde{\bm{\mathcal{U}}}_m = \bm{1} + \sum_{n=1}^\infty
    \left( \frac{\alpha_0}{4\pi} \right)^n\,\widetilde{\bm{\mathcal{U}}}^{(n)}_m \,, \qquad
   \bm{\mathcal{J}}_{\!\!m} = \!\sum_{n=m-1}^\infty\! \left( \frac{\alpha_0}{4\pi} \right)^n
   \bm{\mathcal{J}}_{\!\!m}^{(n)} \,.
\end{equation}

\subsubsection*{Coft functions}

In (\ref{NA_NNLO}) we need the coft functions $\widetilde{\bm{\mathcal{U}}}_1$ to two-loop order and $\widetilde{\bm{\mathcal{U}}}_2$ with one-loop accuracy. For the momentum scaling of coft particles in (\ref{eq:coftscaling}), 
the phase-space constraint allows for emission both inside and outside the jet cones. The energy is only constrained for emissions outside of the jet, because the coft momentum inside the jet is negligible compared to the momentum of the collinear particles. It is therefore dropped in the multipole expansion of the energy-conservation $\delta$-function. Because of this fact, coft functions with all particles inside the jet are scaleless (their energy can be arbitrarily large). Also, a coft particle in the right-moving jet does not see the left-moving jet, since the out-of-left-jet condition is always fulfilled once the multipole expansion is performed. 

\begin{figure}[t!]
\centering
\begin{overpic}[scale=0.7]{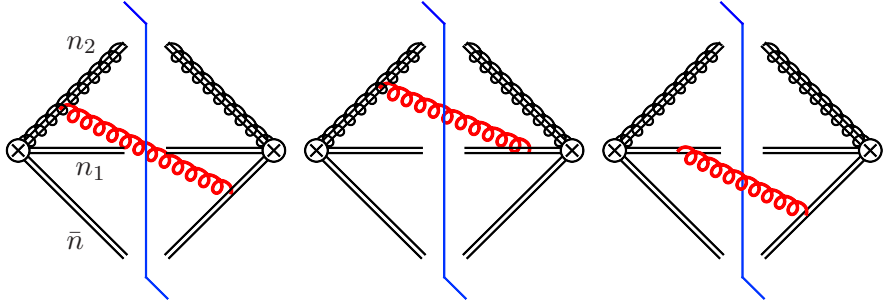}
\put(7,29){$n_2$}
\put(8,14.5){$n_1$}
\put(7,6){$\bar{n}$}
\end{overpic}
\caption{Feynman diagrams contributing to the one-loop coft function $\bm{\mathcal{U}}_{2}$. For each of the three diagrams, there is also an equal, mirrored contribution. We use a double-line notation to represent the Wilson lines.
\label{fig:U2_feyndia}}
\end{figure}

According to the definition \eqref{eq:Um} the coft function $\,\widetilde{\bm{\mathcal{U}}}_1$ contains two Wilson lines, one along the direction $n_1$ of the particle inside the right jet and a second one along the $\bar{n}$ direction, which describes emissions from the left jet. Similarly, the coft function $\,\widetilde{\bm{\mathcal{U}}}_2$ contains three Wilson lines, two along the direction of the particles inside the right jet and a third one along the $\bar{n}$ direction. We first discuss the calculation of a general coft function $\,\widetilde{\bm{\mathcal{U}}}_m$ at one-loop order. The relevant Feynman diagrams contributing for the special case $m=2$ are shown in Figure~\ref{fig:U2_feyndia}. Analogous diagrams can be drawn for $\,\widetilde{\bm{\mathcal{U}}}_1$ and for all higher coft functions $\,\widetilde{\bm{\mathcal{U}}}_m$. The general one-loop expression for \eqref{eq:Um} involves a sum over all pairs of emissions and absorptions from directions $i$ and $j$, such that
\begin{equation}\label{eq:UmNLO}
   \bm{\mathcal{U}}_m =\bm{1} - g_s^2\,\tilde{\mu}^{2\e} \sum_{(ij)}\,\bm{T}_i\cdot\bm{T}_j 
    \int\frac{d^{d-1}k}{(2\pi)^{d-1} 2E_k}\,\frac{n_i\cdot n_j}{n_i\cdot k\,n_j\cdot k}\,
    \theta\!\left( \frac{n\cdot k}{\bar{n}\cdot k}-\delta^2\right) \theta(Q\beta-\bar{n}\cdot k) 
    + \dots \,,
\end{equation}
where $(ij)$ with $i\ne j$ denotes an unordered pair of numbers in the range $0\dots m$, and the scale $\tilde{\mu}$ is defined after \eqref{eq:coupRen}. Here, $\bm{T}_0$ and $n_0=\bar{n}$ are the color generators and the direction of the single Wilson line in the left hemisphere. Since the contribution from radiation inside the jet cone is scaleless, we have restricted the emission to lie outside the cone. For the special cases $m=1,2$ the sum over pairs yields
\begin{align}\label{eq:softOneGluon}
   - \sum_{(ij)}\,\bm{T}_i\cdot \bm{T}_j\,\frac{n_i\cdot n_j}{n_i\cdot k\,n_j\cdot k} 
   &= 2 C_F \frac{\bar{n}\cdot n_1}{\bar n\cdot k\,n_1\!\cdot k}\,\bm{1} \,, \\
   - \sum_{(ij)}\,\bm{T}_i\cdot \bm{T}_j\,\frac{n_i\cdot n_j}{n_i\cdot k\,n_j\cdot k} 
   &= 2 \left[ \left( C_F-\frac{C_A}{2} \right)\! \frac{\bar{n}\cdot n_1}{\bar n\cdot k\,n_1\!\cdot k}
    + \frac{C_A}{2} \!\left( \frac{\bar{n}\cdot n_2}{\bar n\cdot k\,n_2\!\cdot k}
    + \frac{n_1\!\cdot n_2}{n_1\!\cdot k\,n_2\!\cdot k} \right) \right] \!\bm{1} \,, \nonumber
\end{align}
where we used color conservation (i.e., the fact that $\sum_{i=1}^3 \bm{T}_i^a=0$) to express the product of color generators $\bm{T}_i\cdot\bm{T}_j\equiv\sum_a \bm{T}_i^a\,\bm{T}_j^a$ through Casimir invariants. The loop integrals can be expressed in terms of the invariant scalar products of the external light-like reference vectors $n$, $\bar n$, $n_1$ and $n_2$. The result for the coft function $ \bm{\mathcal{U}}_m$ must be invariant under the reparameterization transformation
\begin{equation}
   n^\mu\to \xi\,n^\mu \,, \qquad \bar n^\mu\to \xi^{-1}\,\bar n^\mu \,, \qquad
   \delta\to \xi\,\delta \,, \qquad \beta\to \xi^{-1} \beta \,,
\end{equation}
which leaves the product $n\cdot\bar n=2$ unchanged, as well as under arbitrary rescalings of the light-like vectors $n_i^\mu$. It follows that the answer can be expressed in terms of the invariant variables
\begin{equation}\label{eq:hatthetai}
   \hat\theta_i = \frac{1}{\delta} \sqrt{\frac{n\cdot n_i}{\bar n\cdot n_i}} \,, \qquad 
   \Phi_{ij} = \frac{2}{\delta^2}\,\frac{n_i\cdot n_j}{\bar n\cdot n_i\,\bar n\cdot n_j} \,, \qquad
   \beta\delta \,.
\end{equation}
The product $\beta\delta$ enters the definition of the coft scale. The variables $\hat\theta_i$ are closely related to the polar angles $\theta_i$ of the quark and antiquark with respect to the thrust axis, namely
\begin{equation}
   \hat\theta_i = \frac{1}{\delta}\,\tan\frac{\theta_i}{2}\approx \frac{\theta_i}{2\delta} \,,
\end{equation}
where the last step holds in the small-angle limit. The jet-cone constraint implies that $0\le\hat\theta_i\le 1$. In four dimensions, the variables $\Phi_{ij}$ are related to the azimuthal angle differences $\Delta \phi_{ij}=\phi_{i}-\phi_{j}$ via
\begin{equation}\label{eq:azimuthal}
   \Phi_{ij} = \hat\theta_i^2 + \hat\theta_j^2 - 2 \hat\theta_i \hat\theta_j \cos(\Delta\phi_{ij}) \,.
\end{equation}

Performing the loop integral in terms of these variables and applying the Laplace transformation (\ref{SoftLaplace}), we obtain
\begin{equation}
\begin{aligned}
   \widetilde{\bm{\mathcal{U}}}_1(\hat\theta_1,Q\tau\delta,\e) 
   &= \bm{1} + \frac{C_F\alpha_0}{4\pi}\,e^{-2\,\e L_t}\,u_F(\hat\theta_1)\,\bm{1}
    + \dots \,, \\
   \widetilde{\bm{\mathcal{U}}}_2(\hat\theta_1,\hat\theta_2,\phi_2,Q\tau\delta,\e) 
   &= \bm{1} + \frac{\alpha_0}{4\pi}\,e^{-2\,\e L_t} 
    \left[ C_F\,u_F(\hat\theta_1) + C_A\,u_A(\hat\theta_1,\hat\theta_2,\phi_2) \right] \bm{1}
    + \dots \,.
\end{aligned}
\end{equation}
The Laurent expansions of the coefficient functions to the required order in $\epsilon$ read
\begin{equation}
\begin{aligned}
   u_F(\hat\theta_1) 
   &= - \frac{2}{\epsilon^2} 
    + \frac{2}{\epsilon}\,\ln(1-\hat\theta_1^2) - \frac{\pi^2}{2} + f_0(\hat\theta_1) \\
   &\quad + \epsilon \left[ - \frac{14\zeta_3}{3} + f_1(\hat\theta_1) \right] 
    + \epsilon^2 \left[ - \frac{7\pi^4}{48} + f_2(\hat\theta_1) \right] 
    + {\cal O}(\epsilon^3) , \\
   u_A(\hat\theta_1,\hat\theta_2,\phi_2) 
   &= \frac{1}{\epsilon} \left[ 2\ln(1-\hat\theta_2^2)
    - \ln(1-2\cos\phi_2\,\hat\theta_1\hat\theta_2+\hat\theta_1^2\hat\theta_2^2) \right] \\[2mm]
   &\quad + g_0(\hat\theta_1,\hat\theta_2,\phi_2) + \epsilon\,g_1(\hat\theta_1,\hat\theta_2,\phi_2) 
    + {\cal O}(\epsilon^2)  \,,
\end{aligned}
\end{equation}
where 
\begin{equation}\label{eq:f0g0}
\begin{aligned}
   f_0(\hat\theta_i) &= - 2\ln^2(1-\hat\theta_i^2) - 2{\rm Li}_2(\hat\theta_i^2) \,, \\[2mm]
   g_0(\hat\theta_1,\hat\theta_2,\pi) 
   &= - \ln^2(1-\hat\theta_1^2) - 3\ln^2(1-\hat\theta_2^2)
    + 2 \left[ \ln(1-\hat\theta_1^2) + \ln(1-\hat\theta_2^2) \right] 
    \ln(1+\hat\theta_1\hat\theta_2) \\
   &\quad - 2 \, {\rm Li}_2(\hat\theta_2^2) + 2 \, {\rm Li}_2(-\hat\theta_1\hat\theta_2) 
    - 2\,{\rm Li}_2\bigg( \! - \frac{\hat\theta_1^2+\hat\theta_1\hat\theta_2}{1-\hat\theta_1^2} \bigg) 
    - 2\,{\rm Li}_2\bigg( \! - \frac{\hat\theta_2^2+\hat\theta_1\hat\theta_2}{1-\hat\theta_2^2} \bigg) \,,
\end{aligned}
\end{equation}
and in general $f_i(0)=0$ and $g_i(0,0,\pi)=0$. We also note that 
\begin{align}\label{eq:g1integ}
  g_0(0,\hat\theta_2,\pi) =& -2\ln^2(1-\hat\theta_2^2) \,, \nno \\
  g_1(0,\hat\theta_2,\pi) =& \frac{4}{3} \ln^3(1-\hat\theta_2^2)
   - 4\ln^2\hat\theta_2\,\ln^2(1-\hat\theta_2^2) + \frac{11\pi^2}{6} \ln(1-\hat\theta_2^2) 
   - 4\ln(1-\hat\theta_2^2)\,\text{Li}_2(\hat\theta_2^2) \nno \\
   &- 2\,\text{Li}_3(\hat\theta_2^2) - 8\,\text{Li}_3(1-\hat\theta_2^2) + 8\zeta_3 \,.
\end{align}
From momentum conservation and the definition of the jet axis it follows that the direction of particle~1 (the quark) in the jet function $\bm{\mathcal{J}}_{\!\!1}$ must be aligned with the jet axis (i.e.\ $n_1=n$), which enforces $\hat\theta_1=0$. Likewise, the azimuthal angles of particles~1 (quark) and 2 (gluon) in the jet function $\bm{\mathcal{J}}_{\!\!2}$ must differ by $180^\circ$, which enforces $\phi_2=\pi$ for our choice $\phi_1=0$. These kinematic relations are not encoded in the coft functions, but they will be enforced when we evaluate the convolutions of the coft functions with the jet functions. We can thus simplify the calculation of the coft functions by implementing these constraints from the beginning. It is for this reason that we only need the function $g_0(\hat\theta_1,\hat\theta_2,\pi)$ given in (\ref{eq:f0g0}), and in the expression for $\,\widetilde{\bm{\mathcal{U}}}_1$ we can set $\hat\theta_1=0$, in which case $f_i(0)=0$ for all $i$ and we end up with a simple expression in terms of $\zeta$ values.

In order to obtain the required two-loop contribution to $\,\widetilde{\bm{\mathcal{U}}}_1$ setting $\hat\theta_1=0$ means a significant simplification. We can then exploit the fact that, after a Lorentz boost along the jet axis, the coft function $\,\widetilde{\bm{\mathcal{U}}}_1$ with two opposite Wilson lines in directions $n$ and $\bar n$ can be mapped onto the hemisphere soft function
\begin{equation}
   S_{\rm hemi}(\omega_L,\omega_R) 
   = \frac{1}{N_c} \sum_{X_s}\,\langle 0|\,S_{\bar n}\,S_{n}\,|X_s\rangle\,
    \langle X_s|\,S_{n}^\dag\,S_{\bar n}^\dag\,|0\rangle\,\delta(\omega_L-{\bar n}\cdot p^L_{X_s})\,
    \delta(\omega_R-n\cdot p^R_{X_s}) \,, \quad
\end{equation}
with $p^{L}_{X_s}$ and $p^R_{X_s}$ the total momenta in the respective hemispheres. In our case the energy in one hemisphere (the one corresponding to the inside of the jet) can be arbitrarily large. The hemisphere soft function is known at two-loop accuracy \cite{Schwartz:2014wha}. In order to obtain the two-loop coft function from this result, we first map the plane used to separate the two hemispheres onto the right jet cone via a Lorentz boost along the thrust axis. Under this boost $\omega_L\to\omega_L/\delta$ and $\omega_R\to\omega_R\,\delta$. In other words, computing the hemisphere soft function with $\omega_L<\beta\delta Q$ and arbitrarily large $\omega_R$ is the same as computing the coft function with energy $\beta Q$ outside the cone, i.e.\
\begin{equation}
   \big\langle \,\bm{\mathcal{U}}_1(\hat\theta_1=0,Q\delta\beta,\e) \big\rangle 
   = \int_0^{Q\delta\beta}\!d\omega_L \int_0^{\infty}\!d\omega_R\,
    S_{\rm hemi}(\omega_L,\omega_R,\e) \,.
\end{equation}
The integration over $\omega_R$ needs to be performed before renormalization, since it leads to additional singularities. Taking the Laplace transform as defined in (\ref{SoftLaplace}), we then obtain 
\begin{align}
   \big\langle \,\widetilde{\bm{\mathcal{U}}}_1(0,Q\delta\tau,\epsilon) \big\rangle 
   &= 1 + \frac{\alpha_0}{4\pi}\,e^{-2\epsilon L_t}\,C_F V_F \nonumber\\
   &\quad + \left( \frac{\alpha_0}{4\pi} \right)^2 e^{-4\epsilon L_t} 
    \left( C_F^2 V_{2F} + C_F C_A V_A + C_F T_F n_f V_f \right) + \dots \,,
\end{align}
where $L_t=\ln\frac{Q\delta\tau}{\mu}$ and the explicit expressions for the coefficients $V_i$ are again collected in Appendix~\ref{NarrowCone}.

\subsubsection*{Jet functions}

\begin{figure}[t!]
\centering
\begin{overpic}[scale=0.7]{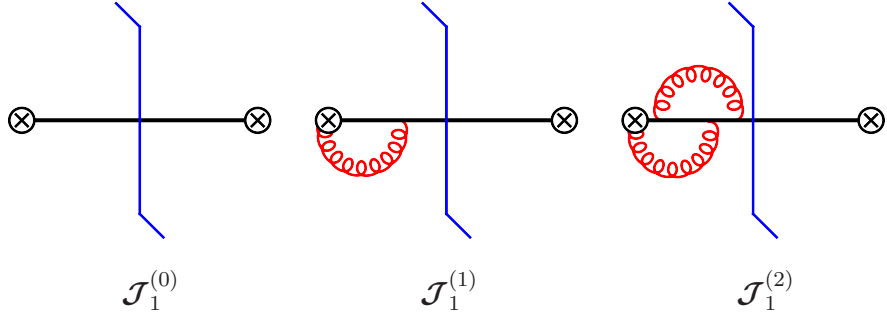}
\put(13,-7){$\bm{\mathcal{J}}_1^{(0)}$}
\put(47,-7){$\bm{\mathcal{J}}_1^{(1)}$}
\put(83,-7){$\bm{\mathcal{J}}_1^{(2)}$}
\end{overpic}
\vspace{0.7cm}
\caption{Sample Feynman diagrams contributing to the jet function $\bm{\mathcal{J}}_1$ at different orders in perturbation theory. Note that only a single propagator is cut. 
\label{fig:J1_feyndia}}
\end{figure}

The jet functions are distribution-valued in the angles of the particles, since they contain additional divergences which arise when particles become collinear. According to the general definition in (\ref{jetfun}), the function $\bm{\mathcal{J}}_{1}$ contains only one parton inside the jet. Examples of corresponding Feynman diagrams are shown in Figure~\ref{fig:J1_feyndia}, where the first diagram represents the LO contribution and the others represent higher-order corrections. Since all loop corrections involve scaleless integrals, which vanish in dimensional regularization, we only need to calculate the LO diagram. It gives
\begin{equation}
   \bm{\mathcal{J}}_{\!\!1}(\{n_1\},Q\delta,\e) 
   = 4\pi\,\bm{1} \!\int dE_1\,E_1^{d-3}\,\delta(Q-\bar{n}\cdot p_1)\, 
    \delta^{(d-2)}(\vec{p}_1^\perp)\,\bar{n}\cdot p_1 
   = 4\pi\,\delta^{(d-2)}(\vec{n}_1^\perp)\,\bm{1} \,, ~
\end{equation}
where $p_1=E_1 n_1$, and we integrate over the energy $E$ of the parton keeping its direction $n_1$ fixed. The jet scale $Q\delta$ does not appear in this result. Introducing angular variables and performing the trivial integration over the azimuthal angle $\phi_1$, we obtain 
\begin{equation}\label{J1res}
   \bm{\mathcal{J}}_{\!\!1}(\hat\theta_1,Q\delta,\e) = \delta(\hat\theta_1)\,\bm{1} 
\end{equation}
for the  jet function in terms of angular variables, as defined in (\ref{JUnewdef}). The convolution with the coft function $\,\widetilde{\bm{\mathcal{U}}}_1$ is now trivially performed and gives
\begin{equation}
   \big\langle \bm{\mathcal{J}}_{\!\!1} \otimes \,\widetilde{\bm{\mathcal{U}}}_1 \big\rangle 
   = \big\langle \,\widetilde{\bm{\mathcal{U}}}_1(0,Q\delta\tau,\epsilon) \big\rangle\,.
\end{equation}

\begin{figure}[t!]
\centering
\begin{overpic}[scale=0.7]{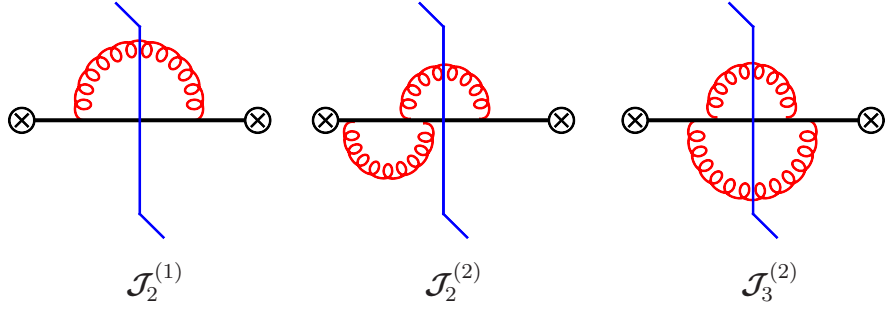}
\put(13.5,-6){$\bm{\mathcal{J}}^{(1)}_{\!\!2}$}
\put(47.5,-6){$\bm{\mathcal{J}}^{(2)}_{\!\!2}$}
\put(83.5,-6){$\bm{\mathcal{J}}^{(2)}_{\!\!3}$}
\end{overpic}
\vspace{0.7cm}
\caption{Sample Feynman diagrams contributing to the jet functions $\bm{\mathcal{J}}_{\!\!2}$ and $\bm{\mathcal{J}}_{\!\!3}$. 
\label{fig:J2_J3_dia}}
\end{figure}

Starting from $\bm{\mathcal{J}}_{2}$ the angular dependence gets nontrivial. The one-loop contribution to $\bm{\mathcal{J}}_2$ contains two collinear partons in the final state and is calculated from the first Feynman diagram shown in Figure~\ref{fig:J2_J3_dia}, integrated over the appropriate phase space. We find 
\begin{align}
   \bm{\mathcal{J}}_{\!\!2}(\{n_1,n_2\},Q\delta,\e) 
   &= \frac{C_F g_s^2 \tilde{\mu}^{2\e}}{(2\pi)^{d-3}}
    \int dE_1\,E_1^{d-3} \int dE_2\,E_2^{d-3}\,
    \theta\!\left(\delta^2 - \frac{n\cdot p_1}{\bar{n}\cdot p_1}\right) 
    \theta\!\left(\delta^2 - \frac{n\cdot p_2}{\bar{n}\cdot p_2}\right) \nonumber\\
   &\quad \times \left[ 
    \frac{4 \left(\bar{n}\cdot p_1\right)^2+(d-2) \left(\bar{n}\cdot p_2\right)^2
          + 4\,\bar{n}\cdot p_1\,\bar{n}\cdot p_2}{p_1\cdot p_2\,\bar{n}\cdot p_2} \right] \nno\\[1mm]
   &\quad\times \delta(Q-\bar{n}\cdot p_1-\bar{n}\cdot p_2)\,
    \delta^{(d-2)}(\vec{p}_{1,\perp}+\vec{p}_{2,\perp})\,\bm{1} \,, \label{eq:J2}
\end{align}
where $p_1=E_1 n_1$ and $p_2=E_2 n_2$ are the momenta of the quark and gluon, respectively. We now split the integration domain into two regions, such that: 
\begin{equation}\label{eq:momReg}
\begin{aligned}
   \mbox{Region I}: &\qquad E_1 > E_2 \quad \Leftrightarrow \quad \hat\theta_1 < \hat\theta_2 \\
   \mbox{Region II}: &\qquad E_2 > E_1  \quad \Leftrightarrow \quad \hat\theta_2 < \hat\theta_1
\end{aligned}
\end{equation}
In region~I the integration generates double poles, because the gluon can be collinear to the quark and it can become soft ($E_2\to 0$), while in region~II only single poles corresponding to a collinear divergence arise. These divergences are overlapping and must be separated using suitable choices of variables. We find it convenient to keep the larger of the two angles $\hat\theta_{\rm max}$ as one variable and use their ratio $y=\hat\theta_{\rm min}/\hat\theta_{\rm max}$ as a second one. Both run from 0 to 1. In terms of these variables we obtain
\begin{align}\label{j2eq}
   \bm{\mathcal{J}}_{\!\!2,\,{\rm I}}^{(1)}(\hat\theta_1,\hat\theta_2,\phi_2,Q\delta,\e) 
   &= C_F\,C_\e\,e^{-2\e L_c} \frac{dy}{d\hat\theta_1}\,\delta(\phi_2-\pi)\,
    \hat\theta_2^{-1-2\epsilon}\,\frac{4y^{-1-2\epsilon}}{\left(1+y\right)^{2-2\e}}
    \left[ 2 + (1-\epsilon)\,\frac{y^2}{1+y} \right] \bm{1} , \nonumber\\
   \bm{\mathcal{J}}_{\!\!2,\,{\rm II}}^{(1)}(\hat\theta_1,\hat\theta_2,\phi_2,Q\delta,\e) 
   &= C_F\,C_\e\,e^{-2\e L_c} \frac{dy}{d\hat\theta_2}\,\delta(\phi_2-\pi)\,
    \hat\theta_1^{-1-2\epsilon}\,\frac{4y^{-2\epsilon}}{\left(1+y\right)^{2-2\e}}
    \left[ 2y + \frac{1-\e}{1+y} \right] \bm{1} \,,
\end{align}
with $L_c=\ln\frac{Q\delta}{\mu}$ and $C_\e=e^{\e\gamma}/\Gamma(1-\e)$. The $1/\epsilon^n$ singularities arise from the regions $y\to 0$ and $\hat\theta_{\rm max}\to 0$ and are non-overlapping. One can thus easily expand the jet function a Laurent series using the standard formula
\begin{equation}  
   x^{-1-\epsilon} = -\frac{1}{\epsilon}\,\delta(x) + \left[ \,\frac{1}{x}\, \right]_+ 
    - \epsilon \left[ \,\frac{\ln x}{x}\, \right]_+ + \dots \,.
\end{equation}
We quote the resulting expression up to ${\cal O}(\e^0)$, reinterpreted as a distribution on functions of the angular variables $\hat\theta_1$, $\hat\theta_2$ and $\phi_2$. We find
\begin{align}
   &\bm{\mathcal{J}}_{\!\!2}^{(1)}(\hat\theta_1,\hat\theta_2,\phi_2,Q\delta,\e) 
    = C_F\,\delta(\phi_2-\pi)\,e^{-2\e L_c} \nonumber\\ 
   &\qquad \times \bigg\{
    \left( \frac{2}{\e^2} + \frac{3}{\e} + 7 - \frac{5\pi^2}{6} + 6\ln 2 \right) 
    \delta(\hat\theta_1)\,\delta(\hat\theta_2) 
    - \frac{4}{\e}\,\delta(\hat\theta_1)\,\bigg[\,\frac{1}{\hat\theta_2}\,\bigg]_+ 
    + 8\,\delta(\hat\theta_1)\,\bigg[\,\frac{\ln\hat\theta_2}{\hat\theta_2}\,\bigg]_+ \nonumber\\
   &\hspace{1.4cm} + 4\,\frac{dy}{d\hat\theta_2}\,\bigg[\,\frac{1}{\hat\theta_1}\,\bigg]_+
    \frac{1+2y+2y^2}{(1+y)^3}\,\theta(\hat\theta_1-\hat\theta_2) \nonumber\\
   &\hspace{1.4cm} + 4\,\frac{dy}{d\hat\theta_1}\,\bigg[\,\frac{1}{\hat\theta_2}\,\bigg]_+
    \left( 2\,\bigg[\,\frac{1}{y}\,\bigg]_+ - \frac{4+5y+2y^2}{(1+y)^3} \right)
    \theta(\hat\theta_2-\hat\theta_1) + {\cal O}(\e) \bigg\}\,\bm{1} \,.
\end{align}
The differentials $dy/d\hat\theta_i$ in the last two lines change one of the integration variables from $d\hat\theta_i$ to $dy$; for example, when applied to a function $F(\hat\theta_1,\hat\theta_2)$ the term in the third line gives
\begin{align}
   \int_0^1\!d\hat\theta_1\,\bigg[\,\frac{1}{\hat\theta_1}\,\bigg]_+
    & \int_0^{\hat\theta_1}\!d\hat\theta_2\,\frac{dy}{d\hat\theta_2}\,
    \frac{1+2y+2y^2}{(1+y)^3}\,F(\hat\theta_1,\hat\theta_2) \nonumber\\
   &= \int_0^1\!d\hat\theta_1\,\bigg[\,\frac{1}{\hat\theta_1}\,\bigg]_+ \int_0^1\!dy\,
    \frac{1+2y+2y^2}{(1+y)^3}\,F(\hat\theta_1,y\hat\theta_1) \,.
\end{align}
The contributions of order $\epsilon$ and $\epsilon^2$ can be obtained in an analogous way but are too lengthy to be presented here.

\subsubsection*{Convolutions}

We finally consider the convolution $\bm{\mathcal{J}}_{\!\!2}\otimes\,\widetilde{\bm{\mathcal{U}}}_2$ in (\ref{NA_NNLO}), which we need to evaluate to $\mathcal{O}(\alpha_s^2)$. Performing the convolution with the tree-level coft function $\bm{\mathcal{U}}_{2}^{(0)}=\bm{1}$, and adding up the contributions from the two regions, we find 
\begin{align}\label{eq:JotimesOne}
   \big\langle \bm{\mathcal{J}}_{\!\!2}^{(1)}\otimes \bm{1} \big\rangle 
   &= C_F\,e^{-2\epsilon L_c} \bigg[ \frac{2}{\epsilon^2} + \frac{3}{\epsilon} 
    + 7 - \frac{5\pi^2}{6} + 6\ln2 
    + \epsilon \left( 14 - \frac{\pi^2}{4} - \frac{44\zeta_3}{3} + 14\ln 2 + 6\ln^2 2 \right)
    \nonumber\\
   &\quad + \epsilon^2 \bigg( 28 - \frac{7\pi^2}{12} - \zeta_3 + \frac{41\pi^4}{720}
    - \frac{4\ln^4 2}{3} + 4\ln^3 2 + 14\ln^2 2 + 28\ln 2 \nonumber\\
   &\hspace{1.4cm} + \frac{4\pi^2\ln^2 2}{3} - \frac{\pi^2\ln 2}{2} - 28\zeta_3\ln 2
    - 32\,\text{Li}_4\Big(\frac{1}{2}\Big) \bigg) \bigg] \,.
\end{align}
Next we need to evaluate the convolution with the one-loop coft function. Doing so, we obtain the NNLO collinear-coft mixing contribution
\begin{align}
    \big\langle \bm{\mathcal{J}}^{(1)}_{\!\!2}\otimes\,\widetilde{\bm{\mathcal{U}}}^{(1)}_2 
     \big\rangle
   = e^{-2\epsilon(L_c+L_t)} \left( C_F^2 M_F + C_F C_A M_A \right) ,
\end{align}
with
\begin{align}
   M_F =& - \frac{4}{\epsilon^4} - \frac{6}{\epsilon^3}
    + \frac{1}{\epsilon^2} \left( - 14 + \frac{2\pi^2}{3} - 12\ln2 \right) 
    + \frac{1}{\epsilon} \left( - 26 - \pi^2 + 10\,\zeta_3 - 32\ln2 \right) \nno \\
   & - 52 - \frac{10\pi^2}{3} -27\zeta_3 + \frac{11\pi^4}{30} - \frac{4}{3}\ln^4 2 - 8 \ln^3 2 
    - 4 \ln^2 2 + \frac{4\,\pi^2}{3}\ln^2 2 \nno \\ 
   & - 52\ln 2 + 4\pi^2\ln 2 - 28\zeta_3 \ln 2 - 32\,{\rm Li}_4\bigg(\frac{1}{2}\bigg) \,, \\
   M_A =& \, \frac{2\pi^2}{3\epsilon^2}
    + \frac{1}{\epsilon} \left( - 2 + \frac{\pi^2}{2} + 12\,\zeta_3 + 6\ln^2 2 + 4\ln 2 \right) 
    - 4 + \frac{7\pi^2}{6} -24\zeta_3 - \frac{\pi^4}{6} + \frac{8}{3} \ln^4 2 \nno \\
   & - 4\ln^3 2 + 6\ln^2 2- \frac{8\pi^2}{3} \ln^2 2 - 4\ln2 + 9\pi^2 \ln 2 
    + 56\zeta_3 \ln 2 + 64\,{\rm Li}_4\bigg(\frac{1}{2}\bigg) \,. \nno
\end{align} 
This result was presented earlier in the supplemental material to \cite{Becher:2015hka}, but the finite terms were given only in numerical form.

\subsubsection*{Higher-order jet functions}

The last unknown ingredients in the factorization formula (\ref{NA_NNLO}) involve the one-loop corrections to the 2-particle jet function as well as the 3-particle jet functions with parton content $qgg$ and $q\bar q'q'$ (summed over flavors $q'$). Their combined contribution to the cross section can schematically be written as
\begin{equation}\label{eq:J2J3}
   2\sigma_0 \left(\frac{\alpha_0}{4\pi}\right)^2 
    \big\langle \bm{\mathcal{J}}_{\!\!2}^{(2)} \otimes \bm{1}
    + \bm{\mathcal{J}}_{\!\!3}^{(2)} \otimes \bm{1} \big\rangle\,.
\end{equation}
Some sample Feynman diagrams for these contributions are shown in Figure~\ref{fig:J2_J3_dia}. We have not computed these contributions individually but have inferred their divergent parts from the finiteness of the cross section. The explicit result is given in Appendix~\ref{NarrowCone}.

\subsection{NNLO cross section}

We now have all the ingredients at hand to obtain the full NNLO result for the cone-jet cross section. The bare functions need to be combined according to the NNLO expansion \eqref{NA_NNLO} of the factorization formula (\ref{factformula}). After coupling renormalization all divergences cancel and we get a finite result for the Laplace-transformed cross section $\widetilde{\sigma}(\tau,\delta)$. Despite the fact that we have not explicitly computed the two-loop divergences of the jet functions, this provides a highly nontrivial check of the factorization formula (\ref{factformula}), since the individual two-loop ingredients all depend on different scales. After expanding in $\e$, the divergences then involve logarithms of the different scales, which must cancel in the cross section. We stress that one would not obtain a finite result starting from the ``standard'' factorization formula \eqref{HJJS} involving only two soft Wilson lines. Beyond one-loop order the nontrivial Wilson-line structure in (\ref{factformula}) becomes an essential feature.

\begin{figure}[t!]
\centering
\hspace{-0.5cm}
\includegraphics[width=\textwidth]{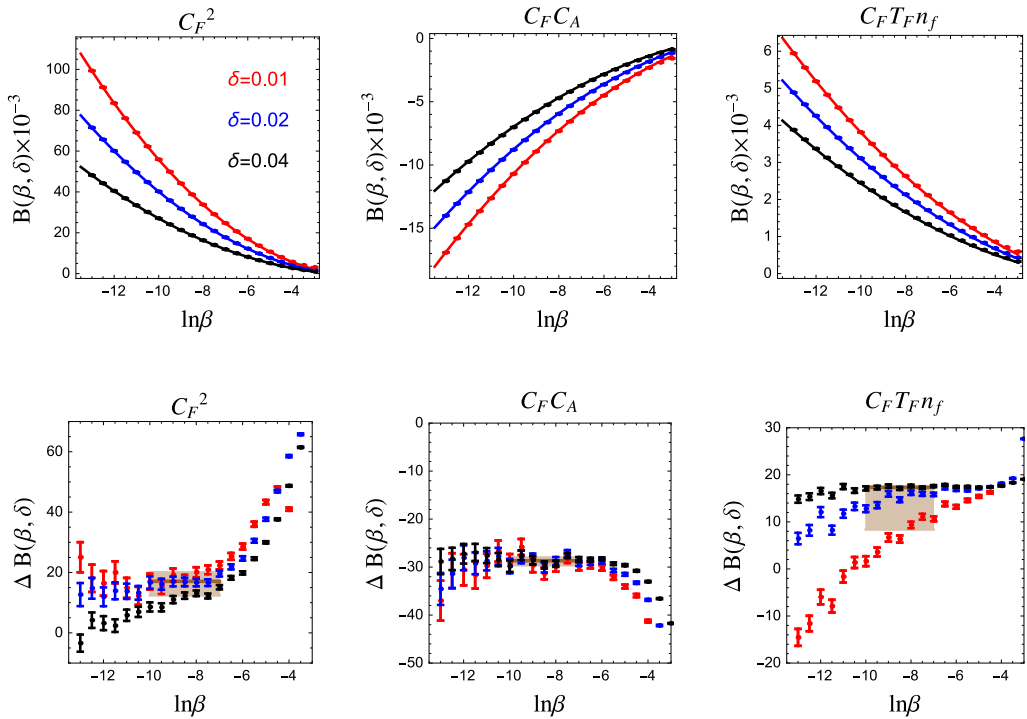}  
\vspace{-0.5cm}
\caption{\label{NJet_Xsec_NLO} 
Comparison of our analytic results (solid lines) for the coefficients of the three color structures in the two-loop coefficient $B(\beta,\delta)$ for $c_2^F=c_2^A=c_2^f=0$ with numerical results (points with invisibly small error bars) obtained using the {\sc Event2} event generator \cite{Catani:1996vz}. In the lower panels we show the difference $\Delta B$ between {\sc Event2} and our result, which for small values of $\beta$ and $\delta$ should be equal to the unknown two-loop constants $c_2^{F,A,f}$. Our fit results for these coefficients are shown by the brown squares. See text for further explanations.}
\end{figure}

Up to the desired order, the Laplace-transformed cross section is a quadratic polynomial in $\ln \tau$. For such a function, the Laplace transformation (\ref{SoftLaplace}) can be inverted by means of the simple substitutions 
\begin{equation}
   \ln\tau \to \ln\beta \,, \qquad \ln^2\tau \to \ln^2\beta - \frac{\pi^2}{6} \,.
\end{equation}
We choose $\mu=Q$ for the renormalization scale of the strong coupling and write
\begin{align}\label{NNLOxsec_narrowjet}
\frac{\sigma(\beta,\delta)}{\sigma_0} = 1 + \frac{\as}{2\pi}\,A(\beta,\delta) + \left(\frac{\as}{2\pi}\right)^2 B(\beta,\delta)+ \dots\, .
\end{align}
We follow the standard convention and define $A(\beta,\delta)$ and $B(\beta,\delta)$ as the coefficients in an expansion in $\alpha_s/(2\pi)$, while we expand in $\alpha_s/(4\pi)$ in the rest of the paper. The explicit result for the one- and two-loop coefficients reads
\begin{align} \label{Bterms}
A(\beta,\delta) &= C_F\Big[- 8\ln\delta\ln\beta - 6\ln\delta - 1 + 6\ln 2 \Big]\, , \nonumber\\
   B(\beta,\delta) &= C_F^2 \left[ 
    \left( 32\ln^2\beta + 48\ln\beta + 18 - \frac{16\pi^2}{3}  \right) \ln^2\delta 
    + \left( - 2 + 10\zeta_3  - 12\ln^2 2 + 4\ln 2 \right) \ln\beta \right. \nonumber\\
   &\hspace{1.2cm}\left. \mbox{}+ \left( (8 - 48\ln2) \ln\beta + \frac{9}{2} + 2\pi^2  - 24\zeta_3
     - 36\ln2 \right) \ln\delta + c_2^F \right] \nonumber\\
   &\hspace{-0.4cm}+ C_F C_A \left[ \left( \frac{44\ln\beta}{3} + 11 \right) \ln^2\delta 
    - \frac{2\pi^2}{3} \ln^2\beta 
    + \left(  \frac{8}{3} - \frac{31\pi^2}{18} - 4\zeta_3  - 6\ln^2 2 - 4\ln 2 \right) \ln\beta 
    \right. \nonumber\\
   &\hspace{1.2cm} \left. \mbox{}+ \left( \frac{44\ln^2\beta }{3} 
    + \left( - \frac{268}{9} + \frac{4\pi^2}{3}  \right) \ln\beta - \frac{57}{2} + 12\zeta_3 
    - 22\ln 2 \right) \ln\delta + c_2^A \right] \nonumber\\
   &\hspace{-0.4cm}+ C_F T_F n_f \left[ \left( - \frac{16\ln\beta}{3} - 4 \right) \ln^2\delta 
    + \left( - \frac{16}{3} \ln^2\beta + \frac{80\ln\beta}{9} + 10 + 8\ln 2 \right) \ln\delta  \right. \nno \\
    &\hspace{1.2cm} \left. \mbox{} + \left( - \frac{4}{3}  + \frac{4\pi^2}{9} \right) \ln\beta + c_2^f \right] .
\end{align}
This is a nontrivial new result, which shows the complete logarithmic dependence on the small parameters $\delta$ and $\beta$ in analytic form. The constants $c_2^F$, $c_2^A$ and $c_2^f$ are directly related to the unknown coefficients $c_2^{J,F}$, $c_2^{J,F}$ and $c_2^{J,f}$ in expression \eqref{eq:JfullConst} for the higher-order jet functions \eqref{eq:J2J3}. In the upper panels in Figure~\ref{NJet_Xsec_NLO} we compare our analytic expressions for the three color structures in $B(\beta,\delta)$ {\em without\/} the contributions from the unknown constants to high-precision numerical results obtained using the event generator {\sc Event2} \cite{Catani:1996vz}. We choose very small values for $\delta$ and $\beta$, so that the logarithmic terms become dominant and the power corrections in $\delta$ and $\beta$ can be ignored. At the scale of these plots we find perfect agreement. In the lower panels we show the differences $\Delta B$ between the numerical results from {\sc Event2} and our predictions, where the vertical scale is now expanded by a factor 1000. For sufficiently small $\delta$ and $\beta$ these differences should be equal to the unknown constants $c_2^{F,A,f}$. Fitting for the values of these constants in an intermediate region of $\ln\beta$ values, where numerical inaccuracies appear to be under control, we obtain
\begin{equation}\label{eq:twoloopcoeff}
   c_2^F = 17.1^{+3.0}_{-4.7} \,, \qquad
   c_2^A = -28.7^{+0.7}_{-1.0} \,, \qquad
   c_2^f = 17.3^{+0.3}_{-9.0} \,.
\end{equation}
The uncertainty on the last constant is fairly large due to numerical instabilities in the four-fermion channel. The same numerical problems were observed for thrust \cite{Becher:2015gsa}, where the analytic result for the constant is known \cite{Monni:2011gb,Kelley:2011ng}.

The above fixed-order expression for the cross section makes it clear that the cross section depends in a most complicated way on ratios of the four relevant and hierarchical physical scales $Q\gg Q\delta\gg Q\beta\gg Q\delta\beta$. Our factorization formula separates these scale and factorizes this complicated dependence to all orders in perturbation theory.

\subsection{Renormalization at NNLO \label{sec:renormNarrow}}

We now perform the renormalization of the jet and coft functions, which are the new ingredients in our factorization formula (\ref{factformula}), and verify the properties we have discussed earlier in Section~\ref{sec:renormFirst}. The jet function $\bm{\mathcal{J}}_{\!\!1}$ in (\ref{J1res}) is trivial and does not require any renormalization. It follows that
\begin{equation}
   \bm{Z}_{11}^J(\hat\theta_1,Q\delta,\e,\mu) = \bm{1}
\end{equation}
to all orders of perturbation theory. The $Z$-factors for the jet functions have been defined in (\ref{eq:ZJdef}), 
and we express them in terms of the same variables as the jet functions themselves. According to the same relation the jet function $\bm{\mathcal{J}}_{\!\!2}$ in (\ref{j2eq}) requires a subtraction proportional to $\bm{\mathcal{J}}_{\!\!1}$. We obtain
\begin{equation}
   \bm{Z}_{12}^J(\hat\theta_1,\hat\theta_2,\phi_2,Q\delta,\e,\mu) 
   = \frac{C_F\alpha_s}{4\pi}\,\delta(\phi_2-\pi)\,\bigg\{
   \left( \frac{2}{\e^2} + \frac{3-4 L_c}{\e} \right) \delta(\hat\theta_2) 
   - \frac{4}{\e}\,\bigg[\,\frac{1}{\hat\theta_2}\,\bigg]_+ \bigg\}\,
   \bm{1} + \dots \,,
\end{equation}
where the dots refer to higher-order terms in $\alpha_s$. The renormalization factor contains logarithmic dependence on the collinear scale $Q\delta$, as is typical for Sudakov problems. The renormalized jet function is now obtained as (recall that $y=\hat\theta_{\rm min}/\hat\theta_{\rm max}$)
\begin{align}
   &\bm{\mathcal{J}}_{\!\!2}(\hat\theta_1,\hat\theta_2,\phi_2,Q\delta,\mu) 
    = \frac{C_F\alpha_s}{4\pi}\,\delta(\phi_2-\pi) \nonumber\\ 
   &\qquad \times \bigg\{
    \left( 4 L_c^2 -6 L_c + 7 - \frac{5\pi^2}{6} + 6\ln 2 \right) 
    \delta(\hat\theta_1)\,\delta(\hat\theta_2) 
    + 8 L_c\,\delta(\hat\theta_1)\,\bigg[\,\frac{1}{\hat\theta_2}\,\bigg]_+ 
    + 8\,\delta(\hat\theta_1)\,\bigg[\,\frac{\ln\hat\theta_2}{\hat\theta_2}\,\bigg]_+ \nonumber\\
   &\qquad\qquad + 4\,\frac{dy}{d\hat\theta_2}\,\bigg[\,\frac{1}{\hat\theta_1}\,\bigg]_+
    \frac{1+2y+2y^2}{(1+y)^3}\,\theta(\hat\theta_1-\hat\theta_2) \nonumber\\
   &\qquad\qquad + 4\,\frac{dy}{d\hat\theta_1}\,\bigg[\,\frac{1}{\hat\theta_2}\,\bigg]_+
    \left( 2\,\bigg[\,\frac{1}{y}\,\bigg]_+ - \frac{4+5y+2y^2}{(1+y)^3} \right)
    \theta(\hat\theta_2-\hat\theta_1) \bigg\}\,\bm{1} + \dots \,.
\end{align}

As we discussed earlier, the product of the jet, hard and soft renormalization factors 
\begin{equation}\label{eq:ZU2l}
   \bm{Z}^U(\{\underline{n}\},Q\delta\tau,\epsilon,\mu) 
   \equiv Z_H^{1/2}(Q,\epsilon,\mu)\,Z_S^{1/2}(Q\tau,\epsilon,\mu)\,
    \bm{Z}^J(\{\underline{n}\},Q\delta,\epsilon,\mu)
\end{equation}
must renormalize the coft functions, see (\ref{eq:renCond}). The factors $Z_H$ and $Z_S$ can be reconstructed from the RG evolution equations \eqref{RGsofthard} or read off from the bare results for the corresponding functions, which are listed in Appendix~\ref{NarrowCone}. To the order we are working, the renormalization of the coft function takes the form
\begin{equation}\label{eq:Uren}
\begin{aligned}
   \widetilde{\bm{\mathcal{U}}}_1(\mu) 
   &= \bm{Z}^U_{11}\,\,\widetilde{\bm{\mathcal{U}}}_1(\epsilon) 
    + \bm{Z}^U_{12}\,\hat{\otimes}\,\,\widetilde{\bm{\mathcal{U}}}_2(\epsilon) 
    + \bm{Z}^U_{13}\,\hat{\otimes}\,\bm{1} + \mathcal{O}(\alpha_s^3) \,, \\
   \widetilde{\bm{\mathcal{U}}}_2(\mu) &= \bm{Z}^U_{22}\,\,\widetilde{\bm{\mathcal{U}}}_2(\epsilon)
    + \bm{Z}^U_{23}\,\hat{\otimes}\,\bm{1} + \mathcal{O}(\alpha_s^2) \,,
\end{aligned}
\end{equation}
where we have used the fact that $\widetilde{\bm{\mathcal{U}}}_3=\bm{1}+\mathcal{O}(\alpha_s)$. For simplicity we only indicate the arguments $\mu$ and $\e$ in order to differentiate between renormalized and bare functions. The off-diagonal contributions depend on the directions of the additional partons, and the symbol $\hat{\otimes}$ indicates that one has to integrate over these. In the first relation above we do not know the two-loop contributions to the renormalization factors $\bm{Z}^J_{12}$ and $\bm{Z}^J_{13}$ which enter in (\ref{eq:ZU2l}). Similarly, in the second relation we do not know the one-loop contribution to the renormalization factors $\bm{Z}^J_{23}$. The corresponding contribution to (\ref{eq:Uren}) is a series of $1/\e^n$ pole terms, whose coefficients depend on the collinear logarithm $L_c$. It is a nontrivial check of our method that in the remaining pole terms all reference to other logarithms cancels out. After some algebra, we find (at $\hat\theta_1=0$)
\begin{equation}
   \widetilde{\bm{\mathcal{U}}}_1(0,Q\delta\tau,\mu) 
   = 1 + \frac{C_F\alpha_s}{4\pi} \left( -4 L_t^2 - \frac{\pi^2}{2} \right)
    + \left(\frac{\alpha_s}{4\pi}\right)^2 \left( C_F^2\,u_1^F + C_F C_A\,u_1^A + C_F T_F n_f\,u_1^f \right) + \dots \,,
\end{equation}
with
\begin{align}
   u_1^F &= 8 L_t^4 + 2\pi^2 L_t^2 + \frac{\pi^4}{8} \,, \nno \\
   u_1^A &= \frac{88 L_t^3}{9} - \frac{268 L_t^2}{9}
    + \left( \frac{844}{27} - \frac{22\pi^2}{9} - 28\zeta_3 \right) \!L_t 
    - \frac{836}{81} - \frac{1139\pi^2}{108} - \frac{187\zeta_3}{9} +\frac{4\pi^4}{5} \,, \nno \\
   u_1^f &= - \frac{32 L_t^3}{9} + \frac{80 L_t^2}{9} 
    + \left( - \frac{296}{27} + \frac{8\pi^2}{9} \right) \!L_t - \frac{374}{81} + \frac{109\pi^2}{27}
    + \frac{68\zeta_3}{9} \,.
\end{align}
To compute the coefficient $u_1^A$ we have made use of relations (\ref{eq:g1integ}). Note that the renormalized coft function only depends on the single coft scale, via $L_t=\ln\frac{Q\delta\tau}{\mu}$, and the two-loop coefficient $u_1^F$ is one half of the square of the one-loop result, as required by non-abelian exponentiation. The one-loop expression for the renormalized coft function $\,\widetilde{\bm{\mathcal{U}}}_2$ reads
\begin{equation}
   \widetilde{\bm{\mathcal{U}}}_2(\hat\theta_1,\hat\theta_2,\phi_2,Q\tau\delta,\mu) 
   = \bm{1} + \frac{\alpha_s}{4\pi} \left[ C_F\,u_F^{\rm ren}(\hat\theta_1)
    + C_A\,u_A^{\rm ren}(\hat\theta_1,\hat\theta_2,\phi_2) \right] \bm{1} + \dots \,,
\end{equation}
where
\begin{align}
   u_F^{\rm ren}(\hat\theta_1) 
   &= - 4 L_t^2 - 4 L_t\,\ln(1-\hat\theta_1^2) - \frac{\pi^2}{2} + f_0(\hat\theta_1) \,, \nonumber\\
   u_A^{\rm ren}(\hat\theta_1,\hat\theta_2,\phi_2) 
   &= - 2 L_t \left[ 2\ln(1-\hat\theta_2^2)
    - \ln(1-2\cos\phi_2\,\hat\theta_1\hat\theta_2+\hat\theta_1^2\hat\theta_2^2) \right]
    + g_0(\hat\theta_1,\hat\theta_2,\phi_2) \,.
\end{align}

\section{Two-loop logarithmic structure for wide-angle jet processes\label{sec:twoLoopWide}}

In this section, we calculate all two-loop ingredients of the cross section for producing two wide-angle jets. This allows us to fully predict all logarithmically enhanced pieces in the cross section also for $\delta \sim 1$. We then verify that the relations \eqref{Sfac} and \eqref{Hfac} among the ingredients in the wide- and narrow-jet cases are indeed fulfilled. Up to two-loop order, the factorized wide-angle cross section \eqref{sigbarefinal} is given by
\begin{align}\label{eq:sigmaNNLOwide}
\sigma(\beta,\delta) =  \big\langle \bm{\mathcal{H}}_2\otimes \bm{\mathcal{S}}_2 + \bm{\mathcal{H}}_3 \otimes \bm{\mathcal{S}}_3 + \bm{\mathcal{H}}_4 \otimes \bm{1} + \dots \big\rangle\,.
\end{align}
There is no need to perform a Laplace transform in this case, because only the soft functions depend on the parameter $\beta$. In analogy to the case of the jet and coft functions discussed in the previous section, $\bm{\mathcal{H}}_m$ and $ \bm{\mathcal{S}}_m$ depend on reference vectors $\{\underline{n}\} = \{n_1, \dots ,n_m\}$, and the symbol $\otimes$ indicates that one has to integrate over the directions of these vectors, see \eqref{eq:otimes}. To do so, one introduces angular integrations as in \eqref{JUnewdef}, but in contrast to the discussion in Section \ref{sec:twoloopNarrow} the angles $\theta_i$ of the particles with respect to the jet axes are now $\mathcal{O}(1)$. The soft function $\bm{\mathcal{S}}_2$ depends on two Wilson-line directions $n_1$ and $n_2$, and the net effect of the convolution with the hard function $\bm{\mathcal{H}}_2$ is that these vectors are set equal to the jet axes $n$ and $\bar{n}$. For higher $m$, the angular integrations become nontrivial.  The wide-angle case is conceptually simpler than the narrow-jet case, because the factorization theorem only involves hard and soft functions. However, the computation of the hard and soft functions is considerably more involved, because they are nontrivial functions of the jet opening angle $\delta$. In the following we will evaluate the cross section \eqref{eq:sigmaNNLOwide} using bare quantities.  Renormalization will be discussed in Section \ref{sec:renorm}.

The perturbative expansion of the hard functions $\bm{\mathcal{H}}_m$ with $m$ partons starts at $\mathcal{O}(\as^{m-2})$ and has the form
\begin{align}
\bm{\mathcal{H}}_m = \sigma_0 \sum_{n=m-2}^{\infty} \left(\frac{\alpha_0}{4\pi}\right)^n\bm{\mathcal{H}}_m^{(n)}\,.
\end{align}
For convenience, we factor out the Born cross section, so that $\bm{\mathcal{H}}_2^{(0)}\otimes {\bm 1} ={\bm 1}$. For the soft function $ \bm{\mathcal{S}}_m$ defined in (\ref{eq:Sn}) the expansion in powers of $\as$ reads 
\begin{align}
\bm{\mathcal{S}}_m = \bm{1} + \frac{\alpha_0}{4\pi}\,\bm{\mathcal{S}}_m^{(1)} + \left( \frac{\alpha_0}{4\pi} \right)^2 \bm{\mathcal{S}}_m^{(2)} + \cdots\,.
\end{align}
Inserting these expansions into \eqref{eq:sigmaNNLOwide}, we find for the NLO coefficient in the cross section  \eqref{NNLOxsec_narrowjet}
\begin{align}
A(\beta,\delta) = \frac{1}{2}\,\big\langle \bm{\mathcal{H}}_2^{(1)}\otimes \bm{1} + \bm{\mathcal{H}}_2^{(0)} \bm{\mathcal{S}}_2^{(1)} + \bm{\mathcal{H}}_3^{(1)} \otimes \bm{1} \big\rangle\,,
\end{align}
and for the NNLO coefficient we obtain
\begin{align}
B(\beta,\delta)=\frac{1}{4}\,\big\langle \bm{\mathcal{H}}_2^{(2)}\otimes \bm{1}  + \bm{\mathcal{H}}_2^{(0)}\otimes\bm{\mathcal{S}}_2^{(2)}+ \bm{\mathcal{H}}_2^{(1)}\otimes \bm{\mathcal{S}}_2^{(1)}+ 
  \bm{\mathcal{H}}_3^{(1)} \otimes \bm{\mathcal{S}}_3^{(1)}+ \bm{\mathcal{H}}_3^{(2)} \otimes \bm{1}+
 \bm{\mathcal{H}}_4^{(2)} \otimes \bm{1} \big\rangle \,.
\end{align}
In the following, we first evaluate the hard function $\bm{\mathcal{H}}_3^{(1)}$ and then describe in detail the calculation of the soft functions $\bm{\mathcal{S}}_2$ and $\bm{\mathcal{S}}_3$.
The finiteness of the cross section can be used to infer the higher-order unknown logarithmic terms in the contribution from the hard functions $\bm{\mathcal{H}}_3^{(2)}$ and $\bm{\mathcal{H}}_4^{(2)}$. 
As in the narrow-angle case, we compare the resulting one- and two-loop coefficients $A(\beta,\delta)$ and $B(\beta,\delta)$ with the numerical results obtained using the event generator {\sc Event2}. 
Finally, we study all the two-loop ingredients in the small-$\delta$ limit and verify the factorization formulas (\ref{Hfac}) and (\ref{Sfac}) at this order.

\subsection{Hard function}

\begin{figure}[t!]
\begin{center}
\begin{overpic}[scale=0.5]{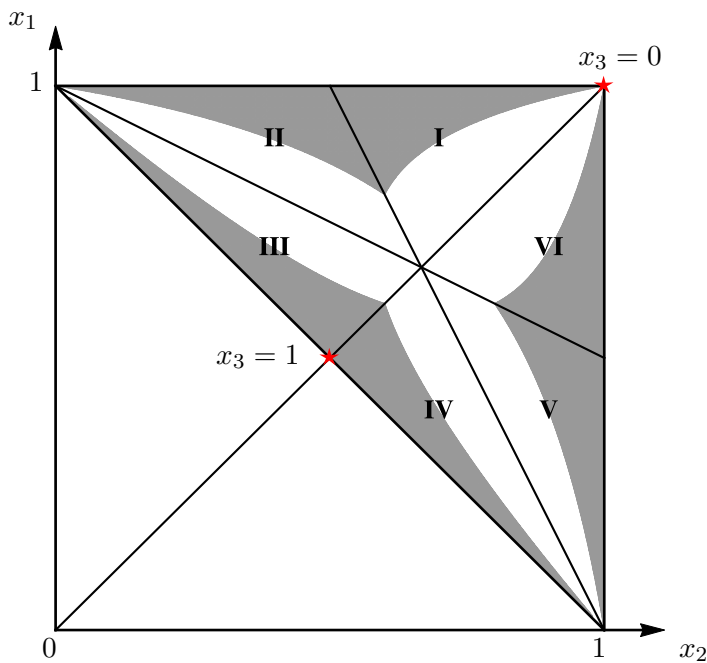}
\put(80,87){\color{black}{$x_3=0$}}
\put(26,42){\color{black}{$x_3=1$}}
\put(95,-2){\color{black}{$x_2$}}
\put(82,-2){\color{black}{$1$}}
\put(0,-2){\color{black}{$0$}}
\put(-2,83){\color{black}{$1$}}
\put(-5,93){\color{black}{$x_1$}}
\end{overpic}
\end{center}
\vspace{0ex}
\caption{Three-body phase space for the $\gamma^\ast(q)\to q(p_1)\,\bar{q}(p_2)\,g(p_3)$ process in different kinematic regions, corresponding to different energy hierarchies among the three partons. The gray region represents the part of phase-space in which the jet-angle constraint is fulfilled.}
\label{fig:H3}
\end{figure}

\begin{figure}[t!]
\begin{center}
\hspace*{0.6cm}
\begin{overpic}[scale=0.85]{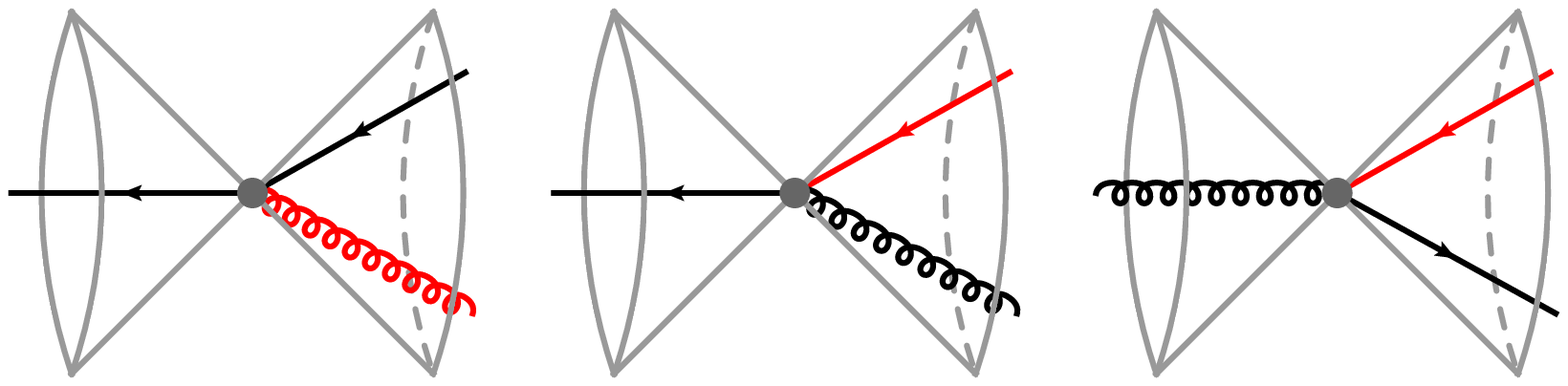}
\put(7,-3){\bf{I}: $x_1>x_2>x_3$}
\put(40.5,-3){\bf{II} : $x_1>x_3>x_2$}
\put(75,-3){\bf{III}:  $x_3>x_1>x_2$}
\put(-6,14){$ q (p_1)$}
\put(29.5,22){$\bar q (p_2)$}
\put(29.5,2){\color{red}{$g(p_3)$}}
\end{overpic}
\end{center}
\vspace{2ex}
\caption{Kinematical configurations in the first three regions defined in (\ref{regions1to6}). Particles with the smallest energy are drawn in red.}
\label{fig:H3_dia}
\end{figure}

Following the operator definition in (\ref{eq:Hm}), the hard function  $ \bm{\mathcal{H}}_3 $ describing the process $\gamma^* \to q(p_1)\,\bar{q}(p_2)\,g(p_3)$ starts at $\mathcal{O}(\as)$ and is given by
\begin{align}
 \frac{\alpha_0}{4\pi}\,\sigma_0 \, \bm{\mathcal{H}}_3^{(1)} = & \frac{1}{2Q^2 (2\pi)^{2-4\e}} \prod_{i=1}^3 \int d E_i \,E_i^{1-2\e} \, | \mathcal{M}_3(\{p_1,p_2,p_3\}) \rangle \langle \mathcal{M}_3(\{p_1,p_2,p_3\}) |  \nno \\
 &\hspace{1cm} \times \delta(Q-E_1-E_2-E_3) \,\delta^{(3-2\e)}(\vec{p}_1+\vec{p}_2+\vec{p}_3)\  \Theta_{\rm in}^{n\bar n}(\{p_1,p_2,p_3\})\,.
\end{align}
The integrations over the energies are performed keeping the directions of the three particles fixed. The crucial point is that for three-particle final states, the thrust axis always points opposite to the direction of the most energetic particle, and the jet cone therefore centers around it.  For this reason, it is natural to decompose
 the phase-space integration into the following regions:
\begin{equation}\label{regions1to6}
\begin{aligned}
  {\rm I}: &\quad x_1 > x_2 > x_3 \,, & {\rm II}: &\quad x_1 > x_3 > x_2 \,, & {\rm III}: &\quad x_3 > x_1 > x_2 \,, \\
  {\rm IV}: &\quad x_3 > x_2 > x_1 \,, & {\rm V}: &\quad x_2 > x_3 > x_1 \,, & {\rm VI}: &\quad x_2 > x_1 > x_3 \,,
\end{aligned}
\end{equation}
where we parameterize the particle energies as $x_i = 2 E_i/Q$. The corresponding three-body phase space is shown in Figure~\ref{fig:H3}. In Figure~\ref{fig:H3_dia} the kinematical configurations in the different regions are illustrated. Regions~I and VI suffer from overlapping soft and collinear divergences, while regions~II and V contain collinear divergences only. Regions~III and IV are infrared safe. Due to the $x_1\leftrightarrow x_2$ symmetry, only the contributions of regions~I, II and III need to be computed.

It is obvious from Figure \ref{fig:H3_dia} that the phase-space constraints for $\bm{\mathcal{H}}_3^{(1)}$  take a relatively complicated form. To perform the relevant integrations, it is convenient to use the energy and angle of the least energetic particle as variables for the hard function. In region~I, the gluon with momentum $p_3$ is the particle with the lowest energy. We parameterize its energy fraction $x_3$ and angle $\theta_3$ with respect to the thrust axis as
\begin{equation}\label{eq:paramOne}
x_3 = \frac{u}{2}\,\frac{1+\delta^2}{1+\delta^2-v\,\delta^2}\, , \qquad
 \sin^2\frac{\theta_{3}}{2} =\frac{v\,\delta^2}{1+\delta^2}\,.
\end{equation}
The variables $u$ and $v$ are integrated from $0$ to $1$ and disentangle overlapping soft and collinear divergences. Using momentum conservation one can express $x_3$ as a function of the angles $\theta_2$ and $\theta_3$, so that the variables $u$ and $v$ play the role of the angular variables $\hat{\theta}_i$ used in the previous section. Indeed, in the small-angle limit the variable $u$ is directly related to the ratio of the angles introduced for the jet function: $y=\hat\theta_{\rm min}/\hat\theta_{\rm max}=u/(2-u)$. Analogous parameterizations are introduced in regions~II and III; see Appendix \ref{widejetapp} for more details. The nontrivial part of the angular integration is then written as
an integral over $u$ and $v$
\begin{align}
 \bm{\mathcal{H}}_{3}(\{n_1,n_2,n_3\},Q,\delta,\e)& \otimes  \bm{\mathcal{S}}_{3}(\{n_1,n_2,n_3\},Q,\delta,\e)\nonumber\\
 &= \int_0^1 \!du \int_0^1\! dv\,\bm{\mathcal{H}}_{3}(u,v,Q,\delta,\e)\,\bm{\mathcal{S}}_{3}(u,v,Q \beta,\delta,\e)\,.
\end{align}
The one-loop hard functions $ \bm{\mathcal{H}}_3^{(1)}$ in the different sectors are given by 
\begin{equation}\begin{aligned}
   \bm{\mathcal{H}}_{3,{\rm I}}^{(1)}(u,v,Q,\delta,\e) 
   &= C_F\,4^\e\,C_\e\,e^{-2\e L_h}\,\delta^{-2\e}\,u^{-1-2\e}\,v^{-1-\e}\,h_3^{\rm I}(u,v,\delta,\e)\,\bm{1} \,, \\
   \bm{\mathcal{H}}_{3,{\rm II}}^{(1)}(u,v,Q,\delta,\e) 
   &= C_F\,4^\e\,C_\e\,e^{-2\e L_h}\,\delta^{-2\e}\,u^{-2\e}\,v^{-1-\e}\,h_3^{\rm II}(u,v,\delta,\e)\,\bm{1} \,, \\
   \bm{\mathcal{H}}_{3,{\rm III}}^{(1)}(u,v,Q,\delta,\e) 
   &= C_F\,4^\e\,C_\e\,e^{-2\e L_h}\,\delta^{2-2\e}\,u^{-2\e}\,v^{-\e}\,h_3^{\rm III}(u,v,\delta,\e)\,\bm{1} \,,
\label{h3eq}
\end{aligned}
\end{equation}
where as previously $L_h=\ln\frac{Q}{\mu}$ and $C_\e=e^{\e\gamma}/\Gamma(1-\e)$. The explicit expressions for $h^i_3(u,v,\delta,\e)$ are listed in Appendix \ref{widejetapp:1}. Convoluting these expressions with the trivial LO soft function $\bm{\mathcal{S}}_3=\bm{1}$ and combining all the kinematic regions, one finds
\begin{align}
\big\langle \bm{\mathcal{H}}_3^{(1)} \otimes \bm{ 1}\big\rangle =& \, C_F \,e^{-2\e L_h} \!\left[ \frac{4}{\epsilon^2} +  \frac{1}{\epsilon} \left(6- 8\ln \delta\right)  + 14 -\frac{5\pi^2}{3} + 12\left(1-\delta^4\right)\ln2 + 8\ln^2\delta - 12 \ln\delta\right. \nno \\
& \left. \hspace{1.4cm} \phantom{\frac{4}{\e^2}} -12 \,\delta^2 + 9\, \delta^4 + 8\,{\rm Li}_2(\delta^2)
+ {\cal O}(\epsilon) \right].
\end{align}

\subsection{Soft function}

\begin{figure}[t!]
\centering
\hspace{-0.5cm}
\begin{overpic}[scale=0.6]{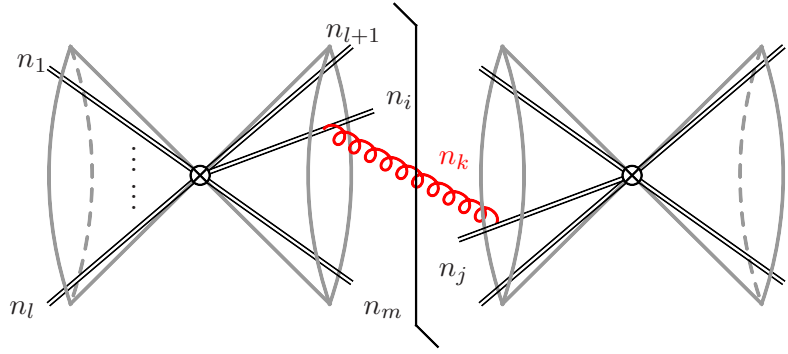}
\put(12,24){$\vdots$}
\put(12,19){$\vdots$}
\put(-3,38){$ n_1$}
\put(-4,5){$ n_l$}
\put(38,42){$ n_{l+1}$}
\put(46,33){$ n_i$}
\put(43,5){$ n_m$}
\put(53,10){$ n_j$}
\put(53,25){$\color[rgb]{1,0,0} {n_k}$}
\end{overpic}
\caption{Sample Feynman diagram for a general one-loop soft function $\bm{\mathcal{S}}_m(\{\underline{n}\},Q \beta)$. The vector $n_k$ points along the direction of the exchanged soft gluon.
\label{fig:S_m}}
\end{figure}

The soft function $ \bm{\mathcal{S}}_m$ is given by the vacuum expectation value of $m$ soft Wilson-line operators, as defined in (\ref{eq:Sn}). The LO contribution is the unit operator $\bm{1}$ in color space. The one-loop soft function $ \bm{\mathcal{S}}_m$ involves a sum of contributions from pairs of different Wilson lines, as shown in Figure \ref{fig:S_m}. Since the one-loop virtual corrections are scaleless, only real-emission diagrams contribute. We first calculate the one-loop correction to the function $\bm{\mathcal{S}}_2$ containing two Wilson lines. In momentum space, the bare soft function is given by
\begin{align}
\frac{\alpha_0}{4\pi}\,\bm{\mathcal{S}}_2^{(1)}(\{n,\bar{n}\},Q\beta,\delta,\epsilon) \!=\! \int \frac{d^d k}{(2\pi)^{d-1}} \, \delta^+(k^2) \, \frac{4 C_F g_{s}^2 \tilde{\mu}^{2\e}}{n\cdot k \, \bar{n}\cdot k} \left[ \Theta^{n\bar n}_{\rm in}(k) + \Theta^{n\bar n}_{\rm out}(k)\,\theta(Q\beta - 2 E_k)\right] \bm{1}\,.
\end{align}
Anticipating the convolution with the hard function we have set the reference vectors $n_1$ and $n_2$ of the Wilson lines equal to $n$ and $\bar{n}$. If the momentum $k$ is clustered into a jet the resulting integral is scaleless. 
Therefore, the one-loop function only receives a non-vanishing contribution from the out-of-jet region. After convolution with the LO hard function, the result reads 
\begin{equation}
   \big\langle  \bm{\mathcal{H}}_3^{(0)} \otimes \bm{\mathcal{S}}_2^{(1)} \big\rangle 
   = C_F\,e^{-2\e L_s} \left[ \frac{8\ln\delta}{\epsilon} - 8\,\text{Li}_2(-\delta^2) - 8\ln^2\delta - \frac{2\pi^2}{3}
    + {\cal O}(\e) \right] ,
\end{equation}
with $L_s=\ln\frac{Q\beta}{\mu}$. The full expression for this function, without expanding in $\e$, can be found in \cite{Kelley:2011aa}. For the two-loop soft function $\bm{\mathcal{S}}_2^{(2)}$ three color structures arise. The color-averaged expression has the form
\begin{align}\label{widejet_s2}
\big\langle \bm{\mathcal{H}}_3^{(0)} \otimes  \bm{\mathcal{S}}_2^{(2)} \big\rangle = e^{-4\e L_s}\left[ C_F^2 \, s_F(\delta,\e) + C_F C_A \, s_A(\delta,\e) + C_F T_F n_f \, s_f(\delta,\e) \right] .
\end{align} 
The $C_F^2$ term in this expression can be derived from the one-loop result by means of non-abelian exponentiation, while the $C_F C_A$ and $C_F T_F n_f$ terms can be extracted from the results given in \cite{Kelley:2011aa}. In Appendix \ref{widejetapp:2} we describe the necessary steps in detail and list the explicit two-loop expressions. 

At NNLO we also need the convolution of the one-loop soft function $\bm{\mathcal{S}}_3$ with the one-loop hard function $\bm{\mathcal{H}}_3$. First we calculate the expression $\bm{\mathcal{S}}_3^{(1)}$. Computing the amplitude squared and integrating over the energy of the gluon, one finds
\begin{align}
\bm{\mathcal{S}}_3^{(1)}\left(\{\underline{n}\},Q\beta,\delta,\e\right) = \frac{2}{\e} \, e^{-2\e L_s}\, e^{\e\gamma} (4\pi)^\e\, \sum_{(ij)} \bm{T}_i \cdot \bm{T}_j \int \frac{d\Omega(n_k)}{4\pi}\,\frac{n_i \cdot n_j}{n_i \cdot n_k \, n_k \cdot n_j} \, \Theta_{\rm out}^{n \bar n}(n_k)\, ,
\end{align}
with $i,j \in \{1,2,3\}$. The divergence arises from the energy integral. The angular integral is finite, since the gluon is always outside the cone and can therefore never become collinear to a Wilson-line vector whose direction tracks a hard parton inside the jet. In region~I the hard function $\bm{\mathcal{H}}_3$ suffers from double divergences. Therefore, in order to obtain all divergent terms after the convolution, we need to calculate $\bm{\mathcal{S}}_3^{(1)}$ up to $\mathcal{O}(\e)$. In contrast, in regions~II and III we need it only up to $\mathcal{O}(1)$ and $\mathcal{O}(\e^{-1})$, respectively. The explicit expressions are listed in Appendix \ref{sec:S3}.

\begin{table}[t]
\begin{center}
\begin{tabular}{|c|c|c|c|}
\hline
 & $\alpha=\pi/3$ &$\alpha=\pi/4$ &$\alpha=\pi/6$  \\
\hline
\hline
$M_F^{[1]}$& $-34.63$ & $-59.98$ & $-112.0$ \\
\hline
$M_A^{[1]}$& $\phantom{-}54.43$ & $\phantom{-}59.06$ & $\phantom{-}66.55$ \\
\hline
\end{tabular}
\end{center}
\caption{Numerical results for the $1/\e$ pole terms in $ \bm{\mathcal{H}}^{(1)}_3 \otimes \bm{\mathcal{S}}^{(1)}_3$ for three different cone sizes $\alpha$.}
\label{H3S3_num}
\end{table}

Performing the convolution of the one-loop hard and soft functions, we obtain
\begin{align}
\big\langle  \bm{\mathcal{H}}^{(1)}_3 \otimes \bm{\mathcal{S}}^{(1)}_3  \big\rangle 
 & = 4^\e\,C_{\e}\,e^{-2\e(L_h+L_s)}\,\delta^{-2\e} \left[ C_F^2 \, M_F(\delta,\e) + C_F \, C_A \, M_A(\delta,\e)\right] ,
\end{align}
where 
\begin{align}
   M_F(\delta,\e) &= \frac{32\ln\delta}{\e^3} + \frac{16}{\e^2} \left[
    \left( 3 - 4\ln2 \right) \ln\delta - 2\ln^2\delta - 2\,{\rm Li}_2(-\delta^2) 
    - \frac{\pi^2}{6} \right] + \frac{M_F^{[1]}(\delta)}{\e} + {\cal O}(\e^0) \,, \nonumber \\
   M_A(\delta,\e) &= \frac{8}{\e^2} \left[ - \rm{Li}_2(\delta^4) + \frac{\pi^2}{6} \right]
    + \frac{M_A^{[1]}(\delta)}{\e} + {\cal O}(\e^0) \,.
\end{align}
We have computed the singly-divergent terms in numerical form. In Table~\ref{H3S3_num} their values for some specific cone sizes are listed. Finite terms do not give rise to large logarithms in the cross section and are thus omitted.

Now we have all two-loop ingredients at hand, except for the purely hard contributions given by the two-loop hard functions $\bm{\mathcal{H}}_3^{(2)} $ and $\bm{\mathcal{H}}_4^{(2)}$. 
Finiteness of the cross section implies that we can obtain the divergent terms of the sum of these two-loop hard functions. These terms only depend on the hard scale $Q$, which
provides a nontrivial check of our computation of the various other ingredients. Explicitly, they must be of the form
\begin{align}\label{eq:H34}
\big\langle  \bm{\mathcal{H}}^{(2)}_3 \otimes \bm{1} + \bm{\mathcal{H}}^{(2)}_4 \otimes \bm{1}  \big\rangle 
= e^{-4\e L_h}\left(C_F^2 h_F + C_F C_A h_A + C_F T_F n_f h_f\right),
\end{align}
where the two-loop coefficients are functions of $\delta$, but independent of $\beta$. Our explicit results, which are presented in Appendix~\ref{sec:H34}, satisfy this condition. 

\subsection{NNLO cross section}

After combining all ingredients and performing charge renormalization, we obtain a finite NNLO cross section. Similar to the narrow-angle case, the coefficients $A$ and $B$ in (\ref{NNLOxsec_narrowjet}) can also be computed numerically using 
the event generator {\sc Event2} \cite{Catani:1996vz}, and a comparison with these results provides a check of the factorization formula at this order. 

\begin{figure}[t!]
\centering
\hspace{-0.5cm}
\includegraphics[width=1\textwidth]{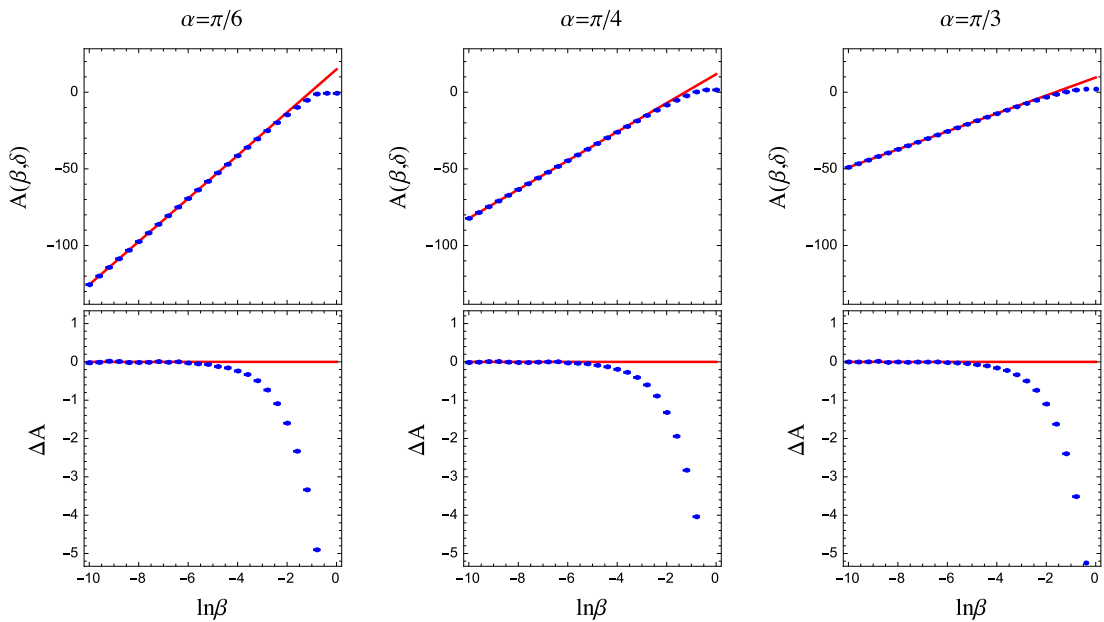}  
\vspace{-0.5cm}
\caption{Comparison of our analytic results (solid lines) for the coefficient $A(\beta,\delta)$ with numerical results (points with invisibly small error bars) obtained using the {\sc Event2} event generator \cite{Catani:1996vz}. In the lower panels we show the difference $\Delta A$ between {\sc Event2} and our result, which should vanish for small values of $\beta$. \label{Wide_Jet_NLO}}
\end{figure}

We find that the one-loop coefficient $A(\beta,\delta)$ reads 
\begin{equation}\label{WA_A}
   A(\beta,\delta) = C_F\,\Big[ - 8\ln\delta \ln\beta - 1 + 6\ln2 - 6\ln\delta - 6\delta^2
    + \left( \frac{9}{2} - 6\ln2 \right) \delta^4 + 4\,{\rm Li}_2(\delta^2) - 4\,{\rm Li}_2(-\delta^2) \Big] \,.
\end{equation}
Since we now count $\delta\sim 1$, this result holds for arbitrary values of $\delta$, up to terms suppressed by powers of $\beta$. In Figure~\ref{Wide_Jet_NLO}, we compare our analytical result for $A(\beta,\delta)$ (red line) to the numerical results obtained using {\sc Event2} (blue dotted). 
As it must be, the difference $\Delta A$ between the logarithmic terms and the full result goes to zero at small values of $\ln\beta$ within the numerical uncertainty of the Monte-Carlo integration. 

\begin{figure}[t!]
\centering
\hspace{-0.5cm}
\includegraphics[width=0.32\textwidth]{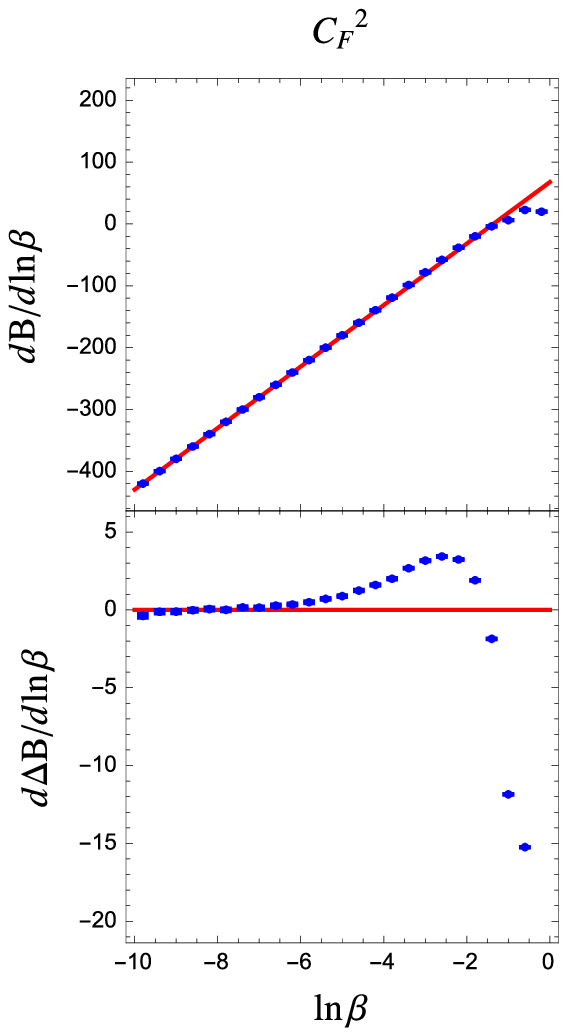}  
\hspace{0.1cm}
\includegraphics[width=0.32\textwidth]{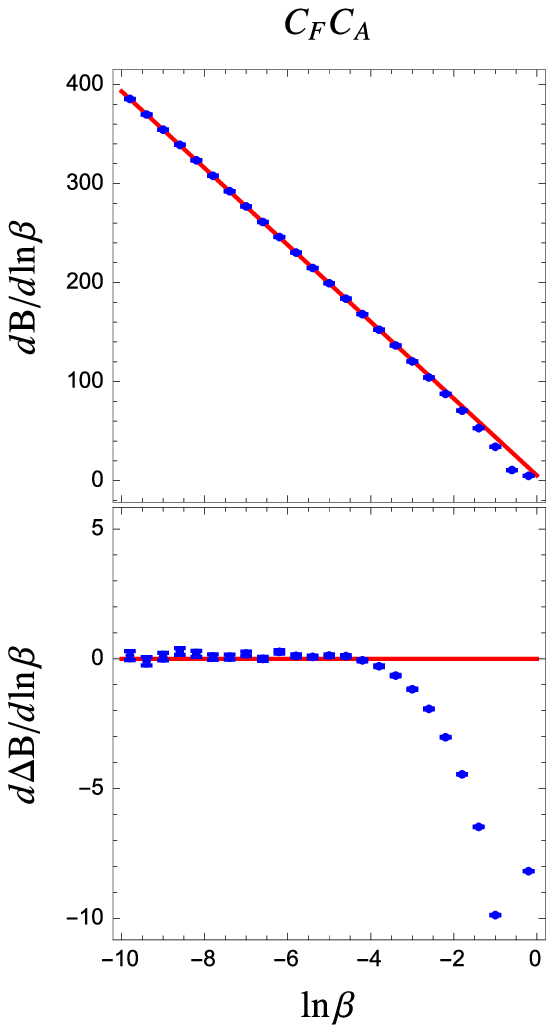} 
\hspace{0.1cm}
\includegraphics[width=0.32\textwidth]{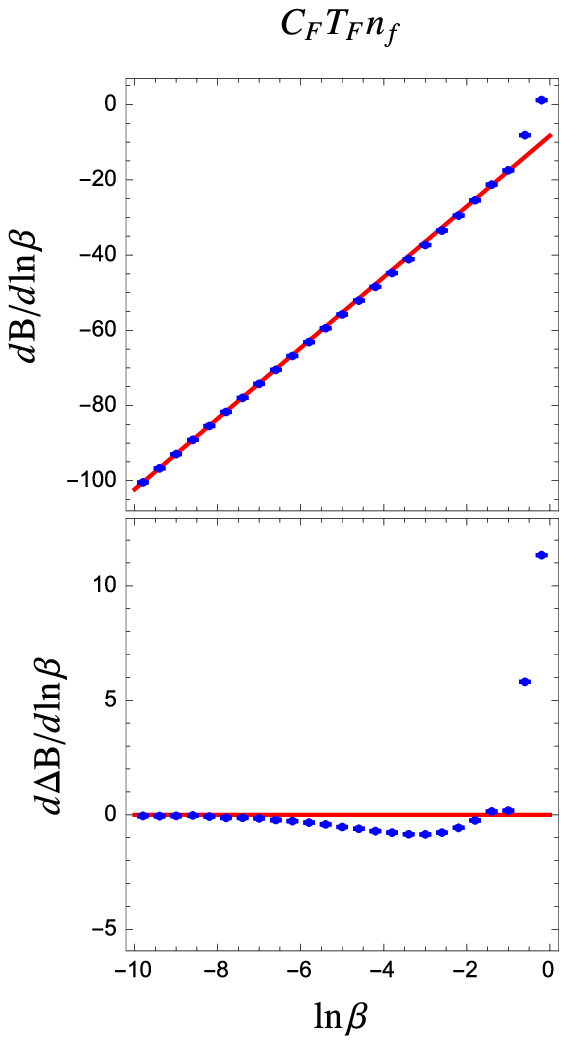}  
\vspace{-0.5cm}
\caption{Comparison of our analytic results (solid lines) for the coefficients of the three color structures in the two-loop coefficient $dB(\beta,\delta)/d\ln\beta$ with numerical results (points with invisibly small error bars) obtained using the {\sc Event2} event generator \cite{Catani:1996vz}. In the lower panels we show the difference $\Delta B$ between {\sc Event2} and our result, which should be equal for small values of $\beta$. The cone size is chosen as $\alpha=\pi/4$, corresponding to $\delta\approx 0.414$.  
\label{Wide_Jet_NNLO}}
\end{figure}

After combining all two-loop ingredients, we obtains the coefficient $B(\beta,\delta)$ at leading power in $\beta$ as 
\begin{align}\label{WA_B}
B(\beta,\delta) = C_F^2 B_F + C_F C_A B_A + C_F T_F n_f B_f \, ,
\end{align}
with 
\begin{align}
B_F = & \,32 \ln^2\delta \ln^2\beta + \frac{8}{3} \Bigg[ 4 \ln ^3\delta +12 \ln2\, \ln ^2\delta +9 \ln^2\delta-6 \ln ^2\left(1+\delta ^2\right) \ln \delta -\pi ^2 \ln
   \left(1+\delta ^2\right) \nno \\
   &  +12 \ln ^2 2 \ln \delta -18 \ln2 \ln \delta -\frac{5}{2} \pi ^2
   \ln \delta +24 \ln \delta -9 \, \text{Li}_2\left(-\delta
   ^2\right) +24 \ln \delta \,
   \text{Li}_2\left(-\delta ^2\right)\nno \\
   &-12 \ln \left(1+\delta ^2\right)
   \text{Li}_2\left(-\delta ^2\right)+12 \ln2 \,\text{Li}_2\left(-\delta
   ^2\right)+6\,\text{Li}_3\left(\frac{\delta ^2}{1+\delta ^2}\right)-6\, \text{Li}_3\left(\frac{1}{1+\delta ^2}\right) \nno \\
   & -\frac{3 \pi ^2}{4}+\pi ^2 \ln2 -\frac{3}{16}\, M_F^{[1]}(\delta)\Bigg] \ln \beta + c_2^F(\delta)\, , \nno \\ 
 B_A = & \,\frac{4}{3}\Bigg[ 11\ln\delta -\frac{\pi^2}{2} + 3\,{\rm Li}_2(\delta^4) \Bigg]\ln^2\beta + \frac{4}{3}\Bigg[11 \ln ^2\delta
   -\frac{67 \ln \delta }{3} +\frac{4 \delta ^4 \ln \delta }{\left(1-\delta
   ^4\right)^2} + \frac{1}{1-\delta ^4}  \nno \\
   &+36 \ln \delta  \ln ^2\left(1-\delta
   ^2\right)-12 \ln \delta  \ln ^2\left(1+\delta ^2\right)+22 \ln \delta  \ln
   \left(1-\delta ^2\right)-5 \pi ^2 \ln \left(1-\delta ^2\right) \nno \\
   &+22 \ln \delta  \ln
   \left(1+\delta ^2\right)-\pi ^2 \ln \left(1+\delta ^2\right) -4 \ln ^3\left(1+\delta ^2\right) +33 \,\text{Li}_2\left(-\delta ^2\right)+22 \,\text{Li}_2\left(\delta
   ^2\right)  \nno \\
   &+48 \ln \delta  \,\text{Li}_2\left(-\delta ^2\right)-12 \ln
   \left(1-\delta ^2\right) \,\text{Li}_2\left(-\delta ^2\right)-36 \ln \left(1+\delta
   ^2\right) \,\text{Li}_2\left(-\delta ^2\right) \nno \\
   & +12 \ln 2 \,\text{Li}_2\left(-\delta
   ^2\right) +24 \ln \delta  \,\text{Li}_2\left(\delta ^2\right) +24 \ln \left(1-\delta
   ^2\right) \,\text{Li}_2\left(\delta ^2\right)+12 \ln 2 \,\text{Li}_2\left(\delta
   ^2\right) \nno \\
   &+12 \ln \left(1-\delta ^4\right) \,\text{Li}_2\left(1-\delta ^2\right) -6 \,\text{Li}_3\left(1-\delta
   ^4\right) +24 \,\text{Li}_3\left(1-\delta ^2\right) -36 \,\text{Li}_3\left(-\delta ^2\right)\nno \\
   &-36 \,\text{Li}_3\left(\delta
   ^2\right)   +24 \,\text{Li}_3\left(\frac{\delta
   ^2}{1+\delta ^2}\right) -12\, \zeta_3-\frac{11 \pi
   ^2}{12}-\frac{1}{2}-\pi ^2 \ln 2 -\frac{3 }{8}M_A^{[1]}(\delta) \Bigg]\ln\beta \nno \\
   & + c_2^A(\delta)\, , \nno \\
B_f = & - \frac{16}{3} \ln\delta \ln^2\beta 
    - \frac{8}{3}\,\Bigg[ \frac{1}{1-\delta^4} + \frac{4\delta^4 \ln\delta}{\left(1-\delta^4\right)^2}  
    + 4 \ln(1-\delta^4) \ln\delta + 2 \ln^2\delta - \frac{10}{3} \ln\delta  \nno \\
   & + 6\,\text{Li}_2(-\delta^2)
    + 4\,\text{Li}_2(\delta^2) - \frac{\pi^2}{6} - \frac{1}{2} \Bigg] \ln\beta + c_2^f(\delta) \,. 
\end{align}
We have chosen $\mu=Q$ for convenience. The quantities $c_2^F$, $c_2^A$ and $c_2^f$ represent the unknown constant (with respect to $\ln\beta$) terms, which are functions of $\delta$. The above expressions extend our earlier result \eqref{Bterms} to arbitrary cone size $\delta$. To the best of our knowledge, we are the first to provide analytical formulas for the logarithmic terms in cone-jet cross sections. Such results can provide useful cross checks on numerical computations. In Figure~\ref{Wide_Jet_NNLO}, we compare our predictions for $dB/d\ln\beta$ with the numerical results from {\sc Event2}. The fact that the results are consistent with each other in the small-$\beta$ region provides a highly nontrivial check of the two-loop logarithmic structure of our analytic expression for the wide-angle jet cross section. 

\subsection{\boldmath The small-$\delta$ limit}
 
As a final check, we now evaluate all two-loop bare ingredients in the small-$\delta$ limit and verify that they fulfill the factorization formulas (\ref{Hfac}) and (\ref{Sfac}). The hard function $\bm{\mathcal{H}}_2$ is independent of the jet opening angle and the factorization (\ref{Hfac}) becomes trivial. 
Interesting relations first arise for the hard function $\bm{\mathcal{H}}_3^{(1)}$. In the small-$\delta$ limit, the contributions from regions { III} and { IV} are power suppressed. For the remaining regions, we indeed find
\begin{align}
\bm{\mathcal{H}}_{3, \,{\rm I}({\rm II})}^{(1)} \equiv \bm{\mathcal{H}}_{2+1, \,{\rm I}({\rm II})}^{(1)}  & =  \bm{\mathcal{J}}^{(0)}_{\!\!1}\,\bm{\mathcal{\overline{J}}}_{\!\!2, \, {\rm I}({\rm II})}^{(1)} \,.
\end{align}
For the two-loop soft function $ \bm{\mathcal{S}}_2$, after performing the Laplace transformation, we obtain
\begin{align}
\widetilde{\bm{\mathcal{S}}}_{2}(Q\tau,\delta) = \widetilde{S}(Q\tau) \,\, \widetilde{\bm{\mathcal{U}}}_1(Q\delta\tau) \,\, \widetilde{\overline{\bm{\mathcal{U}}}}_1(Q\delta\tau)\,.
\end{align} 
This factorization property of the cone-jet soft function was observed also in \cite{Chien:2015cka}. Similarly, for $ \bm{\mathcal{S}}_3$ we find at one-loop accuracy
\begin{align}
\widetilde{\bm{\mathcal{S}}}_{3, \,{\rm I}({\rm II})}(Q\tau,\delta) = \widetilde{S}(Q\tau)\,\,\widetilde{\bm{\mathcal{U}}}_1(Q\delta\tau) \,\, \widetilde{\overline{\bm{\mathcal{U}}}}_{2, \,{\rm I}({\rm II})}(Q\delta\tau) 
\end{align}
in each region. Since all relevant relations are fulfilled, we indeed reproduce the corresponding NNLO cross section (\ref{NNLOxsec_narrowjet}) in the small-$\delta$ limit. We have also verified that in this limit the one-loop coefficient (\ref{WA_A}) coincides with the one in (\ref{NNLOxsec_narrowjet}), and that all the logarithms $\ln\beta$ in the two-loop coefficient (\ref{WA_B}) are the same as those in the narrow-angle coefficient (\ref{Bterms}).

\section{Renormalization-group evolution\label{sec:renorm}}

In this section, we focus on the wide-angle case and check that renormalization works at the one-loop level as described in Section~\ref{sec:renormFirst}. To this end, we determine $\bm{Z}^H_{ij}$, the $Z$-factor for the hard function, and verify that the same matrix renders the soft function $\bm{\mathcal{S}}_m$ with $m$ Wilson lines finite. We then give results for the one-loop anomalous dimensions and show that the lowest-order RG evolution equation is equivalent to the BMS equation. 

\subsection{Renormalization at one-loop order}

Let us write the expansion of the $Z$-factor defined in (\ref{eq:renH}) in the form
\begin{equation}
\bm{Z}^H_{ij}(\{\underline n\},Q,\delta, \epsilon,\mu) =
\left\{
\begin{array}{rl}
\vspace{0.5cm}
&{\displaystyle \sum_{n=j-i}^{+\infty}} \left(\displaystyle\frac{\alpha_s}{4\pi}\right)^{n} \bm{z}_{i,j}^{(n)}(\{\underline n\},Q,\delta, \epsilon,\mu) \,; ~~~{\rm if}~ i\leqslant j\, ,  \\
&0 \,; ~~~{\rm if}~ i>j\, ,
\end{array}
\right.
\end{equation} 
with $z_{i,j}^{(0)}=1$. The entries $\bm{z}_{i,j}$ are matrices in the color space of the partons in the amplitude and its conjugate. We denote the color generators $\bm{T}_i^a$ acting on $i$-th particle in the amplitude on the left-hand-side of $\bm{\mathcal{H}}_m$ in \eqref{eq:Hm} as $\bm{T}_{i,L}^a$, and those acting on the conjugate amplitude on the right-hand side as $\bm{T}_{i,R}^a$. Because of the structure of \eqref{sigbarefinal}, the roles of $\bm{T}_{i,L}^a$ and $\bm{T}_{i,R}^a$ are reversed for the case of the soft function: the generators $\bm{T}_{i,L}^a$ act on the right-hand side of $\bm{\mathcal{S}}_m$. 

Let us now verify that $\bm{Z}_H$, which is introduced to absorb the divergences of the hard function, can indeed be used to renormalize the one-loop soft function. If this is true, we must find that
\begin{equation}\label{Ssub}
\sum_{l\geq m} \bm{Z}^H_{ml}(\{\underline n\},Q,\delta, \epsilon,\mu)\,\hat{\otimes}\, \bm{\mathcal{S}}_l (\{\underline{n}\},Q\beta,\delta, \epsilon)  = \bm{\mathcal{S}}_m (\{\underline{n}\},Q \beta,\delta, \mu) ={\rm finite}\,.
\end{equation}
Due to the structure of the matrix, only the diagonal terms $\bm{z}_{m,m}$, and the terms $\bm{z}_{m,m+1}$ above the diagonal can contribute to the renormalization of  $\bm{\mathcal{S}}_m$ at the one-loop-level. Explicitly, the finiteness condition at one-loop order reads
\begin{align}\label{eq:oneloopFinite}
\frac{\alpha_s}{4\pi} \, \bm{z}^{(1)}_{m,m}(\{\underline{n} \},Q,\delta,\e,\mu) &+ \frac{\alpha_s}{4\pi}
 \int \frac{d\Omega(n_{m+1})}{4\pi}\, \bm{z}^{(1)}_{m,m+1}(\{\underline{n} ,n_{m+1}\},Q,\delta,\e,\mu) \nno\\
&+ \bm{\mathcal{S}}_m (\{\underline{n} \},Q\beta,\delta, \e) = {\rm finite}\,,
\end{align} 
where we have used $\bm{\mathcal{S}}_m=\bm{1}+\mathcal{O}(\alpha_s)$, so that the $Z$-factors multiply the identity matrix. In the second term we integrate over the angle of the additional emission.  

One can easily obtain the divergent part of the one-loop soft functions, since it is given by a sum of exchanges between two legs. A sample Feynman diagram is shown in Figure~\ref{fig:S_m}. We get 
\begin{equation}\label{eq:Sdiv}
\bm{\mathcal{S}}_m(\{\underline{n}\},Q\beta,\delta, \e) = \bm{1} + \frac{\alpha_s}{2\pi\e} \sum_{(ij)} \bm{T}_i \cdot\bm{T}_j \int \frac{d\Omega(n_k)}{4\pi}\, W_{ij}^k\, \Theta_{\rm out}^{n\bar n}(n_k) \,, 
\end{equation}
where we have introduced the dipole radiator
\begin{equation}
W_{ij}^k=\frac{n_i\cdot n_j}{n_i\cdot n_k\, n_j\cdot n_k}\,.
\end{equation}
The function $ \Theta_{\rm out}^{n\bar n}(n_k)=1-\Theta_{\rm in}^{n\bar n}(n_k)$ ensures that the gluon is outside the two jet cones around the $n$ and $\bar{n}$ directions. Note that the angular integral does not suffer from collinear divergences, since the vectors $n_i$ and $n_j$ lie inside the jet cones, while the direction $n_k$ associated with the soft emission points outside the cone. (The soft radiation can also be emitted inside the cone, but as mentioned earlier this contribution is scaleless, since it does not have an upper limit on the energy of the emission.) 

In \eqref{eq:oneloopFinite}, the quantity $\bm{z}_{m,m}$ represents the divergences of the virtual corrections to the amplitude with $m$ legs, while $\bm{z}_{m,m+1}$ gives the divergences  from an additional real emission. Let us now consider the real and virtual corrections together, since all collinear divergences drop out and only a single soft divergence remains. The leading divergence can be obtained by using the soft approximation for the emitted (real or virtual) gluon. In the soft approximation, the real-emission contribution factorizes as
\begin{equation}
-g_s^2\sum_{(ij)} \int \frac{d^{d-1}k}{2E_k(2\pi)^{d-1}}\,\frac{1}{E_k^2}\,W_{ij}^k \, \bm{T}_{i,L}\cdot\bm{T}_{j,R}  \Theta_{\rm in}^{n\bar n}(k)\,\bm{\mathcal{H}}_m(\{\underline{n}\},Q-E_k)\,.
\end{equation}
In this approximation, one can write the virtual correction in the same form as the real-emission contribution, because the principal-value part of the propagator of the emission does not contribute. The virtual correction then reads 
\begin{equation}
+ g_s^2\sum_{(ij)} \int\!\frac{d^{d-1}k}{2E_k(2\pi)^{d-1}} \frac{1}{E_k^2}\,W_{ij}^k\,\frac{1}{2}(\bm{T}_{i,L}\cdot  \bm{T}_{j,L}+\bm{T}_{i,R}\cdot  \bm{T}_{j,R}) \, \bm{\mathcal{H}}_m(\{\underline{n}\},Q-E_k) \!\left[ \Theta_{\rm in}^{n\bar n}(k) +  \Theta_{\rm out}^{n\bar n}(k) \right] .
\end{equation}
The virtual gluon can be either inside or outside the cone. Since the gluon is soft, we can replace $Q-E_k \to Q$ in the hard function. Adding up the two terms and extracting the divergence from the lower end of the energy integration, the total result for the divergent part becomes
\begin{align}\label{ZHint}
\frac{\alpha_s}{4\pi}\,\bm{z}^{(1)}_{m,m}(\{\underline{n} \},Q,\delta,\e,\mu) &+\frac{\alpha_s}{4\pi}\, 
\int \frac{d\Omega(n_{m+1})}{4\pi}\,\bm{z}^{(1)}_{m,m+1}(\{\underline{n} ,n_{m+1}\},Q,\delta,\e,\mu) \nno \\
&=  - \frac{\alpha_s}{2\pi\e} \sum_{(ij)}  \bm{T}_i \cdot\bm{T}_j \int \frac{d\Omega(n_k)}{4\pi}\, W_{ij}^k\,\Theta_{\rm out}^{n\bar n}(n_k) \,.
\end{align}
Since the color factors are contracted with the trivial tree-level soft function, we do not need to distinguish the left and right color generators. Note that inside the cone the real and virtual corrections have cancelled, so that the net result only gets contributions from out-of-cone radiation and precisely cancels against the divergence of the soft function. We see that the renormalization indeed works at the one-loop level. We have repeated the same exercise also for the narrow-jet case, see Appendix \ref{sec:oneloopRenNarrow}. In this case, we can give explicit expressions for the angular integrals. Again, we find that the divergences cancel as they should.

\subsection{Renormalization-group evolution at leading logarithmic level}

We now discuss the anomalous-dimension matrix $\bm{\Gamma}^H$ defined in \eqref{eq:gammaH}, which governs the RG evolution of the hard \eqref{eq:hrdRG} and soft functions \eqref{eq:sftRG}, and verify the agreement between the perturbative expansion of the BMS equation and our RG-based resummation method. In order to resum the leading logarithmic terms, the anomalous-dimension matrix is needed up to $\mathcal{O}(\alpha_s)$. It can be expressed as
\begin{align}
\bm{\Gamma}^H\left(\{\underline{n}\} ,Q,\delta,\mu\right)  = \frac{\alpha_s}{4\pi}\,\bm{\Gamma}^{(1)}\left(\{\underline{n}\},Q,\delta,\mu\right)  + \mathcal{O}(\alpha_s^2)\, ,
\end{align}
where
\begin{equation}\label{eq:gammaOne}
\bm{\Gamma}^{(1)} =  \left(
\begin{array}{ccccc}
   \, \bm{V}_{2} &   \bm{R}_{2} &  0 & 0 & \hdots \\
 0 & \bm{V}_{3} & \bm{R}_{3}  & 0 & \hdots \\
0 &0  &  \bm{V}_{4} &  \bm{R}_{4} &   \hdots \\
 0& 0& 0 &  \bm{V}_{5} & \hdots
   \\
 \vdots & \vdots & \vdots & \vdots &
   \ddots \\
\end{array}
\right).
\end{equation}
It follows from the discussion in the previous section that, in the soft approximation, the corresponding matrix elements are given by
\begin{align}\label{eq:oneLoopRG}
 \bm{V}_m &= \bm{\Gamma}_{m,m}^{(1)} =  2\,\sum_{(ij)}\,(\bm{T}_{i,L}\cdot  \bm{T}_{j,L}+\bm{T}_{i,R}\cdot  \bm{T}_{j,R})  \int \frac{d\Omega(n_k)}{4\pi}\, W_{ij}^k \left[ \Theta_{\rm in}^{n\bar n}(k) +  \Theta_{\rm out}^{n\bar n}(k) \right] , \nonumber\\
 \bm{R}_m &= \bm{\Gamma}_{m,m+1}^{(1)} = -4\,\sum_{(ij)}\,\bm{T}_{i,L}\cdot\bm{T}_{j,R}  \,W_{ij}^{m+1}\,  \Theta_{\rm in}^{n\bar n}(n_{m+1})\,.
\end{align}
The anomalous dimensions $\bm{V}_m$ and $\bm{R}_m$ depend on the directions $\{\underline{n}\}=\{n_1,\dots,n_m\}$ and colors of the hard partons, and the indices $i,j$ in the sum run from $1$ to $m$. The quantities $\bm{R}_m$ also depend on the additional direction $n_{m+1}$ of the real emission. The integration over this direction is performed after the multiplication with the soft function. 
At first sight, the expressions in \eqref{eq:oneLoopRG} look problematic, since the angular integrals involve collinear divergences. However, we know on general grounds that the collinear divergences must cancel when the anomalous-dimension matrix is applied to the soft functions, see Section \ref{sec:renormFirst}. We have observed this cancellation at the one-loop level in the previous section, and we will see explicitly in the following that the same pattern continues at higher orders. The expressions \eqref{eq:oneLoopRG} are therefore valid for the RG evolution of the soft function. To obtain an expression which is suitable for evolving the hard functions, one would need to regularize the angular integrations and extract the collinear divergences. Furthermore, the soft approximation would then not be appropriate to obtain the anomalous-dimension matrix.

At the soft scale $\mu_s\approx Q\beta$ the soft functions do not involve large logarithms, and the higher-order corrections are thus suppressed by powers of $\alpha_s$. These can be neglected at leading logarithmic accuracy, so that the soft functions reduce to the identity matrix
\begin{align}
\bm{\mathcal{S}}_m(\{\underline{n}\}, Q\beta,\delta, \mu_s) = \bm{1} + \dots \,.
\end{align}
To get the resummed cross section at leading logarithmic accuracy, we evolve the soft functions to the hard scale $\mu_h\approx Q$ using (\ref{softRG}). At this scale the hard functions do not involve any large logarithms, and hence all higher-order hard functions are suppressed, $\bm{\mathcal{H}}_m = \mathcal{O}(\alpha_s^{m-2})$. The only non-vanishing contribution arises from the lowest-order function
\begin{equation}
   \bm{\mathcal{H}}_2(\{n_1,n_2 \}) \otimes \bm{\mathcal{S}}_2(\{n_1,n_2 \}, Q\beta,\delta,\mu_h) 
   = \sigma_0\,\bm{\mathcal{S}}_2(\{n,\bar{n} \}, Q\beta,\delta,\mu_h) \,.
\end{equation}
The net effect of the convolution with $\bm{\mathcal{H}}_2$ is that the reference vectors $n_1$ and $n_2$ are set equal to the jet directions $n$ and $\bar n$, together with a  multiplication by the Born cross section. We thus find that at leading logarithmic order the resummed cross section is equal to
\begin{equation}\label{eq:LLresum}
\sigma_{\rm LL}(\delta,\beta) =  \sigma_0\,\big\langle \bm{\mathcal{S}}_2(\{n,\bar{n}\}, Q\beta,\delta,\mu_h)  \big\rangle
=\sigma_0  \sum_{m=2}^\infty \big\langle\bm{U}_{2m}^S(\{\underline{n}\},\delta,\mu_s,\mu_h) \,\hat{\otimes}\, \bm{1} \big\rangle \,,
\end{equation}
where the symbol $\hat{\otimes}$, which was introduced in \eqref{eq:softRen}, indicates that one has to integrate over the additional directions present in the higher-multiplicity anomalous dimensions $\bm{R}_m$ and $\bm{V}_m$.

It is of course highly nontrivial to obtain an explicit form for the formal expression \eqref{eq:US}, but it is easy to write down explicit expressions order by order. To do so we rewrite the exponent of the evolution matrix \eqref{eq:US}  at leading order in RG-improved perturbation theory in the form
\begin{equation}
\int_{\mu_s}^{\mu_h} \frac{d\mu}{\mu}\, \bm{\Gamma}^H_{nm}= \int_{\alpha(\mu_s)}^{\alpha(\mu_h)} \frac{d\alpha}{\beta(\alpha)}\, \frac{\alpha}{4\pi}\,\bm{\Gamma}_{nm}^{(1)} =\frac{1}{2\beta_0}\ln\frac{\alpha(\mu_s)}{\alpha(\mu_h)}\,\bm{\Gamma}_{nm}^{(1)}  \,,
\end{equation}
which corresponds to leading logarithmic accuracy. By using the prefactor
\begin{equation}\label{eq:runt}
t = \frac{1}{2\beta_0}\ln\frac{\alpha(\mu_s)}{\alpha(\mu_h)} =\frac{\alpha_s}{4\pi} \ln\frac{\mu_h}{\mu_s} +\mathcal{O}(\alpha_s^2)
\end{equation}
as our expansion parameter, we automatically include running coupling effects. Using the structure of \eqref{eq:gammaOne} and expanding the exponential in \eqref{eq:US}, we find for the first three coefficients of the expansion in $t$
\begin{align}\label{eq:S123}
\bm{\mathcal{S}}_2^{(1)}= &\,\bm{R}_2 + \bm{V}_2  ,  \nonumber \\
 \bm{\mathcal{S}}_2^{(2)}= &\frac{1}{2!}\left\{ \bm{R}_2\left(\bm{R}_3 + \bm{V}_3 \right) + \bm{V}_2\left(\bm{R}_2 + \bm{V}_2\right) \right\} , \\
 \,\bm{\mathcal{S}}_2^{(3)}=& \frac{1}{3!}\left\{ \bm{R}_2 \left[\bm{R}_3(\bm{R}_4 + \bm{V}_4) + \bm{V}_3(\bm{R}_3 + \bm{V}_3) \right] + \bm{V}_2 \left[\bm{R}_2 (\bm{R}_3 + \bm{V}_3)+ \bm{V}_2 (\bm{R}_2 + \bm{V}_2) \right]\right\} . \nonumber 
\end{align}
As explained after \eqref{eq:LLresum}, we have to integrate over the directions of the additional emissions but for brevity, suppress the integrations in the above expressions. Including the integration, the one-loop term reads
\begin{equation}\label{eq:S1expl}
   \bm{\mathcal{S}}_2^{(1)} = \bm{V}_2 + \bm{R}_2\,\hat{\otimes}\,\bm{1} 
   = \bm{V}_2 + \int\frac{d\Omega(n_3)}{4\pi}\,\bm{R}_2 \,.
\end{equation}
The structure of the result \eqref{eq:S123} is very simple. To obtain the result at the next order, one takes the existing result and adds an additional real emission plus a virtual correction to each term. This type of iterative structure is reminiscent of a parton shower, and it should therefore be possible to solve the evolution equation numerically, using Monte Carlo methods. Indeed such Monte Carlo methods have been used to perform resummations of NGLs, see e.g.\ \cite{Dasgupta:2001sh,Dasgupta:2014yra}. 

\begin{figure}[t!]
\centering
\hspace{1cm}\begin{overpic}[scale=0.85]{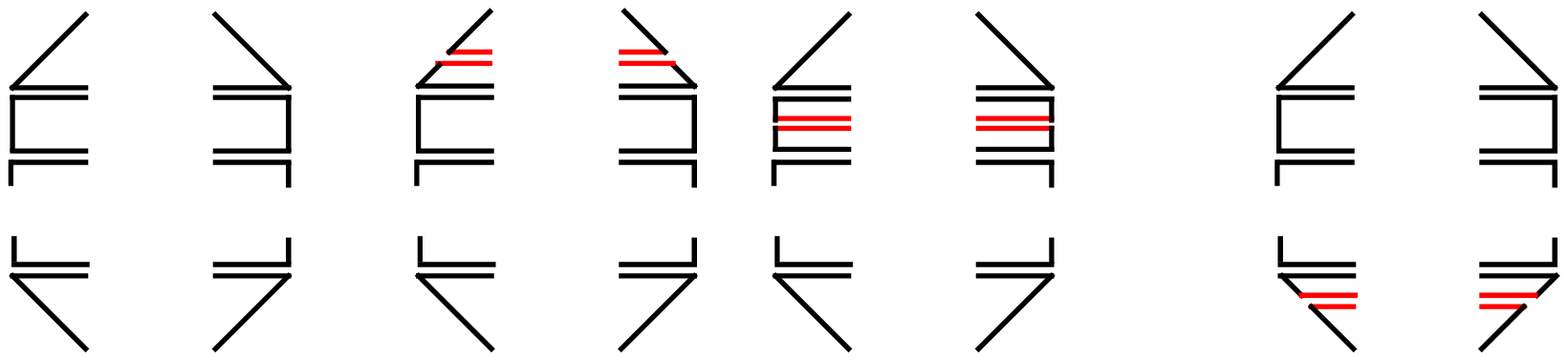}
\put(-7,11){$ \bm{R}_m$}
\put(-3,11){$\Bigg[$}
\put(20,11){$\Bigg]\hspace{0.05cm}=$}
\put(46,11){$+$}
\put(70,11){$+\,\,\cdots\,\,+$}
\put(8.5,22){$1$}
\put(8.5,0){$2$}
\put(8.5,16){$i_3$}
\put(8.5,12.5){$i_4$}
\put(8.5,4.5){$i_m$}
\put(32,19){$\color{red}m+1$}
\put(35,22){$1$}
\put(35,0){$2$}
\put(35,16){$i_3$}
\put(35,12.5){$i_4$}
\put(35,4.5){$i_m$}
\put(8.5,8){$\vdots$}
\put(35,8){$\vdots$}
\put(57.5,8){$\vdots$}
\put(90.5,8){$\vdots$}
\end{overpic}
\caption{The action of the operator $\bm{R}_m$ on an amplitude in the large-$N_c$ limit, where $(i_3, i_4, \cdots, i_m)$ is a permutation of $\{3,4,\cdots,m\}$. The double and single lines represent gluons and quarks, respectively. The sum on the right-hand side represents all the contributions from the planar diagrams.
\label{fig:BMS_dia}}
\end{figure}

\begin{figure}[t!]
\centering
\vspace{0.5cm}
\hspace{1cm}\begin{overpic}[scale=0.85]{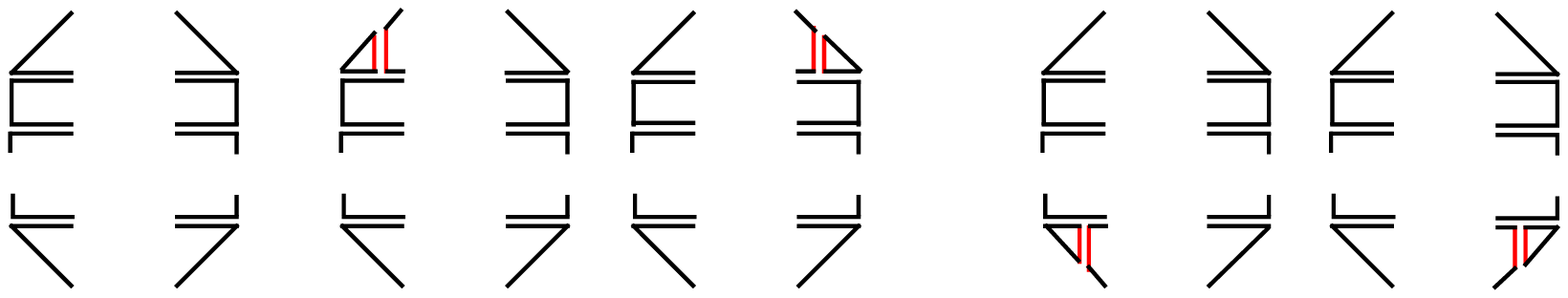}
\put(-8,9){$ 2\bm{V}_m$}
\put(-3,9){$\Bigg[$}
\put(16,9){$\Bigg]\hspace{0.01cm}=$}
\put(37,9){$+$}
\put(55.7,9){$+\,\,\cdots\,\,+$}
\put(82.3,9){$+$}
\put(6.5,18.5){$1$}
\put(6.5,-1){$2$}
\put(6.5,13.5){$i_3$}
\put(6.5,9.5){$i_4$}
\put(6.5,3){$i_m$}
\put(25,15.5){$\color{red}m+1$}
\put(28,18.5){$1$}
\put(28,-1){$2$}
\put(28,13){$i_3$}
\put(28,9.5){$i_4$}
\put(28,3){$i_m$}
\put(6.5,6){$\vdots$}
\put(28,6){$\vdots$}
\put(47,6){$\vdots$}
\put(73.5,6){$\vdots$}
\put(92,6){$\vdots$}
\end{overpic}
\caption{The action of the operator $\bm{V}_m$ on an amplitude in the large-$N_c$ limit.
\label{fig:Vl_LNc_dia}}
\end{figure}

One nontrivial complication is that the anomalous dimensions are matrices in color space and that the color algebra becomes nontrivial for high multiplicities. This difficulty can be avoided by taking the large-$N_c$ limit, which is also useful to compare \eqref{eq:S123} to the result obtained using the BMS equation. To take the large-$N_c$ limit, it is simplest to adopt the trace basis (see for example \cite{Sjodahl:2009wx}), i.e.\ to write the color structure of the $m$-particle amplitudes in the form
\begin{align}
 |\mathcal{M}_m(\{\underline{p}\}) \rangle = \sum_{\sigma \in P(m-2)} (t^{i_{\sigma(1)}} t^{i_{\sigma(2)}}  \,\cdots\, t^{i_{\sigma(m-2)}})_{ba}\, A_\sigma(\{\underline{p}\}) + \dots \,,
\end{align}
with color-ordered amplitudes $A_\sigma(\{\underline{p}\})$. The indices $a$ and $b$ are the color indices of the $\bar{q}q$ pair. We only include single trace terms, since contributions with multiple traces are suppressed at large $N_c$.  At large $N_c$, emissions arise only between nearest-neighbour legs, since all other attachments would lead to non-planar contributions which are suppressed. Based on the above simplification, the effect of $\bm{R}_m$ in the large-$N_c$ limit is shown diagrammatically in Figure \ref{fig:BMS_dia}.  The action of $\bm{V}_m$ simplifies analogously, as shown in Figure \ref{fig:Vl_LNc_dia}. The large-$N_c$ color factor from squaring the amplitudes is simply a factor of $N_c$ for each color loop, and the number of additional color loops is equal to the number of powers of $\alpha_s$, so that the color factor is obtained by switching to the 't Hooft coupling $\lambda=N_c\, \alpha_s$.

We now plug the explicit results \eqref{eq:oneLoopRG} for the anomalous-dimension coefficients $\bm{V}_m$ and $\bm{R}_m$ into the expressions \eqref{eq:S123}. For the coefficients of the expansion in $t$, we then obtain 
\begin{align}
   \bm{\mathcal{S}}_2^{(1)}
   &=  - 4N_c \int_{\Omega}{\bm 3}_{\rm Out}\, W_{12}^3 \,, \nno\\
   \bm{\mathcal{S}}_2^{(2)} &= \frac{\left(4N_c\right)^2}{2!} 
    \int_{\Omega}\Big[ -{\bm 3}_{\rm In}\,{\bm 4}_{\rm Out}  
    \left(P_{12}^{34} - W_{12}^3\,W_{12}^4\right) 
    + {\bm 3}_{\rm Out}\,{\bm 4}_{\rm Out}\,W_{12}^3\,W_{12}^4\Big] \,, \nno\\
   \bm{\mathcal{S}}_2^{(3)} &= \frac{\left(4N_c\right)^3}{3!} 
    \int_\Omega \Big[ {\bm 3}_{\rm In}\,{\bm 4}_{\rm Out}\,{\bm 5}_{\rm Out} 
    \left[ P_{12}^{34}\left( W_{13}^5 + W_{32}^5 + W_{12}^5\right) 
    - 2W_{12}^3\,W_{12}^4\,W_{12}^5 \right] \nonumber \\
    &  \hspace{0.5cm}  - \, {\bm 3}_{\rm In}\,{\bm 4}_{\rm In}\,{\bm 5}_{\rm Out}\,W_{12}^3  \left[ \left( P_{13}^{45} - W_{13}^4\,W_{13}^5 \right)+\left( P_{32}^{45} - W_{32}^4\,W_{32}^5\right) - \left( P_{12}^{45} - W_{12}^4\,W_{12}^5\right) \right]  \nno \\
&  \hspace{0.5cm}  - {\bm 3}_{\rm Out}\,{\bm 4}_{\rm Out}\,{\bm 5}_{\rm Out}\,W_{12}^3\,W_{12}^4\,W_{12}^5 \, \Big] \,,\label{eq:BMS3}
\end{align}
where $\int_{\Omega}{\bm 3}_{\rm Out}=\int\frac{d\Omega(n_3)}{4\pi} \,\Theta^{n\bar n}_{\rm out}(n_3)$, and we have used the abbreviation 
\begin{align}
P_{ij}^{kl} = W_{ij}^k\left(W_{ik}^l + W_{kj}^l\right).
\end{align}
The above expressions include all leading logarithms, i.e.\ the global and non-global logarithmic terms appear together. 

Let us now relate the above expressions to the leading logarithmic resummation of NGLs at large $N_c$, which can be obtained by solving the BMS equation \cite{Banfi:2002hw}
\begin{align}\label{eq:BMS}
\partial_{\hat{L}} G_{kl}(\hat{L}) = 
\int\frac{d\,\Omega(n_j)}{4\pi}\,W_{kl}^j \left[ \Theta_{\rm in}^{n\bar n}(j)\,G_{kj}(\hat{L})\,G_{jl}(\hat{L}) - G_{kl}(\hat{L}) \right] ,
\end{align}
with boundary condition $G_{kl}(0)=1$. The function $G_{kl}(\hat{L})$ depends on two light-like reference vectors $n_k$ and $n_l$. After solving the equation, the resummed soft function is obtained as $\bm{\mathcal{S}}(\{\underline{n}\}, Q\beta,\mu)=G_{12}(\hat{L})$ with $\hat{L}= 4N_c\,t$. While the non-linear integral equation \eqref{eq:BMS} can in general only be solved numerically, it is easy to solve it iteratively, order by order in $\hat{L}$. This was done in \cite{Schwartz:2014wha}, where the resulting non-global terms were listed up to three-loop order. Our expressions \eqref{eq:BMS3} agree with these results. This demonstrates the equivalence of the RG equation \eqref{eq:sftRG} driven by the one-loop anomalous dimensions \eqref{eq:gammaOne} and the BMS equation in the large-$N_c$ limit. However, an important advantage of our RG framework is that it is valid for arbitrary $N_c$ and logarithmic accuracy.

We have verified that the two-loop expression $\bm{\mathcal{S}}_2^{(2)}$ indeed reproduces the leading logarithmic term in the NNLO result \eqref{WA_B}. In the literature one often distinguishes global from non-global logarithms. The global part is defined as the exponentiated one-loop logarithm,
\begin{align}
\frac{\sigma_{\rm GL}^{\rm LL}}{\sigma_0}&= \exp\left(16\, C_F\, t \ln \delta\right) 
= 1 - \frac{\as}{2\pi} \,8 C_F \ln \delta \ln\beta \nno \\
&  + \left( \frac{\as}{2\pi} \right)^2 \left( 32\, C_F^2 \ln^2\delta  + \frac{44}{3}\,C_F C_A \ln\delta - \frac{16}{3}\, C_F T_F n_f \ln\delta \right)\ln^2\beta \,,
\end{align}
where the parameter $t$ was defined in \eqref{eq:runt}. After removing the global piece, we find that the leading non-global terms in \eqref{WA_B} have the form
\begin{align}
\frac{\sigma_{\rm NGL}^{\rm LL}}{\sigma_0} = \left(\frac{\alpha_s}{2\pi}\right)^2 C_F C_A\left[ - \frac{2\pi^2}{3} + 4\, {\rm Li}_2(\delta^4)\right] \ln^2\beta \,.
\end{align}
This agrees with the result of \cite{Dasgupta:2002bw}, which analyzed non-global logarithms in the presence of rapidity gaps, after one relates the rapidity gap $\Delta\eta$ to our cone parameter using $\delta=\exp\left(-\Delta\eta/2\right)$. From our point of view, the separation into global and non-global logarithms is somewhat artificial. Beyond leading logarithmic accuracy it is also not unique: in general, the non-global logarithms are defined by dividing the cross section through the cross section of a global observable, which has the same leading logarithms. Different reference observables have different subleading logarithms, so that the non-global subleading logarithms depend on the choice of the global observable.

The RG evolution approach developed in this work can be compared to the functional RG approach by Simon Caron-Huot \cite{Caron-Huot:2015bja}. In this formalism, unitary matrices $U(n)$ in color space, which are functions of a direction vector $n$, are used to track the contributions from different particle multiplicities. Using these matrices, the author defines the color-density matrix functional $\sigma[U]$, which is given by the cross section modified by multiplying final-state partons along directions $n$ by the matrix $U(n)$. Taking functional  derivatives with respect to these matrices then separates out the contributions to the cross section from particles traveling along the corresponding directions. The one-loop expressions for the anomalous dimensions governing the RG evolution of the color-density matrix defined in this reference are in one-to-one correspondence to the anomalous dimensions in \eqref{eq:oneLoopRG}. At leading logarithmic order, the resummed result is obtained from solving the corresponding RG evolution equation for trivial initial conditions and is the same as in our formalism. Beyond leading logarithmic accuracy, the relation between the two formalisms is less obvious. The approach \cite{Caron-Huot:2015bja} does not distinguish hard and soft partons but multiplies every parton by a matrix $U(n)$, and the hard function contains arbitrary hard radiation outside the jet. Furthermore, the Wilson-line structure, which is an important feature of our approach, is not immediately manifest. A one-to-one mapping should arise after an averaging procedure over $U(n)$, which sets those matrices corresponding to outside radiation to zero, effectively vetoing hard outside radiation, and multiplies the inside partons by soft Wilson lines \cite{simon}. The details of this procedure were not discussed in \cite{Caron-Huot:2015bja}, which was mainly concerned with the computation of the two-loop anomalous dimension. Recently the result for the anomalous dimension was even extended to three-loop order for planar $\mathcal{N}=4$ Super-Yang-Mills theory \cite{Caron-Huot:2016tzz}. Translated into our formalism, these results should provide an important ingredient for the resummation of subleading non-global logarithms.

Let us also compare our method in more detail with the approach put forward by Larkoski, Moult and Neill in \cite{Larkoski:2015zka}. Rather than attempting to directly arrive at a facorization theorem for a non-global observable, these authors imagine performing a set of increasingly differential measurements on the jet. The goal of these measurements is to isolate the regions of phase space where the jet contains soft subjets near the jet boundary, which then give rise to NGLs. The basic strategy is that resolving the substructure of the jet down to lower scales reduces the size of the NGLs. The resummation of the global logarithms associated with the subjet observables then resums part of the NGLs present in the original, inclusive cross section. However, at each stage non-factorized logarithms remain. A complete factorization theorem for a jet cross section with additional measurements would be at least as complicated as the factorization theorem we obtain for the inclusive cross section, and would include multi-Wilson-line operators. By construction, the method of \cite{Larkoski:2015zka} involves increasingly differential measurements, each of which requires a dedicated effective theory containing the relevant modes for the associated factorization theorem. The method thus involves a tower of effective theories with more and more degrees of freedom. This is rather different from our approach, which factorizes the original cross section and allows for the resummation of the NGLs with RG methods. Rather than with a tower of effective theories, we work with a given theory with a fixed number of degrees of freedom. The wide-angle case is especially simple in this regard, since it only involves a single, soft low-energy mode. Instead of soft subjets, we encounter soft Wilson lines along the directions of the energetic particles, and the NGLs get factorized into hard logarithms from vetoing hard emissions outside the jet, and soft logarithms due to the fact that soft radiation is not constrained inside the jet. The physical picture emerging from our approach is rather different from that put forward in \cite{Larkoski:2015zka}: We find that the NGLs arise from ``subjets'' which are generically hard. The leading logarithms can be obtained by using strongly-ordered amplitudes in which all gluons are soft, but this is only a feature of this particular approximation.

Solving the RG evolution equation (\ref{softRG}) to obtain the evolution between the hard and soft scales is obviously complicated. This is not surprising, given that even in the large-$N_c$ limit and at leading logarithmic order our equation is equivalent to the non-linear integral equation derived by BMS. However, from a conceptual point of view our RG resummation method is completely standard, and it is clear which ingredients are required for a given logarithmic accuracy. The expansion in subjets, on the other hand, is not immediately related to logarithmic accuracy. Even for leading logarithmic accuracy, terms with arbitrary many subjets contribute. The authors of \cite{Larkoski:2015zka} argue that the higher-order terms in the expansion are phase-space suppressed, but it is not clear to us if there is any parametric suppression associated with this assertion.

%The problems with traditional global factorization theorems become visible only at NNLO, and we have therefore evaluated all ingredients of our formula to this accuracy and verified that we reproduce the full QCD result. The paper \cite{Larkoski:2015zka} contains NLO computations of some ingredients, but even at this order no cross checks against QCD predictions are performed. Given the complexity of the relevant factorization theorems, we believe that such checks are important. 

\section{Summary and outlook}

In this paper, we have derived all-order factorization theorems for cone-jet cross sections, both in the wide-angle case and in the narrow-jet limit. With a veto on out-of-jet energy, these cross sections suffer from large logarithms of the ratio of the energy outside and inside the jets and, for narrow jets, from large logarithms of the opening angle of the jet cone. Our factorization theorems separate the physics associated with the different energy scales relevant in these processes and provide operator expressions for the various ingredients of the cross section. We believe that this is an important step forward, since it opens the door for higher-logarithmic resummations for such observables. In the context of the effective field-theory framework we are using, this resummation is achieved by solving the RG evolution equations of the relevant operators. 

The low-energy structure we encounter in jet cross sections is significantly more complicated than what is present for event-shape variables such as thrust. In the case of global dijet event shapes, soft emissions are obtained from an operator consisting of two Wilson lines along the jet directions. For cone-jet cross sections, we instead encounter a separate Wilson line for each energetic particle inside the jet. The presence of this multi-Wilson-line structure is quite intuitive for wide-angle jets, where the characteristic angle between the energetic particles is parametrically large and of the same size as the typical angle of soft emissions. The emissions thus resolve the individual energetic partons, and the soft Wilson lines for the particles inside a  jet cannot be combined into a single Wilson line. Interestingly, we find that the same structure also persists for narrow jets, because it turns out that narrow-angle soft radiation gives an important contribution to the cross section. The effective theory in the narrow-jet limit therefore contains a an additional mode, which is simultaneously soft and collinear. Boosted soft modes have appeared in other contexts in SCET; the new mode we find is different, because every component of its momentum is parametrically smaller than the corresponding one of the energetic, collinear particles. It is this basic property which leads to the multi-Wilson structure not present in earlier SCET applications. 

Starting from our factorization theorems, we have computed the full logarithmic structure of the NNLO jet cross sections both in the wide-angle and narrow-jet cases and have verified that they agree with numerical results obtained from the fixed-order event generator {\sc Event2}. We have also shown how the wide-angle cross section can be refactorized in the narrow-angle limit, and we have verified the resulting relations at NNLO. These results demonstrate that our approach indeed captures all logarithms present in these cross sections.

Since our factorization theorem involves operators with an arbitrarily large number of Wilson lines, the relevant RG evolution equation involves an infinite matrix, and it will be quite challenging to solve it. This complexity is perhaps not that surprising, given that jet cross sections are prime examples of non-global observables, and it is well known that even the leading logarithms in such processes have quite a complicated structure, which is captured by the BMS equation. Indeed, we have shown that our RG evolution equation is equivalent to this non-linear integral equation at the leading logarithmic level. We expect that a natural approach to solving our RG equation will be based on using Monte Carlo methods, since its structure is reminiscent of a parton shower. Also, for many practical applications it could be sufficient to predict the first few NGLs instead of resumming the entire series. In any case, as a next step, it will be important to develop methods to solve the RG equations and to obtain the relevant anomalous dimensions needed to resum subleading NGLs. 

Jet observables are the most important observables at colliders and there are countless applications for our framework once the methods for solving the associated RG equations have been developed. In particular, it would be interesting to use RG methods to systematically improve jet substructure predictions, which are currently mostly based on parton-shower programs. In our paper, we have analyzed the simplest example of a jet cross section, but it is clear that the basic multi-Wilson-line structure is present in all non-global observables. Nevertheless, there can be additional difficulties when trying to apply our approach to multi-jet cross sections. For example, it was stressed in \cite{Walsh:2011fz, Kelley:2012zs} that the complexity of the phase-space constraints for recombination jet algorithms might present a stumbling block when trying to derive all-order statements for such cross sections.  While it is interesting to extend the resummation to more complicated observables, it is equally important to find observables whose structure is simple enough to allow for systematic higher-order resummations. For example, in the case we analyzed, it turned out to be advantageous to choose the thrust axis as the jet axis rather than choosing the axis so as to maximize the energy inside the jet.

For simplicity, we have studied dijet cone cross sections in $e^+e^-$ collisions, but it will obviously be important to apply the same methods to hadronic jet cross sections. In fact, since there are no detectors at very large rapidities, {\rm every} hadron collider cross section contains two narrow cone jets along the beam direction. Their angular size is given by $\delta = e^{-y_{\rm max}}$, where $y_{\rm max}$ is the maximum rapidity up to which particles are measured. If a jet veto is imposed in the measured region, then the presence of these beam jets will induce logarithms of the same type as the ones we have studied in this paper. So far, these logarithms were only studied using parton showers \cite{Banfi:2012yh} (a method to mitigate their effects was proposed in \cite{Gangal:2014qda}), and it will be interesting to analyze them within our framework. An interesting complication in the case of hadronic collisions is the presence of Glauber gluons. An comprehensive effective-theory analysis of these has been presented recently \cite{Rothstein:2016bsq}, following earlier exploratory studies \cite{Idilbi:2008vm,Donoghue:2009cq,Bauer:2010cc,Donoghue:2014mpa,Fleming:2014rea}. Glauber effects are interesting in the context of non-global observables, because their presence can lead to enhanced powers of NGLs, the so-called super-leading logarithms \cite{Forshaw:2006fk,Forshaw:2008cq}. We look forward to studying these issues and applying our framework to many phenomenologically relevant problems.

\newpage
\begin{acknowledgments}	
We thank Andreas von Manteuffel and Rob Schabinger for help with the cone-jet soft function. We are grateful to Simon Caron-Huot and Andrew Larkoski for useful discussions. The research of M.N.\ is supported by the Advanced Grant EFT4LHC of the European Research Council (ERC), the Cluster of Excellence {\em Precision Physics, Fundamental Interactions and Structure of Matter\/} (PRISMA -- EXC 1098), and grant 05H12UME of the German Federal Ministry for Education and Research. T.B.\ is supported by the Swiss National Science Foundation (SNF) under grant 200020\_165786, and the research of L.R. is supported by the German Science Foundation (DFG) through the Emmy-Noether Grant No. TA 867/1-1. The authors thank the high-energy theory groups at Harvard and Heidelberg, the MITP Mainz, the INT Seattle and the KITP Santa Barbara for hospitality and support during different phases of this work.
\end{acknowledgments}

\appendix

\section{NNLO ingredients for the narrow-angle cross section}\label{NarrowCone}

Here we collect the bare ingredients entering the factorization formula for the narrow-jet cross section up to NNLO. The expansion coefficients for the hard function are given by
\begin{align}
   h_F &= - \frac{4}{\epsilon ^2} - \frac{6}{\epsilon} - 16 + \frac{7\pi^2}{3}  
    + \epsilon\left( - 32 + \frac{7\pi^2}{2} + \frac{28\zeta_3}{3} \right) 
    + \epsilon^2\left(  - 64 + \frac{28\pi^2}{3}  + 14\zeta_3 - \frac{73\pi^4}{360} \right) , \nno \\
   h_{2F} &= \frac{8}{\epsilon^4} + \frac{24}{\epsilon^3}
    + \frac{1}{\epsilon^2} \left( 82 - \frac{28\pi^2}{3} \right)
    + \frac{1}{\epsilon} \left(  \frac{445}{2}  - 26\pi^2 - \frac{184\zeta_3}{3} \right) \nno\\
    &\quad + \frac{2303}{4} - 86\pi^2 - 172\zeta_3+ \frac{137\pi^4}{45} \,, \nno\\
   h_A &= - \frac{11}{3\epsilon^3} + \frac{1}{\epsilon^2} 
    \left( - \frac{166}{9} + \frac{\pi^2}{3}  \right)
    + \frac{1}{\epsilon} \left( - \frac{4129}{54} + \frac{121\pi^2}{18} + 26\zeta_3 \right) \nno\\
    &\quad - \frac{89173}{324}  + \frac{877\pi^2}{27} + \frac{934\zeta_3}{9} 
     - \frac{8\pi^4}{45} \,, \nno \\
   h_f &= \frac{4}{3\epsilon^3} + \frac{56}{9\epsilon^2} 
    + \frac{1}{\epsilon} \left( \frac{706}{27} - \frac{22\pi^2}{9} \right)  + \frac{7541}{81}
     - \frac{308\pi^2}{27} - \frac{104\zeta_3}{9} \,.
\end{align}
For expansion coefficients of the Laplace-transformed soft function, we find
\begin{align}
   W_F &= \frac{4}{\epsilon^2} + \frac{\pi^2}{3} + \frac{4\zeta_3}{3}\,\epsilon 
    + \frac{\pi^4}{40}\,\epsilon^2  \,,\nonumber\\
   W_{2F} &= \frac{8}{\epsilon^4} + \frac{4\pi^2}{3\epsilon^2}
    + \frac{16\zeta_3}{3\epsilon} + \frac{7\pi^4}{45} \,,\nonumber\\
   W_A &= \frac{11}{3\epsilon^3} + \frac{1}{\epsilon^2}
    \left( \frac{67}{9} - \frac{\pi^2}{3} \right) 
    + \frac{1}{\epsilon} \left(\frac{404}{27}  + \frac{11\pi^2}{18} - 14\zeta_3 \right) \nonumber\\
    &\quad + \frac{2428}{81} + \frac{67\pi^2}{54} + \frac{22\zeta_3}{9}- \frac{\pi^4}{3} \,, \nonumber\\
   W_f &= -\frac{4}{3\epsilon^3} - \frac{20}{9\epsilon^2}
    + \frac{1}{\epsilon} \left( - \frac{112}{27} - \frac{2\pi^2}{9} \right) - \frac{656}{81} 
    - \frac{10\pi^2}{27} - \frac{8\zeta_3}{9} \,.
\end{align}
The expansion coefficients of the Laplace-transformed coft function read
\begin{align}
V_F &=- \frac{2}{\epsilon^2} - \frac{\pi^2}{2} - \frac{14\zeta_3}{3}\,\epsilon 
    - \frac{7\pi^4}{48}\,\epsilon^2\,,\nonumber \\
 V_{2F} &= \frac{2}{\epsilon^4} + \frac{\pi^2}{\epsilon^2} + \frac{28\zeta_3}{3\epsilon} 
    + \frac{5\pi^4}{12} \,, \nonumber\\
   V_A &= - \frac{11}{6\epsilon^3} - \frac{1}{\epsilon^2}
    \left( \frac{67}{18} + \frac{\pi^2}{6} \right) 
    + \frac{1}{\epsilon} \left(  - \frac{211}{27} - \frac{11\pi^2}{36} + 3\zeta_3 \right) \nonumber\\
    &\quad  - \frac{836}{81} - \frac{1139\pi^2}{108} + \frac{31\pi^4}{90} - \frac{341\zeta_3}{9}   \,,
    \nonumber\\
   V_f &= \frac{2}{3\epsilon^3} + \frac{10}{9\epsilon^2}
    + \frac{1}{\epsilon} \left( \frac{74}{27} + \frac{\pi^2}{9} \right)  - \frac{374}{81} + \frac{109\pi^2}{27} 
    + \frac{124\zeta_3}{9} \,.
 \end{align}   
Finally, the purely collinear two-loop contribution is given by
\begin{align}
   &\hspace{-0.2cm}J^{\rm full}_{\rm bare}(L,\epsilon) = \big\langle \bm{\mathcal{J}}_{\!\!1} \otimes \bm{1}+ \bm{\mathcal{J}}_{\!\!2} \otimes \bm{1}
    + \bm{\mathcal{J}}_{\!\!3} \otimes \bm{1} \big\rangle \nonumber\\
&=   1 + \frac{\alpha_0 C_F}{4\pi}\,e^{-2\epsilon L}
    \left[ \frac{2}{\epsilon^2} + \frac{3}{\epsilon} + 7 - \frac{5\pi^2}{6} + 6\ln2 
    + \epsilon \left(  14 - \frac{\pi^2}{4} - \frac{44\zeta_3}{3}  + 6\ln^2 2 + 14\ln 2 \right)
    \right. \nonumber\\
   & \left.+ \epsilon^2 \left( 28 - \frac{7\pi^2}{12}   - \zeta_3 + \frac{41\pi^4}{720}
    - \frac{4\ln^4 2}{3} + 4\ln^3 2 + 14\ln^2 2 + \frac{4\pi^2\ln^2 2}{3}   + 28\ln 2 - \frac{\pi^2\ln 2}{2}  \right. 
    \right. \nonumber\\
   &\quad\quad \left.\left. - 28\zeta_3 \ln 2 - 32\,\text{Li}_4\Big(\frac{1}{2}\Big)   \right) \right] 
    + \left( \frac{\alpha_0}{4\pi} \right)^2 e^{-4\epsilon L} 
    \left( C_F^2 J_F + C_F C_A J_A + C_F T_F n_f J_f \right) ,
\end{align}
with $L=\ln\frac{Q\delta}{\mu}$ and
\begin{align}\label{eq:JfullConst}
   J_F &= \frac{2}{\epsilon^4} + \frac{6}{\epsilon^3}
    + \frac{1}{\epsilon^2} \left( \frac{37}{2} - \frac{5\pi^2}{3} + 12\ln2 \right) 
    + \frac{1}{\epsilon} \left(  \frac{191}{4} - 4\pi^2 - \frac{22\zeta_3}{3}   + 50\ln 2 \right) 
    + c_2^{J,F} \,, \nonumber\\
   J_A &= \frac{11}{6\epsilon^3}
    + \frac{1}{\epsilon^2} \left( \frac{83}{9} - \frac{\pi^2}{2} \right) 
    + \frac{1}{\epsilon} \left( \frac{3985}{108} - \frac{139\pi^2}{36}  - 21\zeta_3 
    - 6\ln^2 2 + 18\ln 2 \right) + c_2^{J,A} \,, \nonumber\\
   J_f &= - \frac{2}{3\epsilon^3} - \frac{28}{9\epsilon^2} 
    + \frac{1}{\epsilon} \left( - \frac{335}{27} + \frac{11\pi^2}{9} - 8\ln 2 \right) 
    + c_2^{J,f} \,.
\end{align}
The unknown constants $c_2^{J,F}$, $c_2^{J,F}$ and $c_2^{J,f}$ in this expression are directly related to the coefficients $c_2^F$, $c_2^A$ and $c_2^f$ entering the two-loop coefficient $B(\beta,\delta)$ in \eqref{Bterms}, which we have determined numerically in \eqref{eq:twoloopcoeff}.

\section{NNLO ingredients for the wide-angle jet cross section}
\label{widejetapp}

\subsection{\boldmath One-loop hard function $\mathcal{H}_3$}
\label{widejetapp:1}

Here we describe the computation of the hard function $\bm{\mathcal{H}}_3$ and give the results for the functions $h_3^{\rm I}(u,v)$ arising in \eqref{h3eq}. The amplitude squared for $\bm{\mathcal{H}}_3$ has the form
\begin{align}\label{eq:M3}
\big| \mathcal{M}_3\big|^2  = \big| \mathcal{M}^{(0)}_2\big|^2 \, \frac{2 C_F \, g_s^2\, \tilde{\mu}^{2\e}}{Q^2}  \,\frac{x_1^2+x_2^2 - \e \, x_3^2}{(1-x_1)(1-x_2)}\,,
\end{align}
where $x_i = 2 E_i/Q$.
In region {I} the thrust axis is along the direction opposite to $p_1$. In terms of angles $\theta_{i}$ with respect to this axis, the final-state particle momenta can thus be written as
\begin{equation}\label{eq:momparam}
\begin{aligned}
p_1 & = E_1(1,0,\cdots,0, -1)\,, \\
p_2 & = E_2(1,0,\cdots,0, - \sin\theta_{2}, \cos\theta_{2})\,,\\
p_3 &= E_3(1,0,\cdots,0, \sin\theta_{3}, \cos\theta_{3})\,.
\end{aligned}
\end{equation}
The three vectors lie in a plane and we have chosen its azimuthal angle to be zero. Due to momentum conservation, only two of the remaining variables are independent and the three-body phase space
\begin{equation}\label{eq:PS3}
\int d\Pi_3 = \prod_{i=1}^3 \int\!\! \frac{d^{d-1}p_i}{2E_i (2\pi)^{d-1}}\, (2\pi)^d\, \delta^{(d)}(q-p_1-p_2-p_3)
\end{equation}
can be parametrised as
\begin{align}
\int d\Pi_3  = \int dx_3 \, d\cos\theta_{3} \left(\frac{2}{Q}\right)^{4\e} \frac{(2\pi)^{-3+2\e}\,Q^2}{16 \, \Gamma(2-2\e)} \,\frac{\left(\sin^2\theta_{3}\right)^{-\e}x_1^{1-2\e}\, x_3^{1-2\e} }{2-x_3-x_3 \cos\theta_{3}}\, ,
\end{align}
where $x_1$ must be expressed in terms of $x_3$ and $\theta_{3}$ using the relation
\begin{equation}
x_1 =\frac{1 - x_3}{1 -  x_3 \cos^2\! \left(\frac{\theta_3}{2}\right)}\,.
\end{equation}
To obtain the cross section, one integrates \eqref{eq:M3} over the phase space and multiplies by the flux factor $1/(2Q^2)$. To extract the one-loop correction to the cross section one has to divide by the $d$-dimensional Born-level cross section, which is given by $|\mathcal{M}^{(0)}_2|^2$ multiplied by the two-particle phase space.

To evaluate the phase-space integrals in region I, one changes variables to $u$ and $v$ introduced in \eqref{eq:paramOne}. In regions II and III it is convenient to instead parameterize the phase space \eqref{eq:PS3} in terms of $x_2$ and $\cos\theta_{2}$. In this case, the variables $u$ and $v$ are introduced via
\begin{equation}
x_2 = \frac{u}{2} \frac{1+\delta^2}{1+\delta^2-\delta^2v} \,, \qquad
\sin^2\!\Big(\frac{\theta_{2}}{2}\Big) =\frac{ \delta^2  \, v}{1+\delta^2} \,.
\end{equation}
Combining the phase-space measure with the matrix element \eqref{eq:M3}, the explicit expressions for the ingredients of the hard function $\bm{\mathcal{H}}_3^{(1)}$  in \eqref{h3eq} then read
\begin{align}
h_3^{\rm  I}= & \frac{1}{2} (1+\delta^2)(2-u)^{-2+2\e}\left[1+\delta^2(1-v)\right]^{3\e-3}\left[2-u+\delta^2(2-u-2v)\right]^{-2\e} \left[ u^4 (1-\epsilon ) \right. \nno \\
& \left.  +\,4 u^3 (\epsilon -2)-4 u^2 (\epsilon -7)+\delta ^4
   \left(u^4 (1-\epsilon )+u^3 (8 v+4 (\epsilon -2))  +u^2 \left(16
   v^2-40 v \right. \right.\right. \nno \\
   &\left. \left. \left. - \, 4 (\epsilon -7)\right)+u \left(-32 v^2+80 v-48\right)+32
   v^2-64 v+32\right)+\delta ^2 \left(u^4 (2-2 \epsilon ) \right. \right. \nno \\
   &\left. \left.+ \, u^3 (8 v+8
   (\epsilon -2))+u^2 (-40 v-8 (\epsilon -7))+u (80 v-96)-64
   v+64\right)-48 u+32 \right], \nno \\
 h_3^{\rm  II}=  & \frac{1}{2} (1+\delta^2)(2-u)^{-3+2\e}\left[1+\delta^2(1-v)\right]^{3\e-3}\left[2-u+\delta^2(2-u-2v)\right]^{-2\e} \left[u^4 (1-\epsilon ) \right. \nno \\
 &\left. +u^3 (8 \epsilon -4)+u^2 (8-24 \epsilon )+\delta ^4
   \left(u^4 (1-\epsilon )+u^3 (-8 v \epsilon +8 \epsilon -4)+u^2
   \left(-16 v^2 \epsilon +40 v \epsilon \right.\right. \right. \nno \\
   &\left.\left.\left. -\,24 \epsilon +8\right)+u
   \left(32 v^2 \epsilon +v (16-64 \epsilon )+32 \epsilon
   -16\right)-16 v^2 (\epsilon -1)+32 v (\epsilon -1) \right.\right.  \nno \\
   &\left.\left. -16 (\epsilon
   -1)\right)+\delta ^2 \left(u^4 (2-2 \epsilon )+u^3 (-8 v \epsilon
   +16 \epsilon -8)+u^2 (40 v \epsilon -48 \epsilon +16) \right.\right.  \nno \\
   &\left.\left. +u (v (16-64
   \epsilon )+64 \epsilon -32)+32 v (\epsilon -1)-32 (\epsilon
   -1)\right)+u (32 \epsilon -16)-16 (\epsilon -1)\right], \nno \\
 h_3^{\rm  III}=  &  (1+\delta^2)(2-u)^{-3+2\e}\left[1+\delta^2(1-v)\right]^{3\e-3}\left[2-u+\delta^2(2-u-2v)\right]^{-1-2\e}  \left[ 2 u^4-12 u^3 \right. \nno \\
 & \left. -4 u^2 (\epsilon -7)+\delta ^4 \left(2 u^4+u^3 (8
   v-12)+u^2 \left(16 v^2-40 v-4 (\epsilon -7)\right)+u \left(-32
   v^2 \right.\right. \right. \nno \\
   &\left.\left.\left. -16 v (\epsilon -4)+16 (\epsilon -2)\right)-16 v^2 (\epsilon
   -1)+32 v (\epsilon -1)-16 (\epsilon -1)\right)+\delta ^2 \left(4
   u^4 \right.\right. \nno \\
   &\left.\left. +u^3 (8 v-24)+u^2 (-40 v-8 (\epsilon -7))+u (32 (\epsilon -2)-16
   v (\epsilon -4))+32 v (\epsilon -1) \right.\right.  \nno \\
   &\left.\left. -32 (\epsilon -1)\right)+16 u
   (\epsilon -2)-16 (\epsilon -1) \right] .
\end{align}

\subsection{\boldmath Two-loop soft function $\mathcal{S}_2$ }
\label{widejetapp:2}

In order not to be scaleless, diagrams contributing to the soft function must involve at least one gluon outside the jet cone. Up to NNLO, the soft function $\mathcal{S}_2$ thus receives three types of non-vanishing contributions,
\begin{align}
\mathcal{S}_2^{(2)} = \mathcal{S}_2^{\rm R-V} + \mathcal{S}_2^{\rm In-Out} + \mathcal{S}_2^{\rm Out-Out}\,.
\end{align}
The first term $ \mathcal{S}_2^{\rm R-V}$ represents the real-virtual contributions, in which a gluon is radiated off a one-loop virtual correction. The other two terms involve two real gluons, and following \cite{Kelley:2011aa} we split them into ``In-Out" and ``Out-Out" configurations with one and two gluons outside the jet cone.

The real-virtual corrections $ \mathcal{S}_2^{\rm R-V}$ can be written in terms of the one-loop soft function $ \mathcal{S}_2^{(1)}(Q\beta,\e)$. Explicitly, we have 
\begin{align}
\mathcal{S}_2^{\rm R-V}(Q\beta,\e) &=  C_A \, \mathcal{S}_2^{(1)}(Q\beta,\e)|_{\e \to 2\e}\, 
\frac{\pi \, \Gamma(2+\e) \, \Gamma(1-2\e) \cot(\pi\e)\, \Gamma(-\e)^2}{\Gamma(1-\e) \Gamma(-2\e)(1+\e)}\,.
\end{align}
The remaining two contributions can be extracted from \cite{Kelley:2011aa}, where the two-loop soft function for jet thrust was computed.  This quantity is defined as
\begin{align}
S_R(k_L,k_R,\lambda) = \frac{1}{N_c} \sum_{X_s}\, \langle 0 |\, S_{\bar n}\, S_{n}\, \mathcal{M}(k_L,k_R,\lambda) \,| X_s \rangle\, \langle X_s | \,S_{ n}^\dag\, S_{\bar n}^\dag\, | 0 \rangle \,,
\end{align}
where the measurement function is given by
\begin{align}
\mathcal{M}(k_L,k_R,\lambda) \,| X_s \rangle =  \delta(k_L - {\bar n}\cdot p^L_{X_s}) \,\delta(k_R - n\cdot p^R_{X_s}) \,\delta(\lambda -E^{\rm out}_{X_s})  \,| X_s \rangle \,.
\end{align}
The soft radiation inside the right (left) jet has momentum $P^R_{X_s}$ ($P^L_{X_s}$), and $E^{\rm out}_{X_s}$ represents the total out-of-jet energy. Therefore, after integration over $k_L$, $k_R$ and $\lambda$ our two-loop function is obtained. Explicitly, we find
\begin{align}
\mathcal{S}_2^{\rm In(Out)-Out}(Q\beta,\e) &= \int_{0}^{Q\beta/2}\!d\lambda \int_0^{\infty}\! dk_L \int_0^{\infty}\! dk_R \, S_{R}^{r_3(r_4)}(k_L,k_R,\lambda,\mu)\,,
\end{align}
where $S_{R}^{r_3 (r_4)}$ is given in \cite{Kelley:2011aa}. When  performing the integral for $\mathcal{S}_2^{\rm In-Out}$, we need to split the $k_{L(R)}$ integration at $k_{L(R)}=Q\beta$. Combining all contributions, we obtain for the two-loop coefficients defined in (\ref{widejet_s2})
\begin{align}
s_A(\delta,\e) = & \frac{1}{\e^2}\left[ -4\, \text{Li}_2\!\left(\delta^4\right)+16\, \text{Li}_2(\delta^2)+8\,
   \text{Li}_2\!\left(\frac{1}{1+\delta^2}\right)-8\, \text{Li}_2\!\left(\frac{\delta^2}{1+\delta^2}\right)-16\, \ln
   (1+\delta^2) \ln \delta \right.\nno \\
   & \hspace{-1cm} \left. +\frac{44 \ln\delta}{3}-2 \, \pi ^2 \right] + \frac{1}{\e} \left[ -\frac{88\, \text{Li}_2\!\left(\delta^4\right)}{3}+4 \, \text{Li}_3\!\left(\delta^4\right)-8\,
   \text{Li}_3\!\left(\frac{\delta^4}{\delta^4-1}\right)+8 \, \text{Li}_2\!\left(\delta^4\right) \ln (1-\delta^2) \right. \nno \\
   & \hspace{-1cm}\left.-\,8\,
   \text{Li}_2\!\left(\delta^4\right) \ln \delta -8\,  \text{Li}_2\!\left(\delta^4\right) \ln
   (1+\delta^2)+\frac{88 \, \text{Li}_2(\delta^2)}{3}-32\, \text{Li}_3(1-\delta^2)-32\,
   \text{Li}_3\!\left(\frac{\delta^2}{1+\delta^2}\right) \right. \nno \\
   & \hspace{-1cm}\left. -\,32\, \text{Li}_2(\delta^2) \ln (1-\delta^2)-32\, \text{Li}_2(\delta^2)
   \ln \delta +32\, \text{Li}_2(\delta^2) \ln (1+\delta^2)-32\, \text{Li}_2\!\left(\frac{1}{1+\delta^2}\right) \ln\delta \right. \nno \\
   &\hspace{-1cm}\left. +32 \, \text{Li}_2\!\left(\frac{1}{1+\delta^2}\right) \ln (1+\delta^2)+32\,
   \text{Li}_2\!\left(\frac{\delta^2}{1+\delta^2}\right) \ln \delta-32\,
   \text{Li}_2\!\left(\frac{\delta^2}{1+\delta^2}\right) \ln (1+\delta^2)\right. \nno \\
   &\hspace{-1cm}\left.+\frac{2\left(1+\delta^4\right)}{3 \left(\delta^4-1\right)}-\frac{16\, \delta^4 \ln\delta}{3
   \left(\delta^4-1\right)^2}+\frac{4}{3} \ln ^3(1-\delta^2)+\frac{20}{3} \ln ^3(1+\delta^2)-32 \ln \delta
   \ln ^2(1-\delta^2) \right. \nno \\
   &\hspace{-1cm}\left.+4 \ln (1+\delta^2) \ln ^2(1-\delta^2)+4 \ln ^2(1+\delta^2) \ln (1-\delta^2)-\frac{88 \ln
   ^2\delta}{3}-64 \ln \delta \ln ^2(1+\delta^2) \right. \nno \\ 
   & \hspace{-1cm}\left. +64 \ln ^2\delta \ln (1+\delta^2) -\frac{88}{3} \ln \delta
   \ln (1-\delta^2)+\frac{16}{3} \pi ^2 \ln (1-\delta^2)+\frac{20}{3} \pi ^2 \ln \delta+\frac{268
   \ln \delta}{9} \right. \nno \\ 
   &\hspace{-1cm}\left.-\frac{88}{3} \ln \delta \ln (1+\delta^2)-\frac{16}{3} \pi ^2 \ln (1+\delta^2)+24\,
   \zeta_3 \right] + s_2^A(\delta)\,, \nno\\
s_f(\delta,\e) = & -\frac{16\,\ln \delta}{3\,\e^2} + \frac{1}{\e}\left[ \frac{32 \, \text{Li}_2\!\left(\delta^4\right)}{3}-\frac{32 \, \text{Li}_2(\delta^2)}{3}+\frac{4 \,\delta^4}{3
   \left(1-\delta^4\right)}+\frac{4}{3 \left(1-\delta^4\right)}+\frac{32\, \delta^4 \ln \delta}{3
   \left(1-\delta^4\right)^2} \right. \nno \\
   & \hspace{-1cm}\left.  +\frac{32}{3} \ln \delta \ln \left(1-\delta^4\right)+\frac{32 \ln
   ^2\delta}{3}-\frac{80 \ln \delta}{9} \right]+ s_2^f(\delta)\,.
\end{align}

\subsection{\boldmath One-loop soft function $\mathcal{S}_3$ \label{sec:S3}}

In terms of $L_s=\ln\frac{Q\beta}{\mu}$, the one-loop soft function $\bm{\mathcal{S}}_3$ takes the form
\begin{align}
 \bm{\mathcal{S}}_{3}(u,v,Q \beta,\delta,\e)  = e^{-2\e L_s} \left[ C_F\, s_3^F(u,v,\delta,\e) + C_A\, s_3^A(u,v,\delta,\e) \right]  \bm{1}\, . 
\end{align}
In what follows, we present its expression in regions I, II and III, respectively. In region~I the hard function $\bm{\mathcal{H}}_3$ suffers from double poles, so that $\bm{\mathcal{S}}_3$ need to be expanded up to $\mathcal{O}(\e)$ terms in order to be able to obtain all divergent contributions to the cross section, which are related to the logarithms we are interested in. We write the expansion in the form
\begin{equation}
\begin{aligned}
   s_3^F &= \frac{1}{\e}\,F^{[-1]}_{\rm I}(u,v,\delta) + F^{[0]}_{\rm I}(u,v,\delta)
    + \e\,F^{[1]}_{\rm I}(u,v,\delta) \,, \\
   s_3^A &= \frac{1}{\e}\,A^{[-1]}_{\rm I}(u,v,\delta) + A^{[0]}_{\rm I}(u,v,\delta)
    + \e\,A^{[1]}_{\rm I}(u,v,\delta) \,,
\end{aligned}
\end{equation}
with
\begin{align}
F^{[-1]}_{\rm I}(u,v,\delta)= & \, 8\ln\delta + 2\ln\left[ (1+\bar{u})^2(1+\delta^2\bar{v}) - u^2 v \right] - 2\ln\left[ (1+\bar{u})^2(1+\delta^2\bar{v})-\delta^4 u^2 v \right], \nno\\
F^{[0]}_{\rm I}(u,0,\delta)= & \, F^{[0]}_{\rm I}(0,v,\delta)= - \frac{2\pi^2}{3} - 8\ln^2\delta + 4\ln^2(1+\delta^2) + 8 {\rm Li}_2\left( \frac{\delta^2}{1+\delta^2}\right), \nno \\
F^{[1]}_{\rm I}(0,0,\delta)=& \, \frac{16}{3}\ln^3\delta -\ln\delta\left[ \frac{2\pi^2}{3} + 16\ln^2(1+\delta^2) + 16\,{\rm Li}_2\left( \frac{\delta^2}{1+\delta^2}\right) \right] + 8 \ln ^3\left(1+\delta ^2\right) \nno\\
& -\frac{4}{3} \pi ^2 \ln \left(1+\delta
   ^2\right)-8 \text{Li}_3\!\left(\frac{1}{1+\delta ^2}\right)+8
   \text{Li}_3\!\left(\frac{\delta ^2}{1+\delta ^2}\right) \nno \\
   &+\,16 \ln \left(1+\delta
   ^2\right) \text{Li}_2\!\left(\frac{\delta ^2}{1+\delta ^2}\right)
 \end{align} 
 and
 \begin{align} 
A^{[-1]}_{\rm I}(u,v,\delta)= & \, 2\ln(1-v) - 2\ln(1-\delta^2v) - 2\ln\left[ 2 - u\, \bar{v} + \delta^2\left(1+\bar{u}\right)\bar{v} \right]  \nno \\
& + 2\ln\left[ (1+\bar{u})(1+\delta^2\bar{v})+\delta^4 u \,v \right], \nno \\
A^{[0]}_{\rm I}(0,v,\delta)= & \ln\delta\left[ 2 \ln \left(1-\delta ^2\right)+8 \ln \left(1+\delta ^2\right)+2 \ln \left(1-\delta
   ^2 v\right)-8 \ln \left(1+\delta ^2 \bar{v}\right)-4 \ln \bar{v}   \right]  \nno \\
   & + 3 \ln ^2\left(1-\delta ^2 v\right)-\ln v\ln \left(1-\delta ^2 v\right)-\ln
   \left(1-\delta ^2\right) \ln \left(1-\delta ^2 v\right) \nno \\
   &  -\ln\bar{v} \ln
   \left(1-\delta ^2 v\right) -4 \ln \left(1-\delta ^2 v\right) \ln \left(1+\delta ^2
   \bar{v}\right)-\ln v \ln \bar{v}+\ln ^2 v \nno\\
   & +2 \ln ^2\left(1+\delta ^2
   \bar{v}\right)+\ln \left(1-\delta ^2\right) \ln \bar{v}  -2 \ln^2\bar{v} + 8 \text{Li}_2\!\left(-\delta ^2\right)+\text{Li}_2\!\left(\delta
   ^2\right)\nno \\
   &-\text{Li}_2\!\left( \delta ^2v\right)+\text{Li}_2\!\left(\frac{1-\delta ^2}{1-
   \delta ^2v}\right)+\text{Li}_2\!\left(\frac{\bar v}{1-\delta ^2v}\right) +2 \,
   \text{Li}_2\!\left(-\frac{\bar v}{v}\right)-\text{Li}_2(v) \nno \\
   &-4\,
   \text{Li}_2\!\left(- \delta ^2 \bar{v}\right)+4 \, \text{Li}_2\!\left(\frac{\delta
   ^2}{1+ \delta ^2\bar{v}}\right), \nno\\
   A^{[0]}_{\rm I}(u,0,\delta)=& 0\,, \nno \\
   A^{[1]}_{\rm I}(0,0,\delta)=& 0\,.  
\end{align}
In region II we only need the results up to $\mathcal{O}(1)$. They are  
\begin{align}
F^{[-1]}_{\rm II}(u,v,\delta) = & \,8\ln\delta + 2\ln \bar v - 2\ln(1-\delta^2v)\,, \nno\\
F^{[0]}_{\rm II}(u,0,\delta) = &  - \frac{2\pi^2}{3} - 8\ln^2\delta + 4\ln^2(1+\delta^2) + 8 {\rm Li}_2\left( \frac{\delta^2}{1+\delta^2}\right),\nno\\
A^{[-1]}_{\rm II}(u,v,\delta) = & -2 \ln\left[(1+\bar u)^2 \left(1+\delta ^2 \bar v\right)-\delta ^4 u^2 v\right]+2 \ln \left[(1+\bar u)^2 \left(1+\delta ^2
   \bar v \right) - u^2 v\right] \nno \\
   & +2 \ln \left[(1+\bar u) \left(1+\delta ^2 \bar v\right) +\delta ^4 u \, v\right]-2 \ln \left[2-u {\bar v}+\delta ^2 (1+\bar{u}){\bar v}\right], \nno\\
A^{[0]}_{\rm II}(u,0,\delta) = & \, 0\,.
\end{align}
Finally, in region III only the divergent terms are needed, and we find
\begin{align}
F^{(-1)}_{\rm III}(u,v,\delta) =&2 \ln \left[(1+\bar u)^2 \left(1+\delta ^2 \bar v \right)-u^2 v\right]-2 \ln
   \left[(1+\bar u )^2 \left(1+\delta ^2 \bar{v}\right)-\delta ^4 u^2 v\right] \nno\\
   & +4\ln \left[(1+\bar u) \left(1+\delta ^2 \bar v\right) + \delta ^4 u v\right]-4
   \ln \left[2-u {\bar v}+\delta ^2 (1+\bar u) \bar v\right] +2 \ln (1-v) \nno \\
   & -2 \ln\left(1-\delta ^2 v\right), \nno\\
A^{(-1)}_{\rm III}(u,v,\delta) =&  \, 8\ln\delta  + 2\ln\left[ 2 - u\, \bar{v} + \delta^2\left(1+\bar{u}\right)\bar{v} \right]  - 2\ln\left[ (1+\bar{u})(1+\delta^2\bar{v})+\delta^4 u \,v \right].
\end{align}
In the above expressions we have defined $\bar u=1-u$ and $\bar v=1-v$.

\subsection{\boldmath Two-loop hard function $\mathcal{H}_3$ and $\mathcal{H}_4$ \label{sec:H34}}

Here we list the two-loop coefficients for the hard function $\langle \bm{\mathcal{H}}^{(2)}_3 \otimes \bm{1} + \bm{\mathcal{H}}^{(2)}_4 \otimes \bm{1} \rangle$ defined in \eqref{eq:H34}. They read
\begin{align}
h_F = & - \frac{8}{\e^4} -\frac{24}{\e^3} +\frac{1}{\e^2} \left[  \phantom{\frac{1}{2}} \!\!\!\!  -16 \ln ^2\left(1+\delta ^2\right)+32 \ln ^2\delta +32\,
   \text{Li}_2\!\left(\delta ^4\right)-96\, \text{Li}_2\!\left(-\delta
   ^2\right)-64 \, \text{Li}_2\!\left(\delta ^2\right) \right. \nno \\
   & \left. - \, 32 \,
   \text{Li}_2\!\left(\frac{\delta ^2}{1+\delta ^2}\right)+\frac{28 \pi
   ^2}{3}-82\right] + \frac{1}{\e} \left[32 \ln ^3\left(1+\delta ^2\right)-64 \ln ^2\left(1+\delta ^2\right)
   \ln \delta \right. \nno \\
   & \left.  - \, 32 \ln 2 \ln ^2\left(1+\delta ^2\right)+24 \ln
   ^2\left(1+\delta ^2\right)-\frac{16}{3} \pi ^2 \ln \left(1+\delta
   ^2\right)-\frac{128}{3} \ln ^3\delta +64 \ln 2 \ln
   ^2\delta  \right. \nno \\
   & \left. + \, 48 \ln ^2\delta +64 \ln ^2 2 \ln \delta -96
   \ln 2 \ln \delta -\frac{56}{3} \pi ^2 \ln \delta +128 \ln
   \delta +48 \, \text{Li}_2\!\left(\frac{\delta ^2}{1+\delta
   ^2}\right)  \right. \nno \\
   & \left. -\,32\, \text{Li}_3\!\left(\frac{1}{1+\delta ^2}\right)+32\,
   \text{Li}_3\!\left(\frac{\delta ^2}{1+\delta ^2}\right)-64 \ln
   \delta \,  \text{Li}_2\!\left(\frac{\delta ^2}{1+\delta ^2}\right)  \right. \nno \\
   & \left. +\,64
   \ln \left(1+\delta ^2\right) \text{Li}_2\!\left(\frac{\delta
   ^2}{1+\delta ^2}\right)-64 \ln 2 \, \text{Li}_2\!\left(\frac{\delta
   ^2}{1+\delta ^2}\right)+\frac{184\, \zeta_3}{3}+22\, \pi
   ^2-\frac{445}{2}  \right. \nno \\
   & \left. +\frac{16}{3} \pi ^2 \ln 2 - M_F^{[1]}(\delta)  \right]  + h_2^F(\delta),  \nno \\
h_A = & \, \frac{11}{3\,\e^3} + \frac{1}{\e^2}\left[ -\frac{44 \ln \delta }{3}-4 \, \text{Li}_2\!\left(\delta ^4\right)+16 \,
   \text{Li}_2\!\left(-\delta ^2\right)+16 \, \text{Li}_2\!\left(\delta
   ^2\right)-\pi ^2+\frac{166}{9}\right]   + \frac{1}{\e}\left[ \phantom{\frac{1}{2}} \!\!\!\!   \, 33 \, \delta ^4 \right. \nno \\
   & \left. +\,\frac{4}{3 \left(1-\delta ^4\right)} - 44\, \delta ^4 \ln
   2-\frac{16 \ln \delta }{3 \left(1-\delta
   ^4\right)}+\frac{16 \ln \delta }{3 \left(1-\delta
   ^4\right)^2}-44\, \delta ^2-\frac{4}{3} \ln ^3\left(1-\delta
   ^2\right)  \right. \nno \\
   & \left.  -\,\frac{20}{3} \ln ^3\left(1+\delta ^2\right)+32 \ln
   \delta  \ln ^2\left(1-\delta ^2\right)-4 \ln \left(1-\delta
   ^2\right) \ln ^2\left(1+\delta ^2\right)  \right. \nno \\
   & \left. -\,4 \ln ^2\left(1-\delta
   ^2\right) \ln \left(1+\delta ^2\right)+\frac{88}{3} \ln \delta 
   \ln \left(1-\delta ^2\right)-\frac{16}{3} \pi ^2 \ln
   \left(1-\delta ^2\right)+\frac{88}{3} \ln \delta  \ln
   \left(1+\delta ^2\right)  \right. \nno \\
   & \left. +\,\frac{88 \ln ^2\delta
   }{3}+\frac{4}{3} \pi ^2 \ln \delta -\frac{664 \ln \delta
   }{9}+\frac{44 \, \text{Li}_2\!\left(\delta ^4\right)}{3}-4 \,
   \text{Li}_3\!\left(\delta ^4\right)+8 \, \text{Li}_3\!\left(\frac{\delta
   ^4}{-1+\delta ^4}\right)  \right. \nno \\
   & \left. +\,24 \ln \delta \,  \text{Li}_2\!\left(\delta
   ^4\right)+16 \ln 2 \, \text{Li}_2\!\left(\delta ^4\right)+\frac{88 \,
   \text{Li}_2\!\left(\delta ^2\right)}{3}+32 \, \text{Li}_3\!\left(1-\delta
   ^2\right)+32 \, \text{Li}_3\!\left(\frac{\delta ^2}{1+\delta
   ^2}\right)  \right. \nno \\
   & \left. -\,32 \ln \delta\,  \text{Li}_2\!\left(\delta ^2\right)+32
   \ln \left(1-\delta ^2\right) \text{Li}_2\!\left(\delta ^2\right)+32
   \ln \left(1+\delta ^2\right) \text{Li}_2\!\left(\delta
   ^2\right) \right. \nno \\
   & \left.  -\,8 \ln \left(1-\delta ^2\right)
   \text{Li}_2\!\left(\delta ^4\right)-24 \ln \left(1+\delta ^2\right)
   \text{Li}_2\!\left(\delta ^4\right)-50 \, \zeta_3 - \frac{121 \pi
   ^2}{18}+\frac{3697}{54}-\frac{8}{3} \pi ^2 \ln 2  \right. \nno \\
   & \left. +\,44 \ln 2  - M_A^{[1]}(\delta)  \phantom{\frac{1}{2}} \!\!\!\! \right] 
    + h_2^A(\delta) \,, \nno \\
h_f = &  -\frac{4}{3\e^3} + \frac{1}{\e^2}\left( \frac{16}{3}\ln\delta - \frac{56}{9} \right) + \frac{1}{\e}\left[\, 12 \left(1-\delta ^4\right) - \frac{8}{3 \left(1-\delta
   ^4\right)} + \frac{32 \ln \delta }{3 \left(1-\delta
   ^4\right)}-\frac{32 \ln \delta }{3 \left(1-\delta ^4\right)^2} \right. \nno \\
   & \left. - \, 16
   \left(1-\delta ^4\right) \ln 2 + 16 \, \delta ^2-\frac{32}{3} \ln
   \delta  \ln \left(1-\delta ^2\right)-\frac{32}{3} \ln \delta 
   \ln \left(1+\delta ^2\right) -\frac{32 \ln ^2\delta }{3} +\frac{224 \ln \delta
   }{9}\right. \nno \\
   & \left. -\frac{16 \, \text{Li}_2\!\left(\delta ^4\right)}{3}-\frac{32\,
   \text{Li}_2\!\left(\delta ^2\right)}{3}+\frac{22\, \pi
   ^2}{9}-\frac{922}{27}   \phantom{\frac{1}{2}} \!\!\!  \right]+ h_2^f(\delta)\,. 
\end{align}

\section{One-loop renormalization for the narrow-jet cross section \label{sec:oneloopRenNarrow}}

For the narrow-jet case, the one-loop finiteness condition has the form
\begin{align}\label{WAfin}
\frac{1}{2}\,H^{(1)} + \frac{1}{2}\,\widetilde{S}^{(1)}+ \big\langle\bm{z}_{m,m}^{(1)} + \bm{z}_{m,m+1}^{(1)} + \widetilde{\bm{\mathcal{U}}}_{m}^{(1)}\big\rangle = {\rm finite}\,,
\end{align}
where the divergent parts of the hard and soft functions are given by 
\begin{equation}
\begin{aligned}
   \frac{1}{2}\,H^{(1)}(Q,\e) &= C_F \left( -\frac{2}{\e^2} - \frac{3}{\e}
    + \frac{4}{\e} \ln\frac{Q}{\mu} \right) , \\
   \frac{1}{2}\,\widetilde{S}^{(1)}(Q\tau,\e) 
   &= C_F \left( \frac{2}{\e^2} - \frac{4}{\e} \ln\frac{Q\tau}{\mu} \right) .
\end{aligned}
\end{equation}

\begin{figure}[t!]
\centering
\hspace{-0.0cm}
\begin{overpic}[scale=0.42]{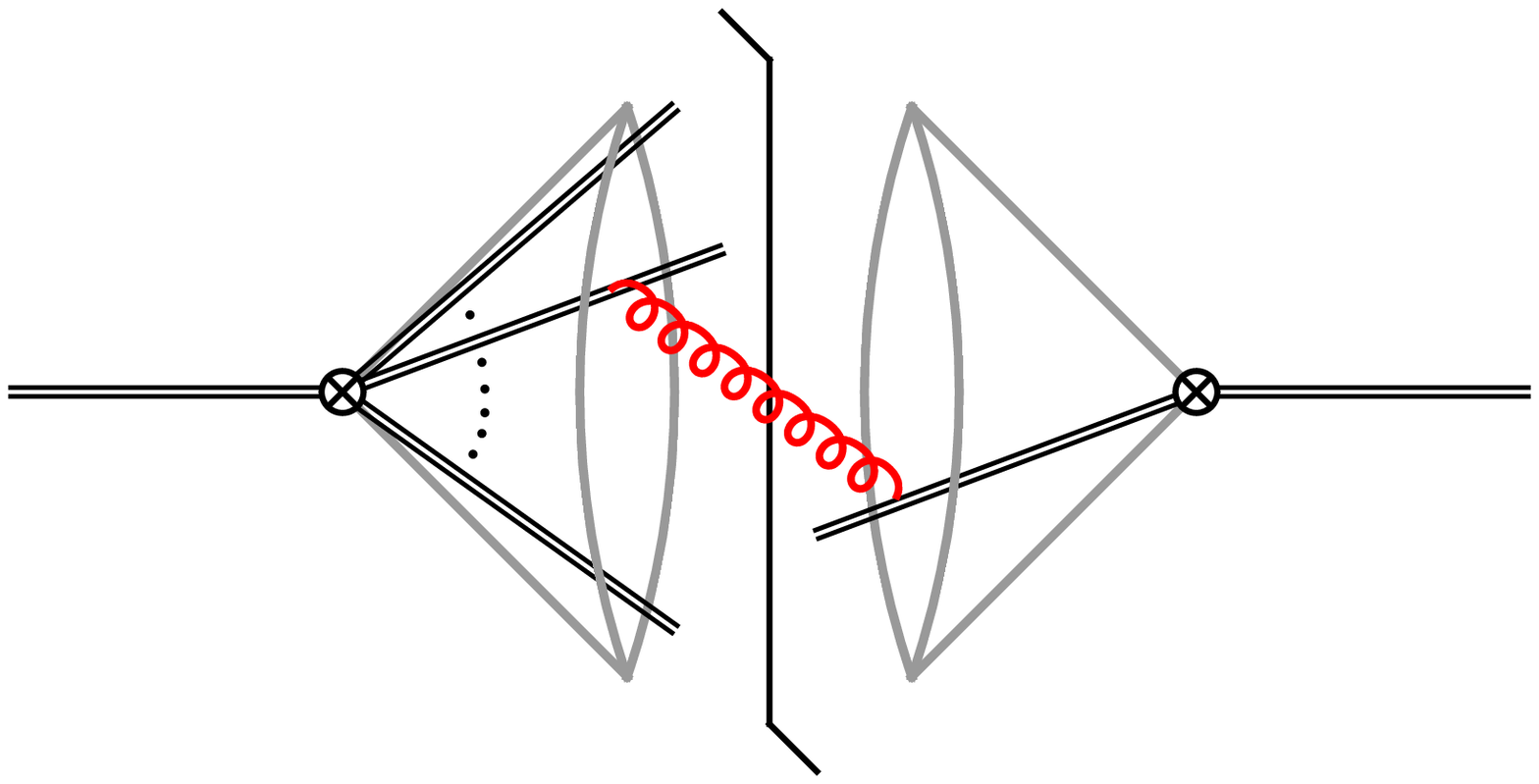}
\put(0,30){$ \bar{n}$}
\put(44,45){$ n_1$}
\put(44,37){$ n_i$}
\put(42,5){$ n_m$}
\put(52,10){$ n_j$}
\put(95,30){$ \bar{n}$}
\end{overpic}
\hspace{0.2cm}
\begin{overpic}[scale=0.42]{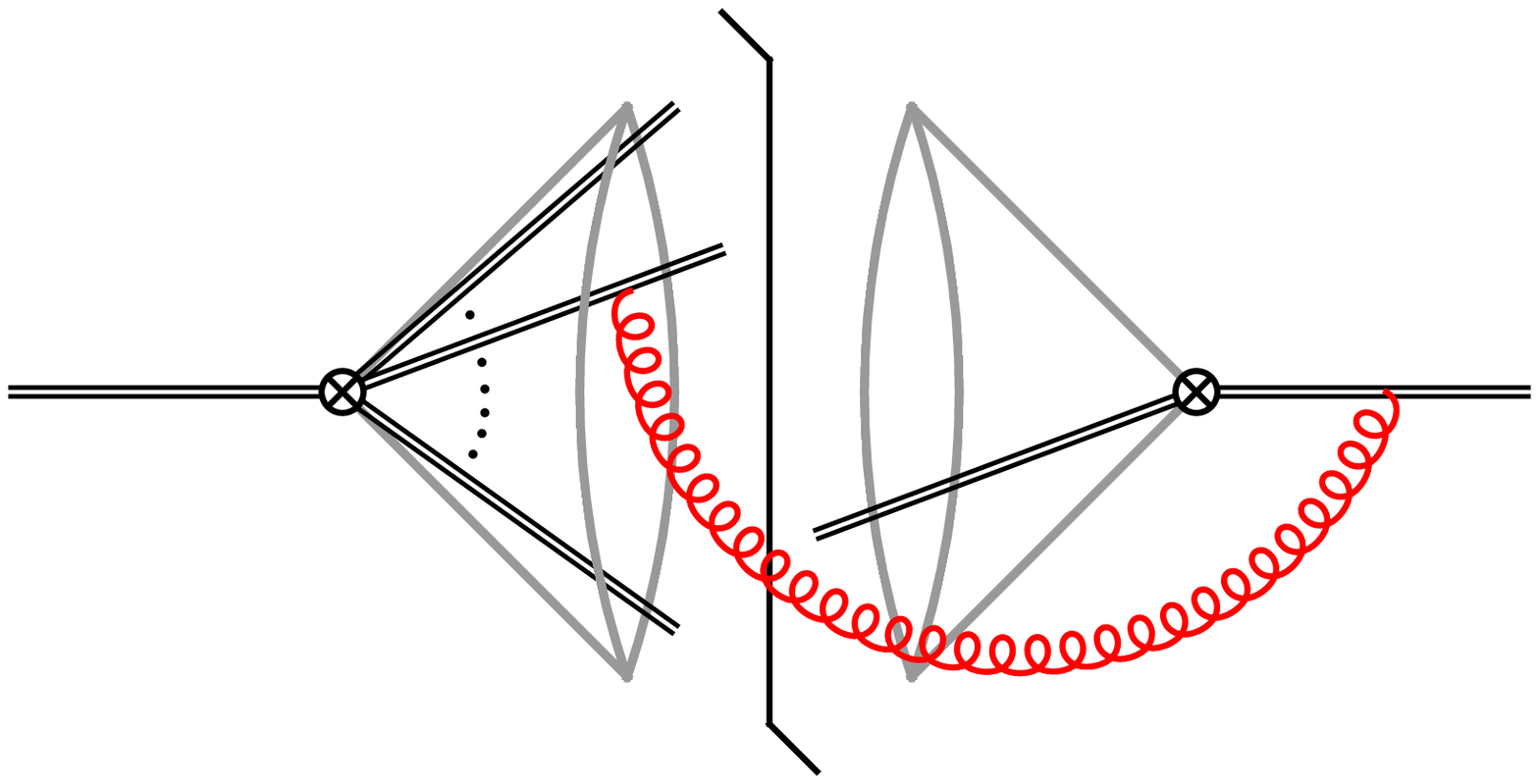}
\end{overpic}
\caption{Sample Feynman diagrams for general one-loop coft function $\bm{\mathcal{U}}_m$.
\label{fig:U_l}}
\end{figure}

The one-loop coft function is obtained as a sum of exchange diagrams shown in Figure~\ref{fig:U_l}. We can distinguish two types of contributions: (i) exchanges between two Wilson lines inside the jet shown on the left side of the figure, and (ii) exchanges between an internal Wilson line and the $\bar{n}$ Wilson line along the other jet, displayed on the right side. Contributions of the first kind only involve a single divergence, which comes from the energy integration, while the second type suffers from a double divergence, arising in the angular integration when the gluon becomes collinear to the $\bar{n}$ direction. Explicitly, the one-loop coft function reads
\begin{align}
\bm{\mathcal{U}}_{m} (\{\underline{n}\},\e) =& \, \bm{1}  + \frac{\as}{2\pi\e} \sum_{[ij]} \bm{T}_i \cdot\bm{T}_j \int\frac{d\Omega(n_k)}{4\pi}\,W_{ij}^k \, \Theta_{\rm out}^{n}(n_k) \nno \\
&-g_{s}^2 {\tilde \mu}^{2\e} \sum_{i=1}^m \left[ \bm{T}_0 \cdot\bm{T}_i + \bm{T}_i \cdot\bm{T}_0 \right] \int \frac{d^{D-1}k}{(2\pi)^{D-1} E_k^3}\,W_{i0}^k \, \Theta^{n}_{\rm out}(n_k)\,\theta(Q\beta - \bar{n}\cdot k)\,,
\end{align}
where we have extracted the single divergence in the energy integral in the first line. In the second line, we have separated out the terms involving $n_0=\bar{n}$ in the sum in \eqref{eq:UmNLO}. Here the notation $[ij]$ indicates unordered pairs with $i,j>0$. The dipole in the second line is
\begin{align}
W_{i0}^k = \frac{\bar{n} \cdot n_i}{\bar{n}\cdot n_k \, n_k \cdot n_i}\,.
\end{align}
Evaluating this expression in the narrow angle limit, one obtains the divergent terms as
\begin{align}\label{Ul}
\widetilde{\bm{\mathcal{U}}}_{m}^{(1)}(\{\underline{n}\},\e) =&\, - \frac{1}{\e} \sum_{[ij]} \bm{T}_i \cdot\bm{T}_j \left[\, \ln(1 - \hat{\theta}_i^2) + \ln(1 - \hat{\theta}_j^2) 
- \ln\big(1 -2 \cos\Delta\phi_{ij} \,\hat \theta_i  \hat \theta_j  + \hat \theta_i^2 \hat \theta_j^2 \big)\right] \nno \\
& - \frac{2}{\e} \sum_{i=1}^m \bm{T}_0 \cdot\bm{T}_i  \ln(1 - \hat{\theta}_i^2) + \bm{T}_0 \cdot\bm{T}_0 \left( - \frac{2}{\e^2} + \frac{4}{\e} \, \ln\frac{Q\tau\delta}{\mu}\right),
\end{align}
where we normalize the polar angles as $\hat \theta_i = \theta_i/(2 \delta)$. Alternatively, we can use the relation (\ref{facHS}) to derive the divergent part of one-loop coft function $\widetilde{\bm{\mathcal{U}}}_{m}^{(1)}$. Specifically, we use the factorization equation
\begin{align}
\lim_{\delta\to0}\widetilde{\bm{\mathcal{S}}}_{1+m}(\{\bar{n}, \underline{n}\}, Q\tau,\delta) &= \widetilde{S}(Q\tau)\cdot  \,\widetilde{\bm{\mathcal{U}}}_m(\{\underline{n}\}, Q\delta\tau) \cdot \,\widetilde{\overline{\bm{\mathcal{U}}}}_1(\{\bar{n}\},Q\delta\tau)
\end{align} 
and then calculate the one-loop divergent terms for $\bm{S}^{(1)}_{1+m}$ in the small angle limit $\delta\to0$.  After subtracting the divergent parts for the one-loop function $S^{(1)}$ and $ \overline{\bm{\mathcal{U}}}_1^{(1)}$, we get the same results as (\ref{Ul}). 

Similarly, starting from the factorization expression for the hard function in  (\ref{facHS}),
\begin{align}
\lim_{\delta\to0}\bm{\mathcal{H}}_{1+m}( \{\bar n,\underline{n}\},Q,\delta) &= \bm{\mathcal{H}}_2(Q)\cdot  \bm{\mathcal{J}}_m(\{\underline{n}\}) \cdot \overline{\bm{\mathcal{J}}}_1(Q\delta, \{\bar n\})\,,
\end{align} 
one can calculate the integral \eqref{ZHint} in small-angle limit and obtain the one-loop $Z$-factor for the jet function as
\begin{align}
\bm{z}_{m,m}^{(1)} + \bm{z}_{m,m+1}^{(1)}  =&\,  -  \frac{2}{\e}  \sum_{(ij)} \bm{T}_i \cdot\bm{T}_j \int\frac{d\Omega(n_k)}{4\pi}\, W_{ij}^k \, \Theta^{n\bar{n}}_{\rm out}(n_k) - H^{(1)}\cdot  \bm{1} -\bm{z}_{1,1}^{(1)} - \bm{z}_{1,2}^{(1)}\,.
\end{align}
After integrating and expanding in the narrow-angle limit, we have
\begin{align}
\bm{z}_{m,m}^{(1)} + \bm{z}_{m,m+1}^{(1)} = &\, \frac{1}{\e} \sum_{[ij]} \bm{T}_i \cdot\bm{T}_j 
\left[ \ln(1 - \hat{\theta}_i^2) + \ln(1 - \hat \theta_j^2) - \ln\big(1 -2 \cos\Delta\phi_{ij} \,\hat\theta_i \hat\theta_j  + \hat\theta_i^2 \hat\theta_j^2 \big)\right] \nno \\
& + \frac{2}{\e} \sum_{i=1}^m \bm{T}_0 \cdot\bm{T}_i  \ln(1 - \hat \theta_i^2) + \bm{T}_0 \cdot\bm{T}_0 \left( \frac{2}{\e^2} + \frac{3}{\e}- \frac{4}{\e} \ln\frac{Q\delta}{\mu} \right).
\end{align}
Using these results, we can immediately verify that all divergences in \eqref{WAfin} cancel out.

\end{document}